\documentclass[10pt,twocolumn,letterpaper]{article}

\usepackage{iccv}              %

\newcommand{\genfacelength}{0.195} %
\newcommand{\gencolPelength}{0.15} %
\newcommand{\gencolimglength}{0.85} %

\usepackage[utf8]{inputenc}
\usepackage[T1]{fontenc}

\usepackage{cuted}
\usepackage{listings}
\usepackage{longtable}
\usepackage{xcolor}

\definecolor{vscBackground}{RGB}{255,255,255}   %
\definecolor{vscKeyword}{RGB}{0,0,255}         %
\definecolor{vscString}{RGB}{163,21,21}        %
\definecolor{vscComment}{RGB}{0,128,0}         %
\definecolor{vscClass}{RGB}{128,0,128}         %
\definecolor{vscFunction}{RGB}{128,0,128}      %
\definecolor{vscNumber}{RGB}{255,102,0}        %
\definecolor{vscOperator}{RGB}{0,0,0}          %
\definecolor{vscBorder}{RGB}{200,200,200}      %

\usepackage{microtype}
\usepackage{graphicx}
\usepackage{subcaption}
\usepackage{booktabs}
\usepackage{tabularx}
\usepackage{natbib}
\usepackage{balance}
\usepackage{colortbl}
\usepackage[column=0]{cellspace}
\newcolumntype{P}[1]{>{\centering\vspace*{\fill}}p{#1}<{\vspace*{\fill}}}

\usepackage{amsmath}
\usepackage{amssymb}
\usepackage{multirow}
\usepackage{mathtools}
\usepackage{amsthm}
\usepackage{bm}
\usepackage{xspace}
\usepackage{pgf} %
\usepackage{tikz}
\usepackage{graphicx} %
\usetikzlibrary{positioning,arrows.meta, matrix}
\usepackage{ragged2e,array}

\usepackage{float}
\usepackage{placeins} %
\usepackage{stfloats}

\usepackage{array, booktabs}
\usepackage[demo,
            export]{adjustbox}  %
\usepackage[skip=1ex]{caption}

\usepackage{algorithm}
\usepackage{algpseudocode}
\usepackage{adjustbox}

\newcommand{\NN}{\boldsymbol\Psi}

\definecolor{iccvblue}{rgb}{0.21,0.49,0.74}
\usepackage[pagebackref,breaklinks,colorlinks,allcolors=iccvblue]{hyperref}

\usepackage{xcolor}

\usepackage{comment}
\usepackage{booktabs}
\usepackage{tabularx}
\usepackage{array}
\newcolumntype{Y}{>{\centering\arraybackslash}X}
\usepackage{flushend}

\def\paperID{10125} %
\def\confName{ICCV}
\def\confYear{2025}

\title{Beyond Blur: A Fluid Perspective on Generative Diffusion Models}

\author{
Grzegorz Gruszczynski$^{1,2}$ \and
Jakub Meixner$^{1,3}$ \and
Michal Wlodarczyk$^{1,4}$ \and
Przemyslaw Musialski$^{1,5,6}$
}

\begin{document}
\twocolumn[{
\maketitle
\begin{center}
$^1$IDEAS NCBR \quad
$^2$Samsung AI Center Warsaw \quad
$^3$Polish Academy of Sciences \\
$^4$Warsaw University of Technology \quad
$^5$IDEAS Research Institute \quad
$^6$New Jersey Institute of Technology \\
\texttt{g.gruszczyns@samsung.com}, \quad \texttt{\{kubameixner, mwlodarzc\}@gmail.com}, \quad \texttt{przem@njit.edu}
\end{center}
\captionsetup{type=figure}

    \centering
    \begin{minipage}[t]{0.32\textwidth}
        \centering
        \begin{tikzpicture}[scale=0.95, transform shape] %
            \node (img1) at (-1.8,0) {\includegraphics[trim=210 80 180 280, clip, width=0.25\textwidth]{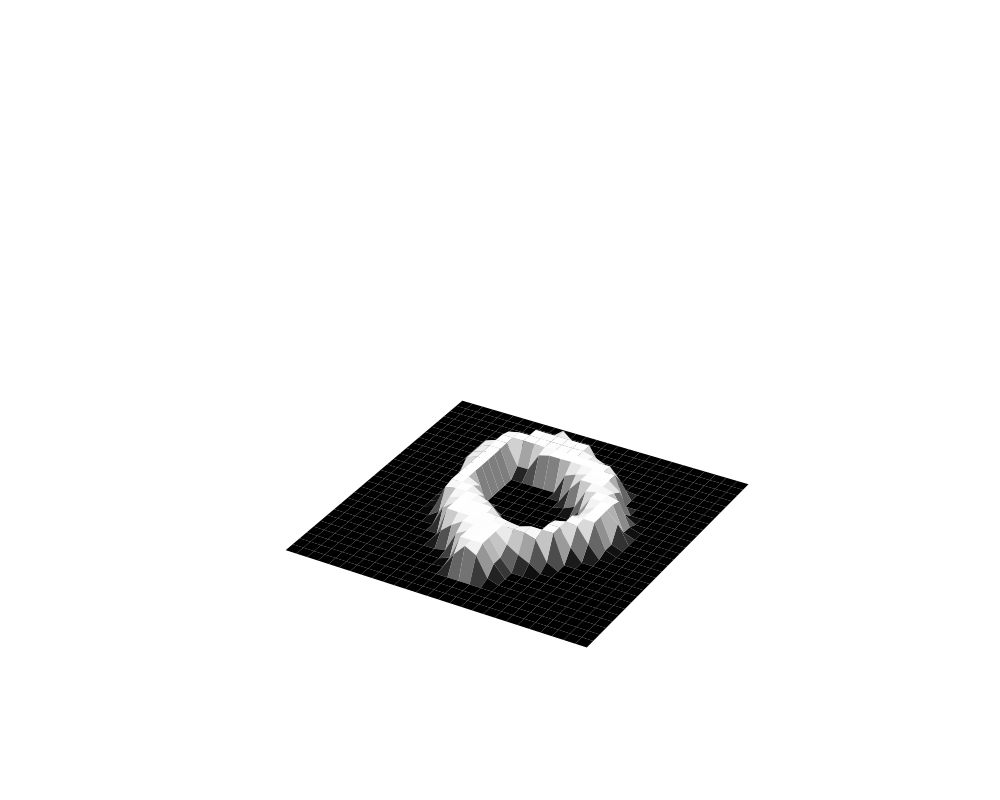}};
            \node (img2) at (0,0) {\includegraphics[trim=210 80 180 280, clip, width=0.25\textwidth]{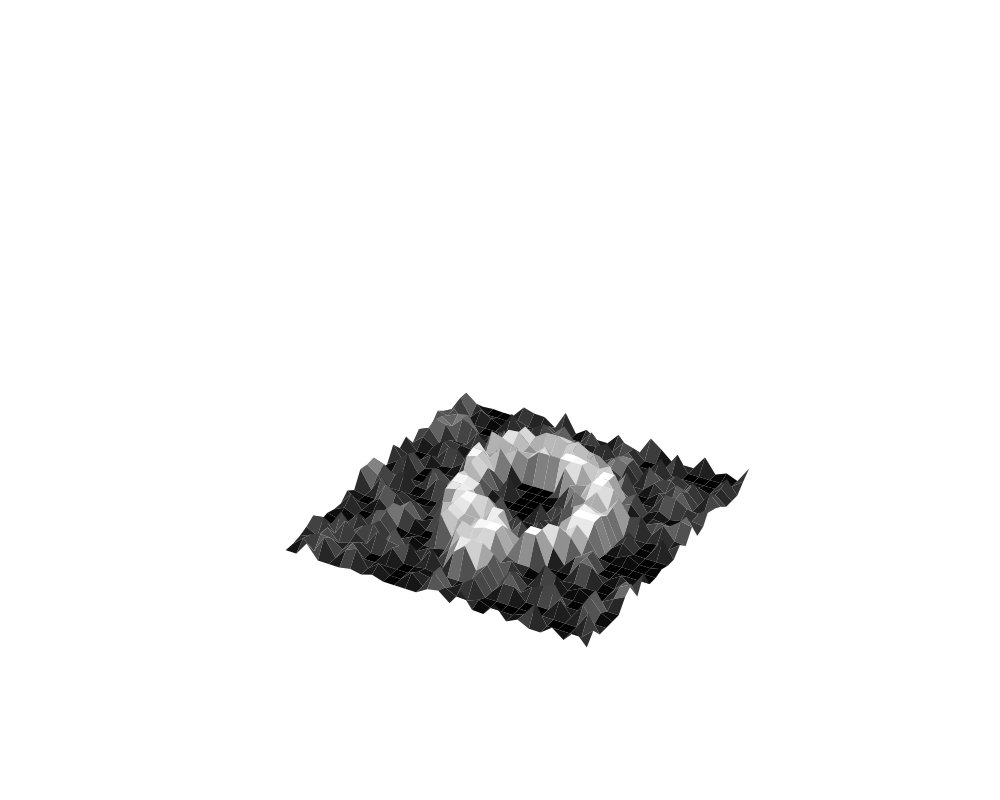}};
            \node (img3) at (1.8,0) {\includegraphics[trim=200 120 180 180, clip, width=0.25\textwidth]{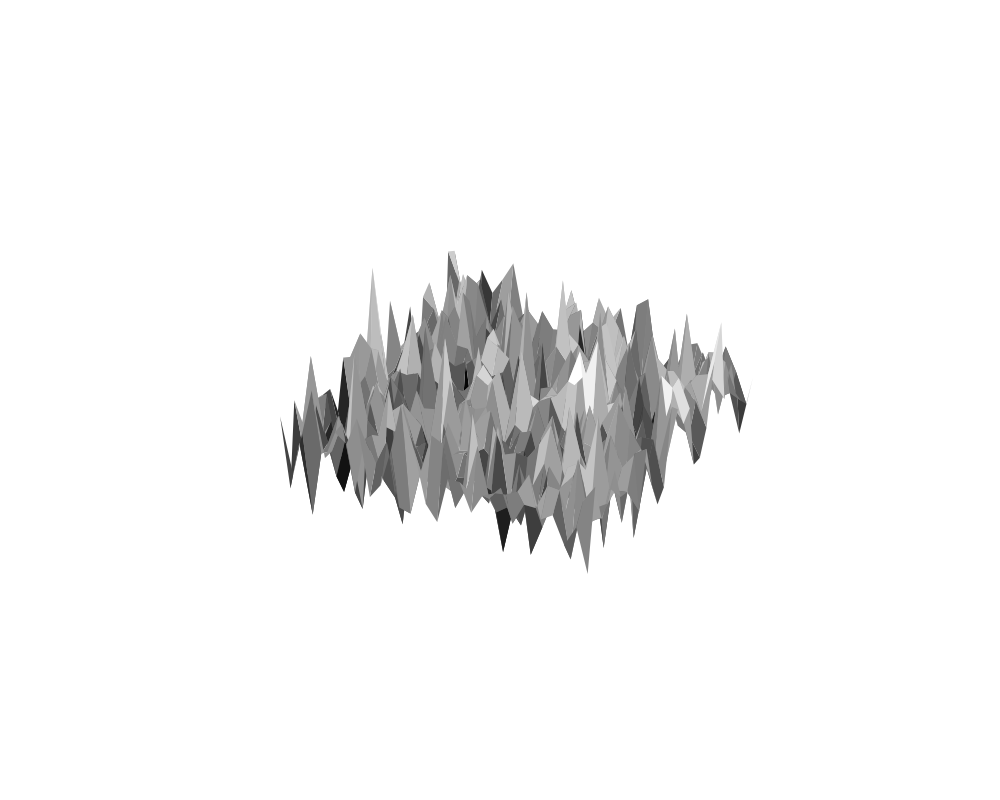}};

            \node[font=\large\bfseries] at (0, 1.5) (title) {Standard diffusion model};
            \node[below=0.1cm of title, font=\small, align=center] (subtitle) {non-invertible forward process};

            \draw[->] ([yshift=3mm]img1.east) -- ([yshift=3mm]img2.west);
            \draw[->] ([yshift=3mm]img2.east) -- ([yshift=3mm]img3.west);

            \draw[<-] ([yshift=-3mm]img1.east) -- ([yshift=-3mm]img2.west);
            \draw[<-] ([yshift=-3mm]img2.east) -- ([yshift=-3mm]img3.west);

            \node[below=0.0cm of img2, font=\small] (generative_text) {generative reverse process};

            \node[below=0.1cm of generative_text] (imgPlaceholder) {
                \begin{tikzpicture}
                    \matrix[matrix of nodes, column sep=0.0cm, nodes={inner sep=0}] {
                        \includegraphics[trim=160 60 140 60, clip, width=0.16\textwidth]{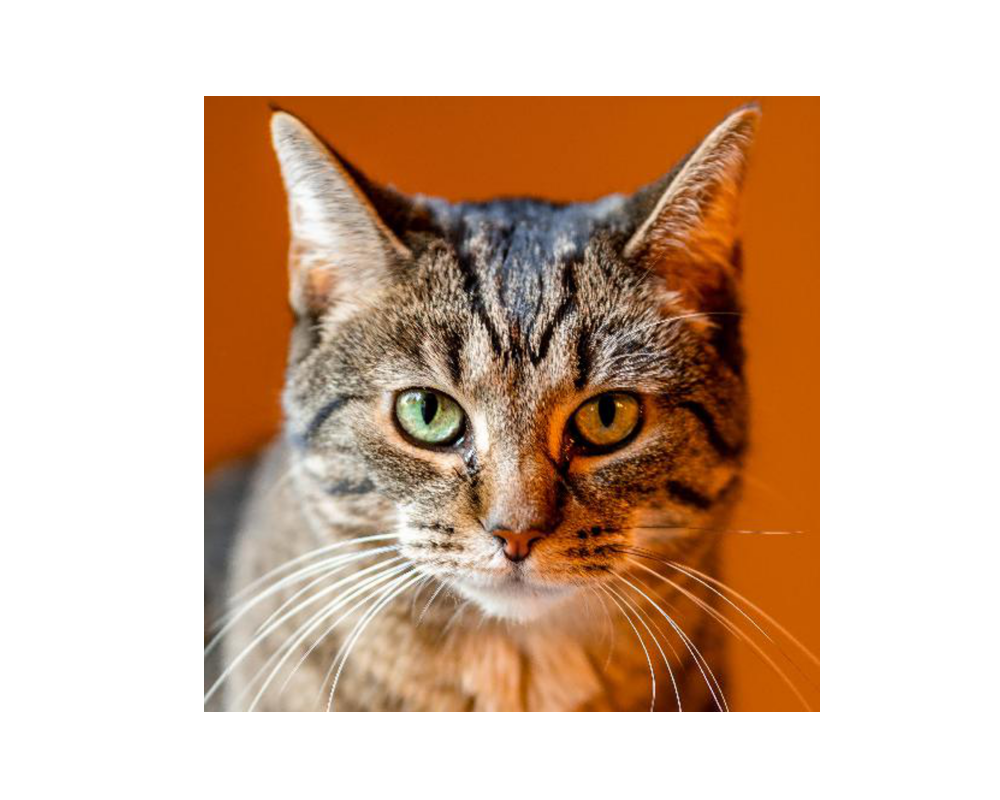} &
                        \includegraphics[trim=160 60 140 60, clip, width=0.16\textwidth]{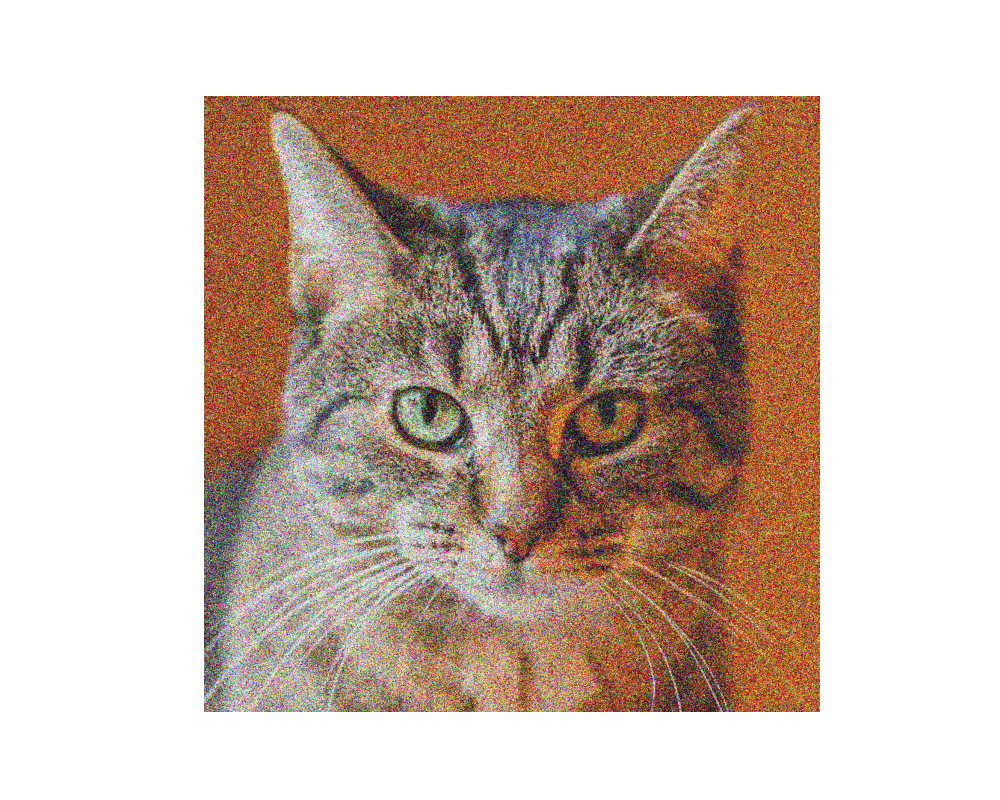} &
                        \includegraphics[trim=160 60 140 60, clip, width=0.16\textwidth]{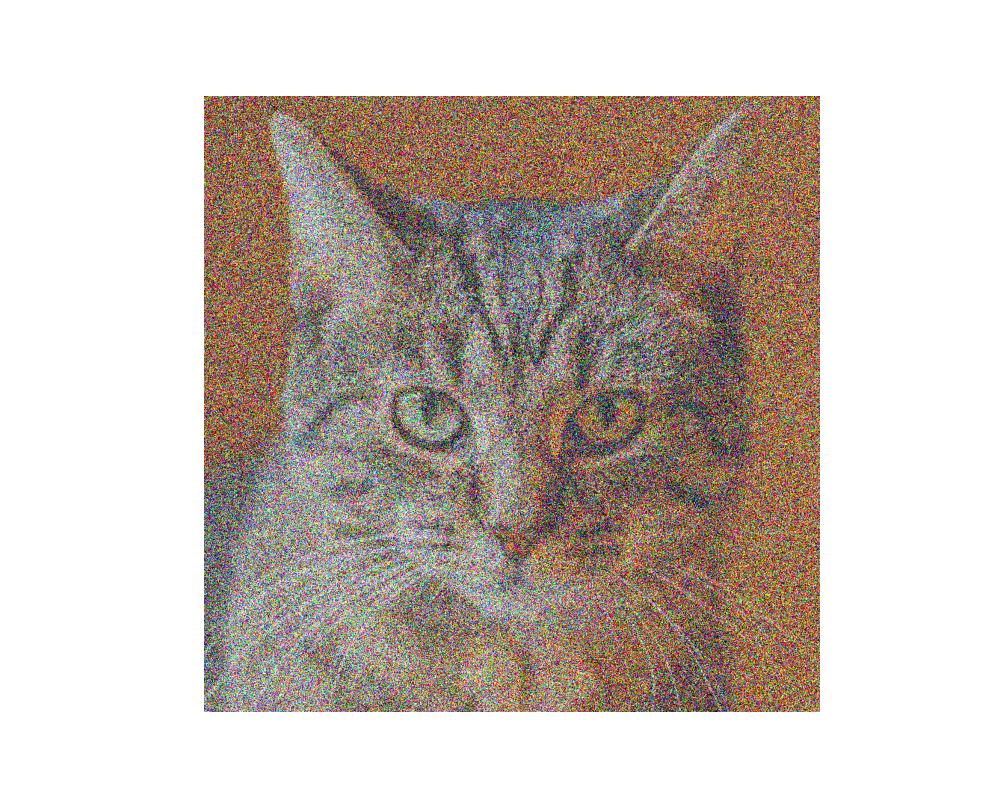} &
                        \includegraphics[trim=160 60 140 60, clip, width=0.16\textwidth]{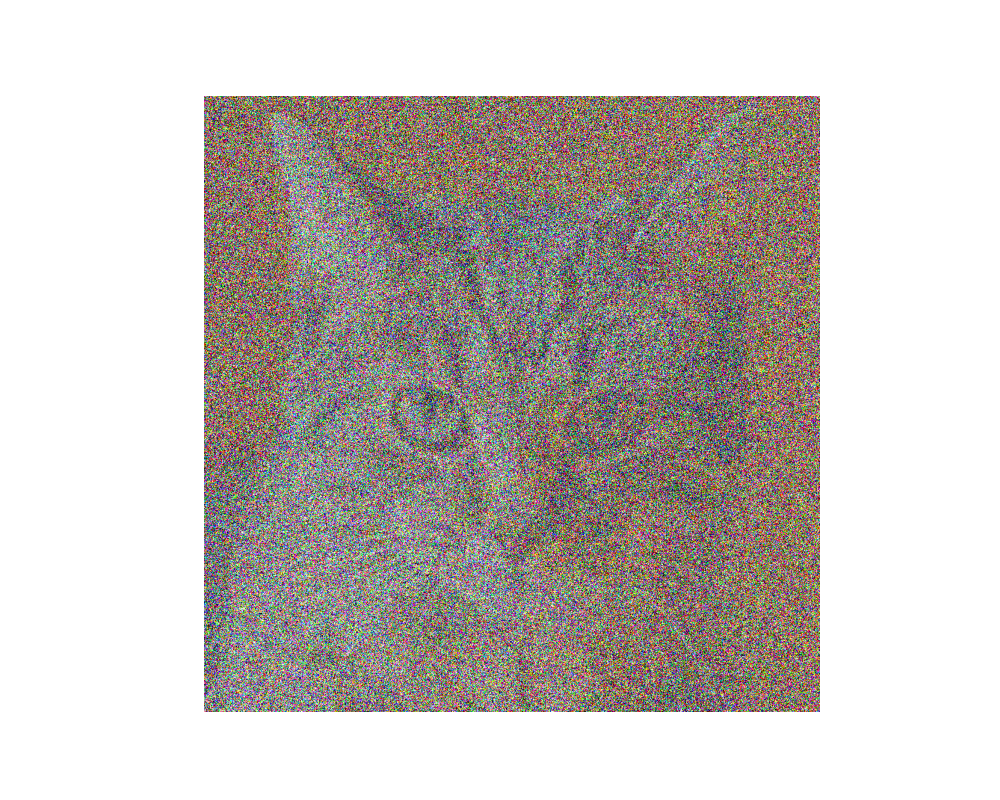} &
                        \includegraphics[trim=160 60 140 60, clip, width=0.16\textwidth]{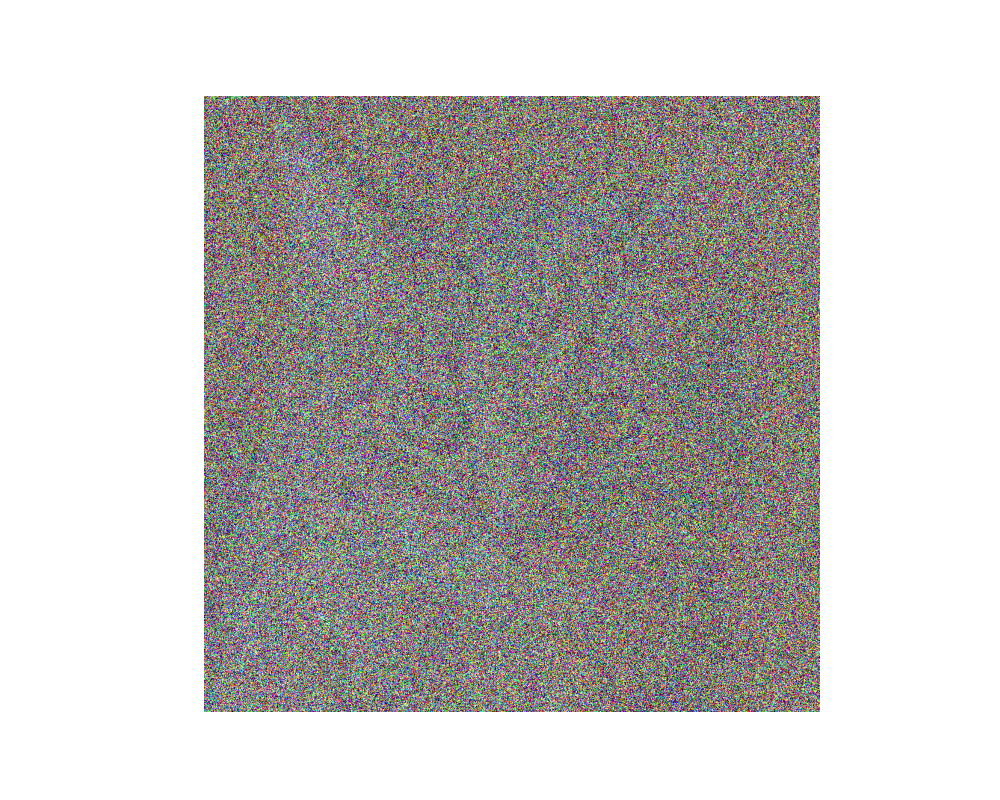} &
                        \includegraphics[trim=160 60 140 60, clip, width=0.16\textwidth]{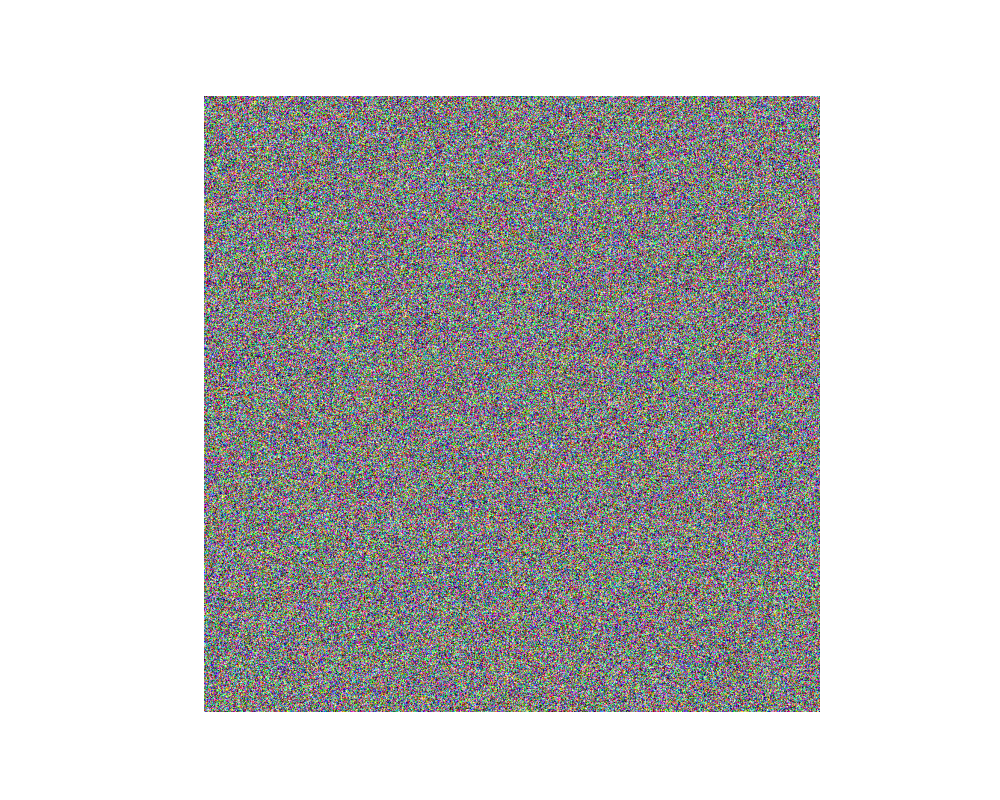} \\
                    };
            \end{tikzpicture}
        };
        \end{tikzpicture}
    \end{minipage}
    \hfill
    \begin{minipage}[t]{0.32\textwidth}
        \centering
        \begin{tikzpicture}[scale=0.95, transform shape] %
            \node (img1) at (-1.8,0) {\includegraphics[trim=210 80 180 280, clip, width=0.25\textwidth]{figures/results/DIAGRAMS/DDPM-ADE-method/clear.png}};
            \node (img2) at (0,0) {\includegraphics[trim=210 80 180 280, clip, width=0.25\textwidth]{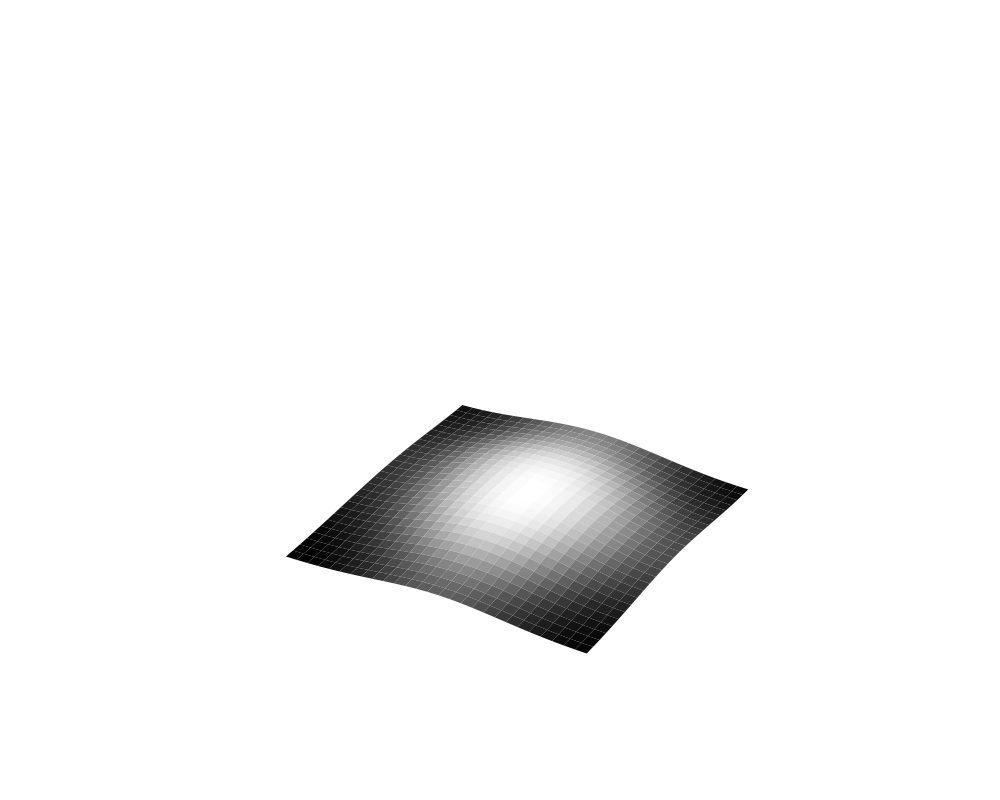}};
            \node (img3) at (1.8,0) {\includegraphics[trim=210 80 180 280, clip, width=0.25\textwidth]{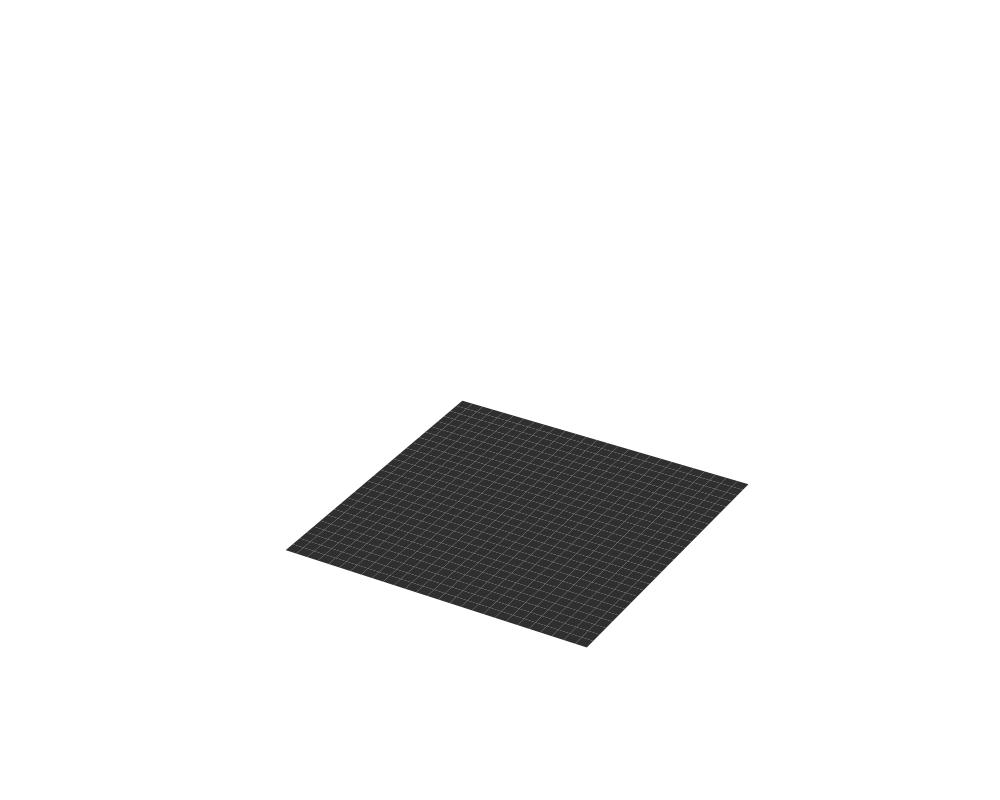}};

            \node[font=\large\bfseries] at (0, 1.5) (title) {Inverse heat dissipation model};
            \node[below=0.1cm of title, font=\small, align=center] (subtitle) {non-invertible forward process};

            \draw[->] ([yshift=3mm]img1.east) -- ([yshift=3mm]img2.west);
            \draw[->] ([yshift=3mm]img2.east) -- ([yshift=3mm]img3.west);

            \draw[<-] ([yshift=-3mm]img1.east) -- ([yshift=-3mm]img2.west);
            \draw[<-] ([yshift=-3mm]img2.east) -- ([yshift=-3mm]img3.west);

            \node[below=0.0cm of img2, font=\small] (generative_text) {generative reverse process};

            \node[below=0.1cm of generative_text] (imgPlaceholder) {
                \begin{tikzpicture}
                    \matrix[matrix of nodes, column sep=0.0cm, nodes={inner sep=0}] {
                        \includegraphics[trim=160 60 140 60, clip, width=0.16\textwidth]{figures/results/DIAGRAMS/DDPM-ADE-method/IHD_example/cat.png} &
                        \includegraphics[trim=160 60 140 60, clip, width=0.16\textwidth]{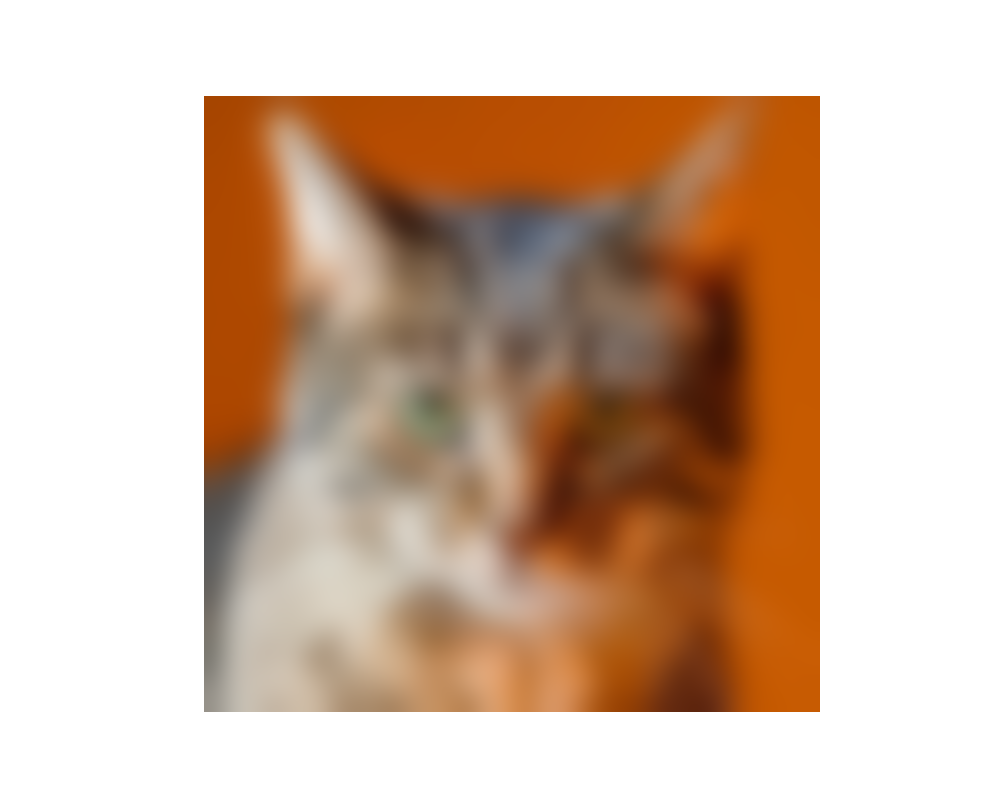} &
                        \includegraphics[trim=160 60 140 60, clip, width=0.16\textwidth]{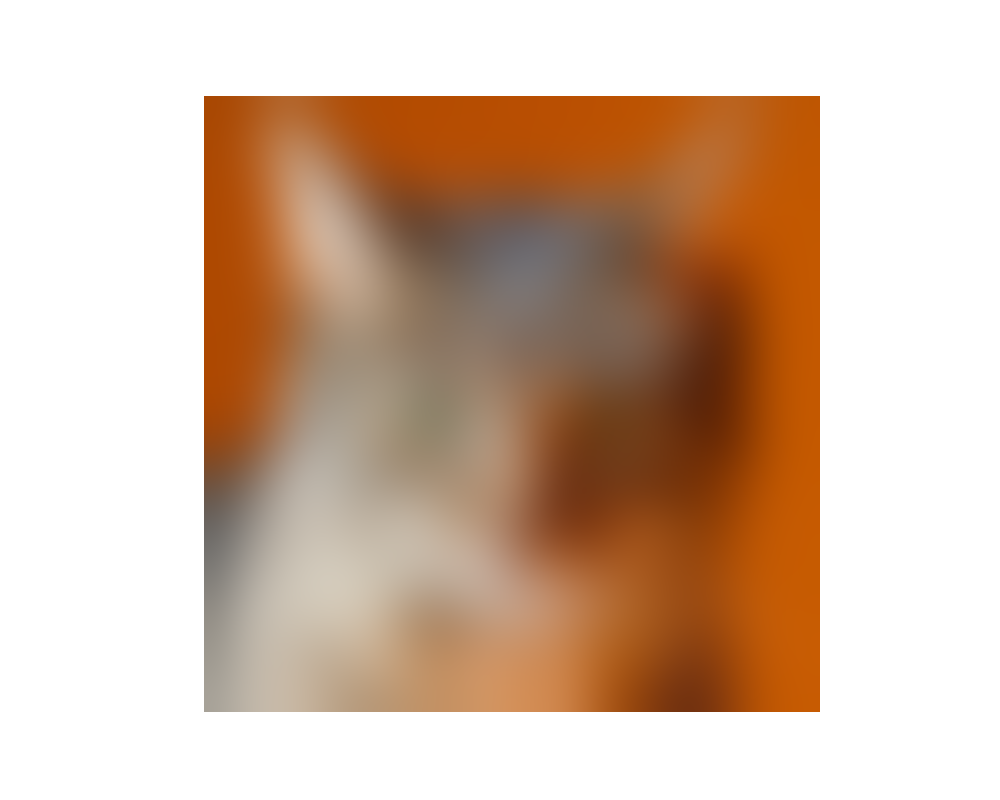} &
                        \includegraphics[trim=160 60 140 60, clip, width=0.16\textwidth]{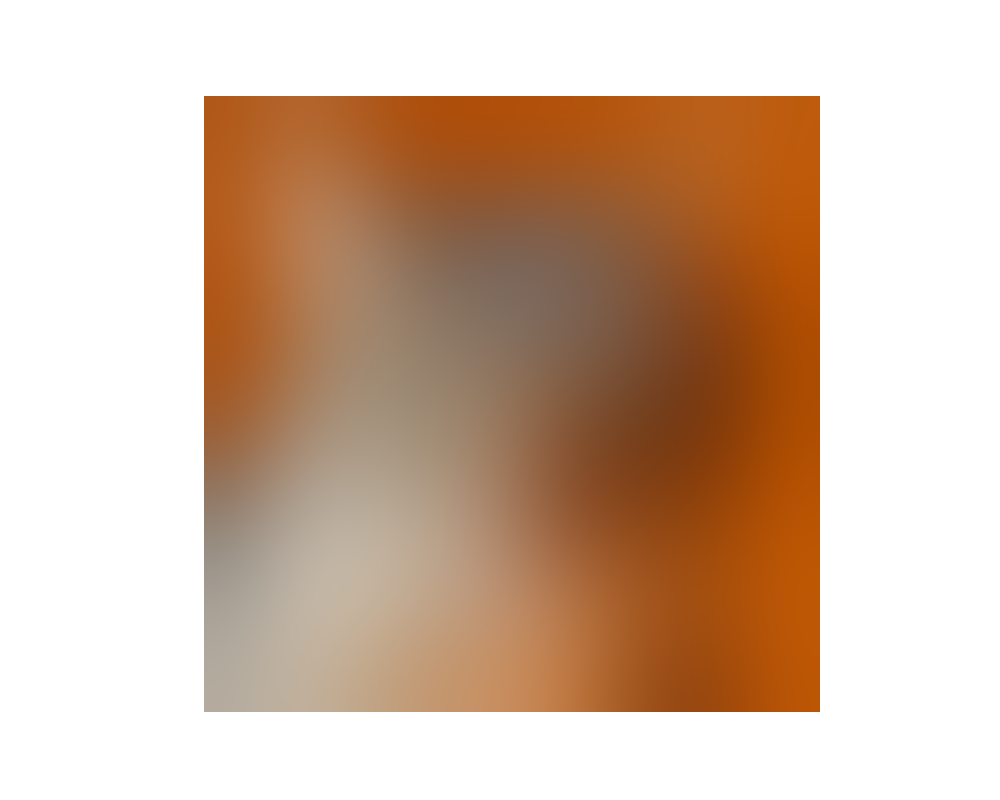} &
                        \includegraphics[trim=160 60 140 60, clip, width=0.16\textwidth]{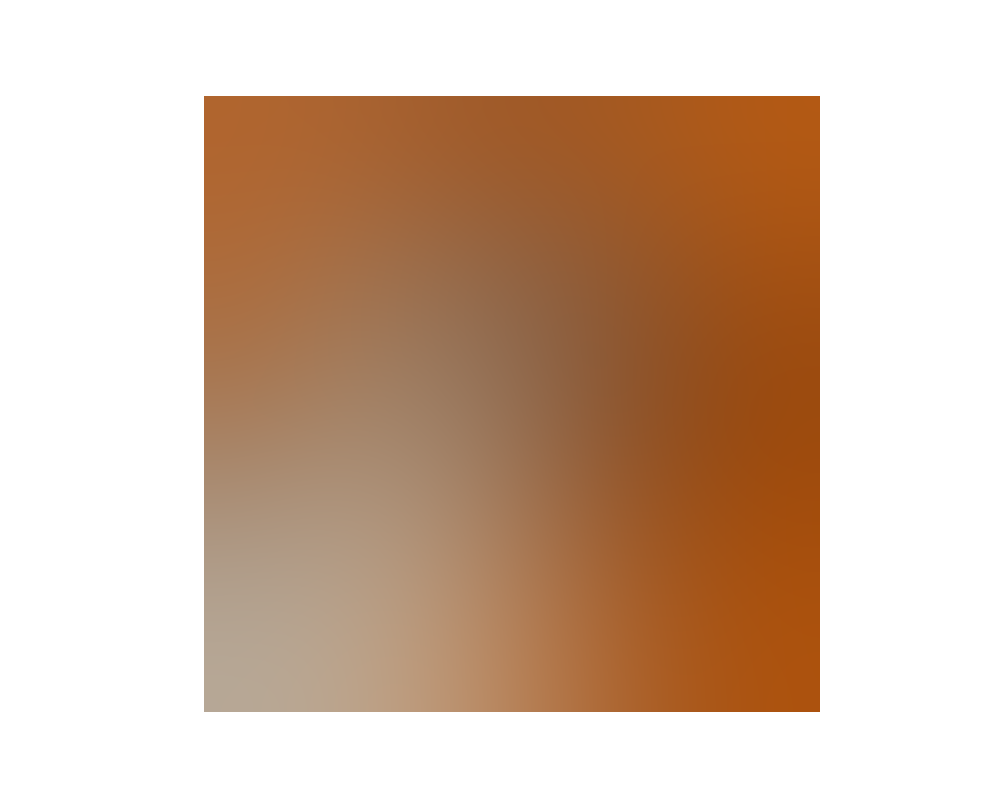} &
                        \includegraphics[trim=160 60 140 60, clip, width=0.16\textwidth]{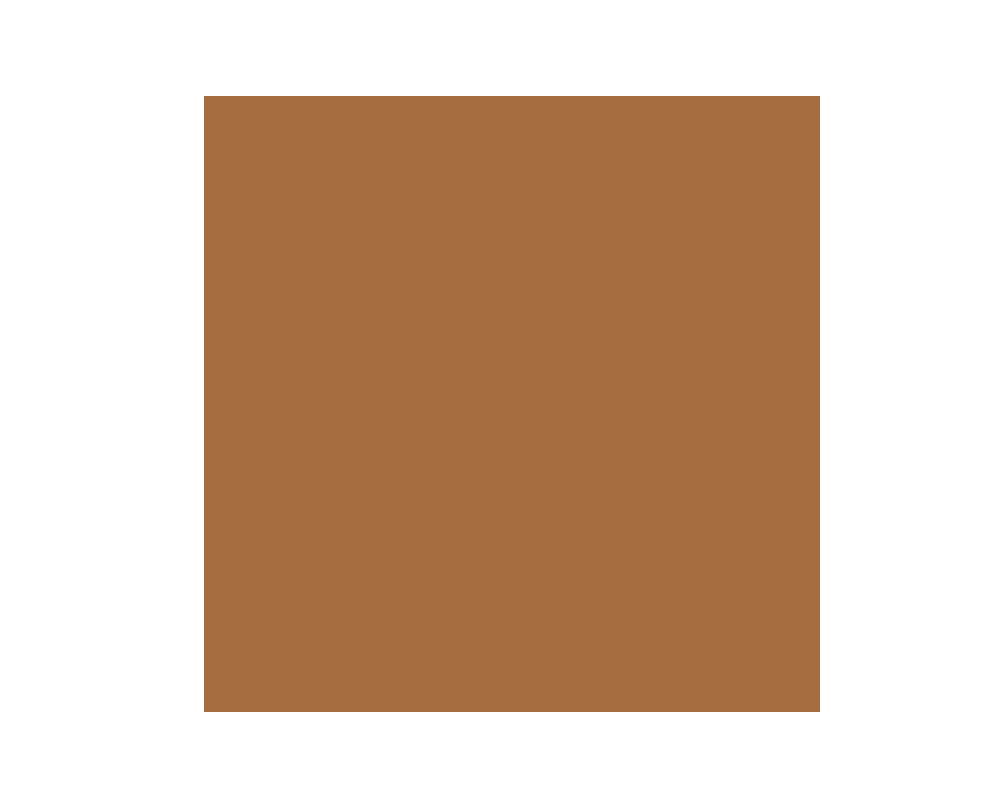} \\
                    };
            \end{tikzpicture}
        };

        \end{tikzpicture}
    \end{minipage}
    \hfill
    \begin{minipage}[t]{0.32\textwidth}
        \centering
        \begin{tikzpicture}[scale=0.95, transform shape] %
            \node (img1) at (-1.8,0) {\includegraphics[trim=210 80 180 280, clip, width=0.25\textwidth]{figures/results/DIAGRAMS/DDPM-ADE-method/clear.png}};
            \node (img2) at (0,0) {\includegraphics[trim=210 80 180 280, clip, width=0.25\textwidth]{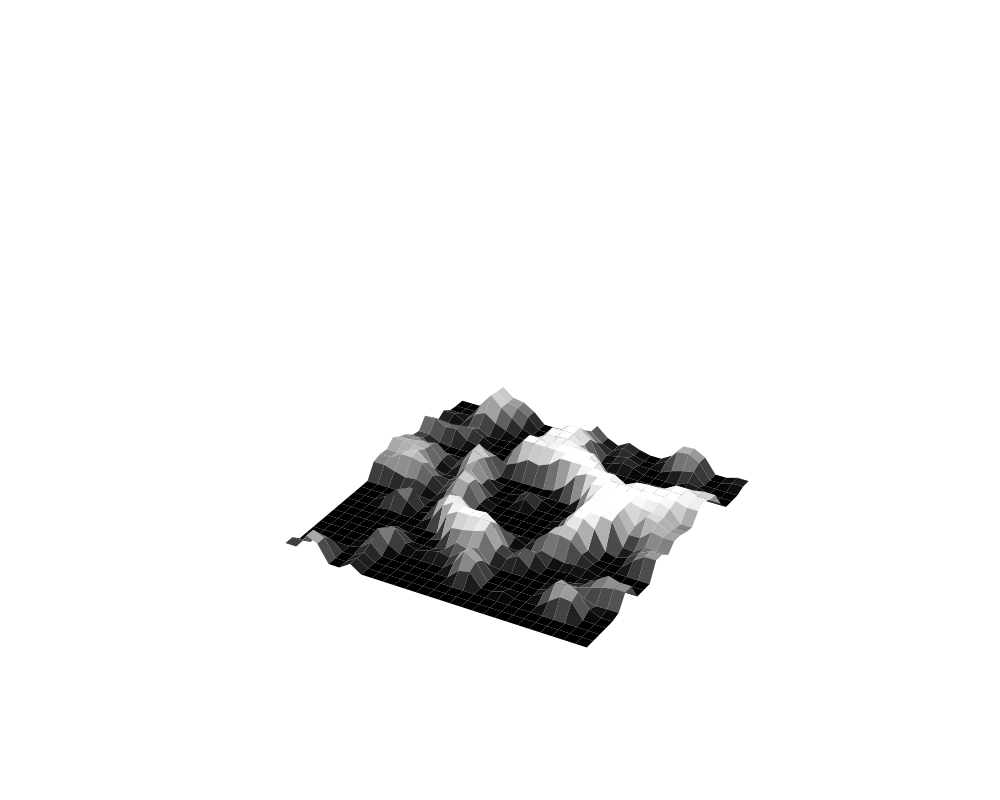}};
            \node (img3) at (1.8,0) {\includegraphics[trim=210 80 180 280, clip, width=0.25\textwidth]{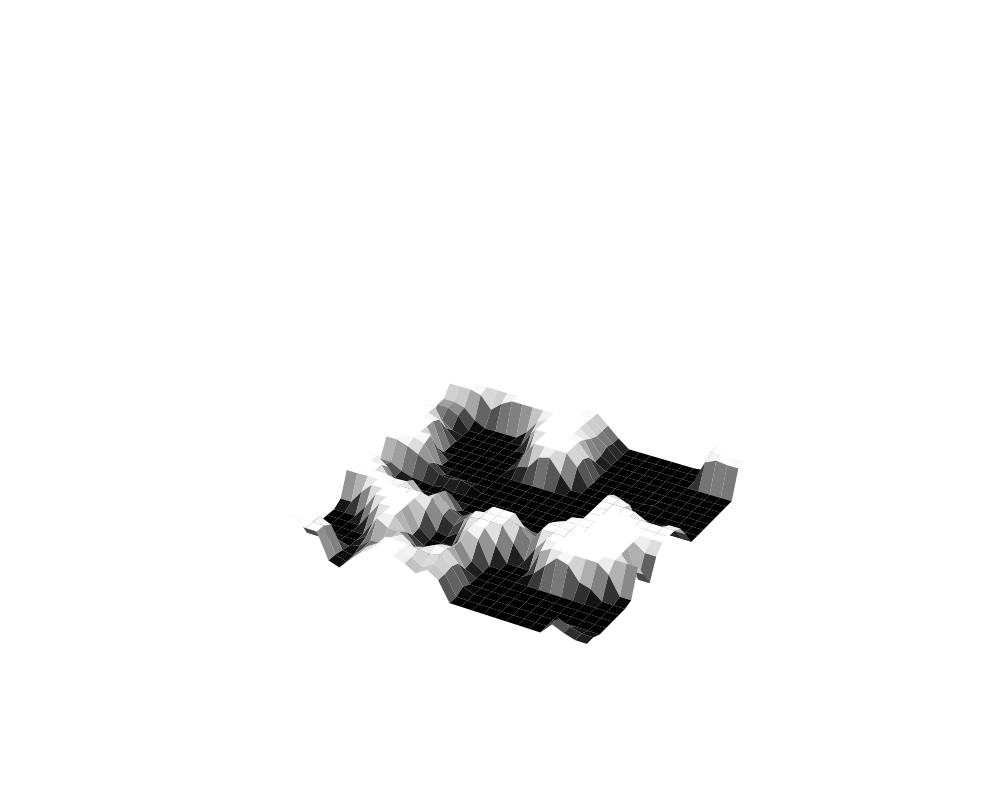}};

            \node[font=\large\bfseries] at (0, 1.65) (title) {Advection diffusion model (Our)};
            \node[below=0.1cm of title, font=\small, align=center] (subtitle) {non-invertible forward process};

            \draw[->] ([yshift=3mm]img1.east) -- ([yshift=3mm]img2.west);
            \draw[->] ([yshift=3mm]img2.east) -- ([yshift=3mm]img3.west);

            \draw[<-] ([yshift=-3mm]img1.east) -- ([yshift=-3mm]img2.west);
            \draw[<-] ([yshift=-3mm]img2.east) -- ([yshift=-3mm]img3.west);

            \node[below=0.0cm of img2, font=\small] (generative_text) {generative reverse process};

        \node[below=0.0cm of generative_text] (imgPlaceholder) {
            \begin{tikzpicture}
                \matrix[matrix of nodes, column sep=0.0cm, nodes={inner sep=0}] {
                    \includegraphics[trim=160 60 140 60, clip, width=0.16\textwidth]{figures/results/DIAGRAMS/DDPM-ADE-method/IHD_example/cat.png} &
                    \includegraphics[trim=160 60 140 60, clip, width=0.16\textwidth]{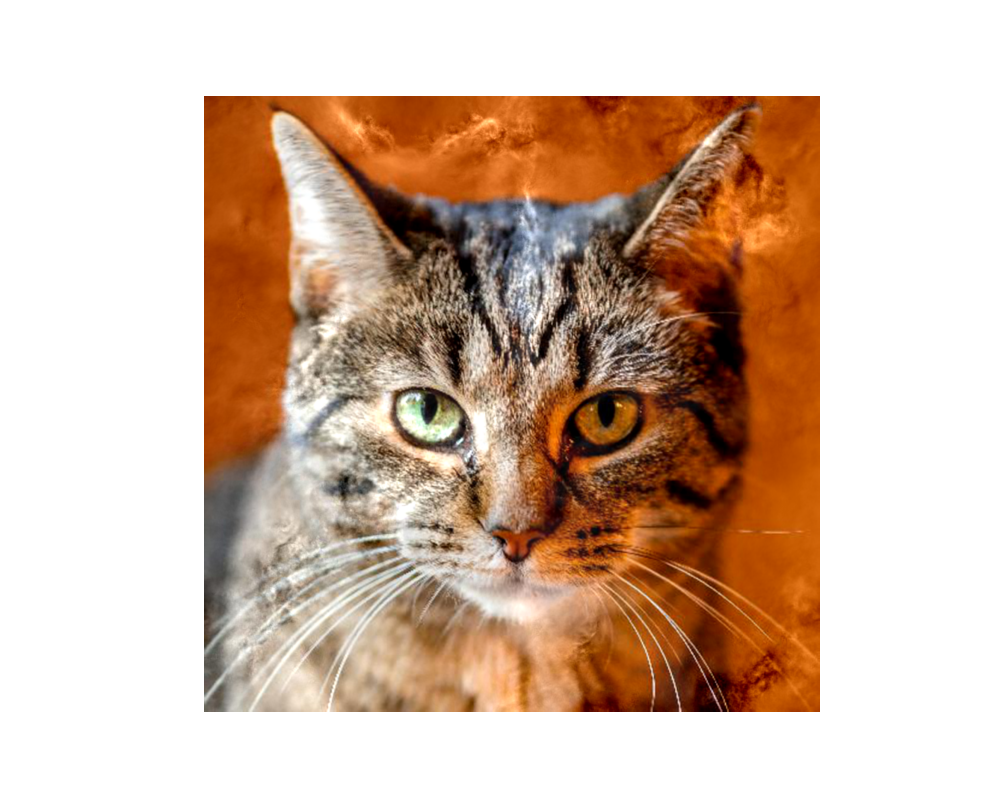} &
                    \includegraphics[trim=160 60 140 60, clip, width=0.16\textwidth]{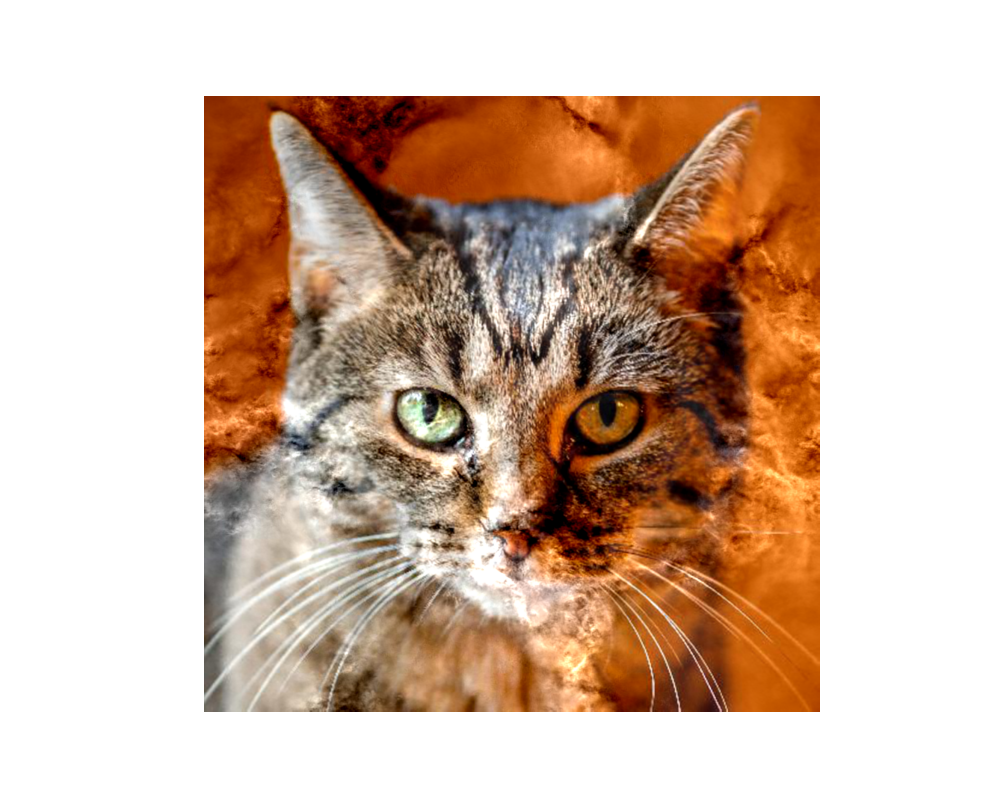} &
                    \includegraphics[trim=160 60 140 60, clip, width=0.16\textwidth]{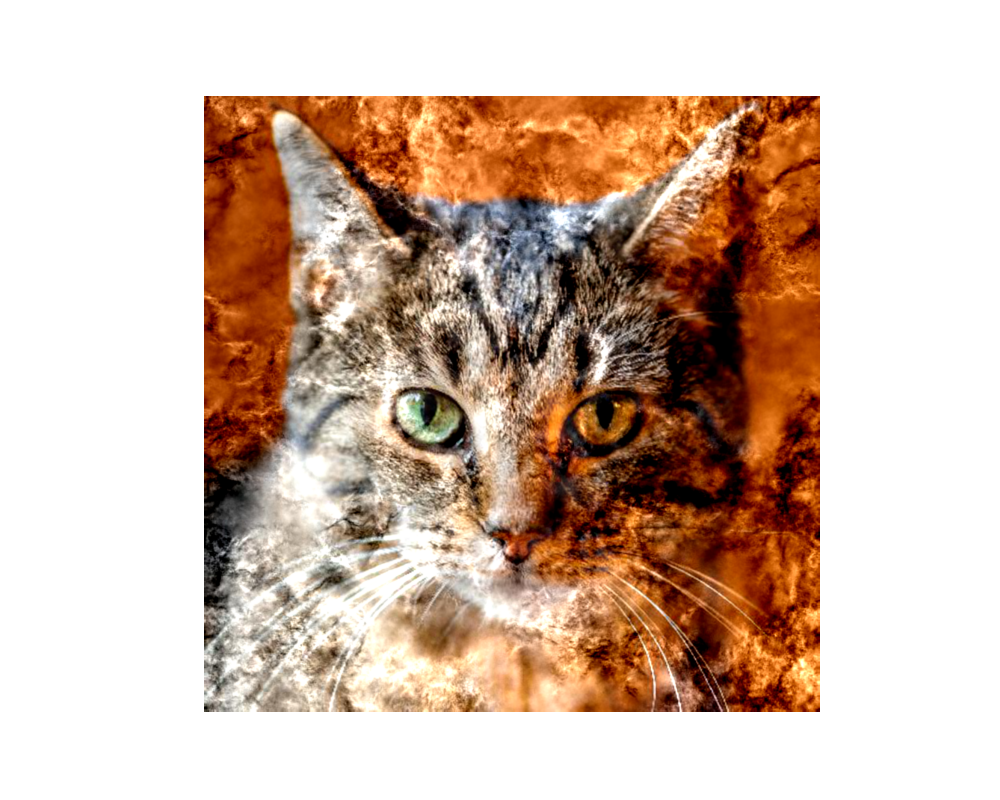} &
                    \includegraphics[trim=160 60 140 60, clip, width=0.16\textwidth]{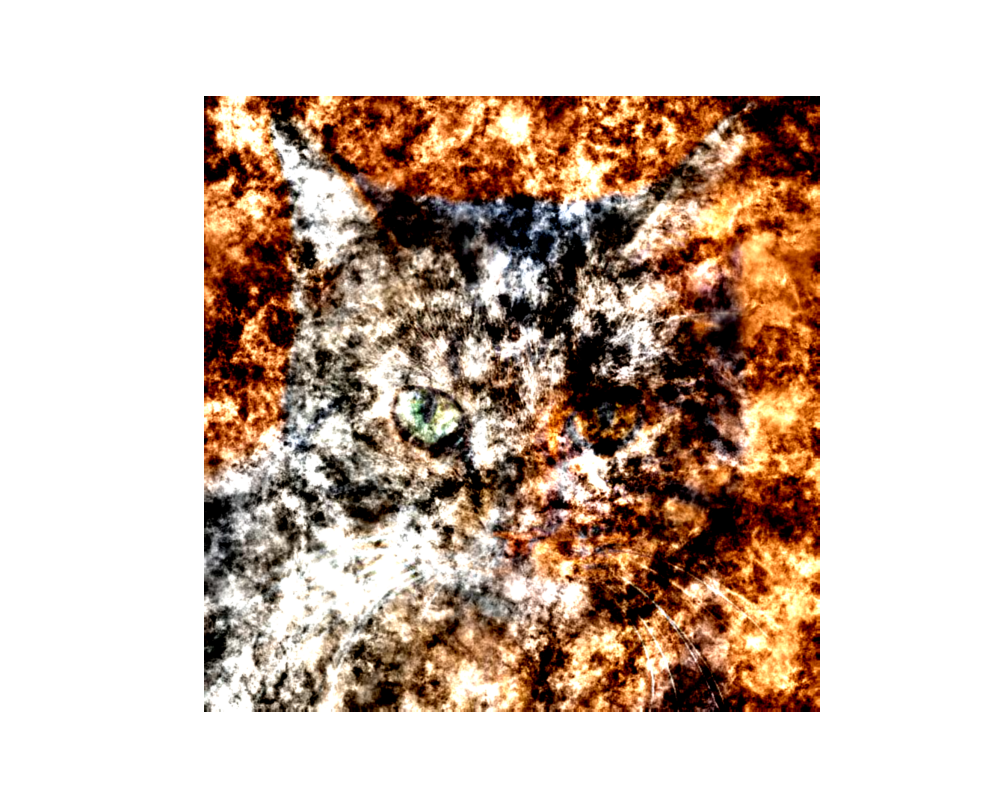} &
                    \includegraphics[trim=160 60 140 60, clip, width=0.16\textwidth]{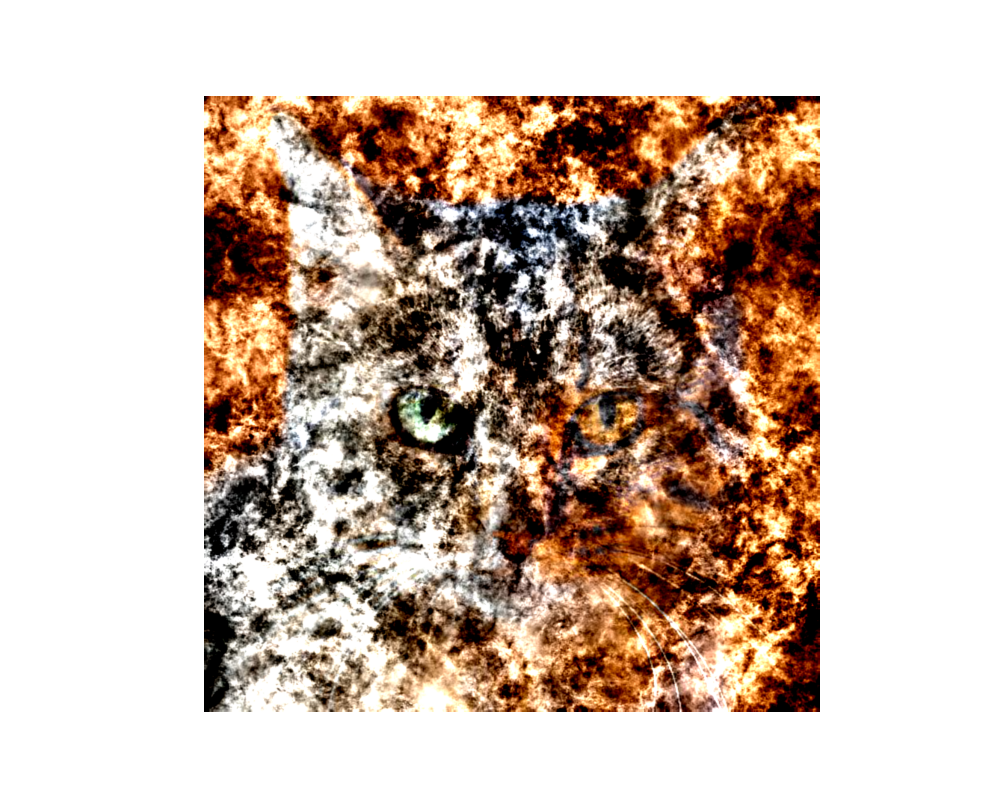} \\
                };
            \end{tikzpicture}
        };
        \end{tikzpicture}
    \end{minipage}
    \caption{The standard diffusion model (DDPM, left) induces Gaussian nose for image corruption, inverse heat dissipation blurs the image using the heat equation (middle), and our proposed advection-diffusion method adds both blur and translation of pixels (right). }

\label{fig:teaser} \vspace{5mm}
}]

\begin{abstract} \vspace{-1pt} %
We propose a novel, Partial Differential Equation (PDE) driven, corruption process for generative image synthesis which generalizes existing PDE-based approaches. 
Our forward pass formulates image corruption via a physically motivated PDE that couples directional advection with isotropic diffusion and Gaussian noise, controlled by dimensionless numbers. 
We solve this PDE numerically through a GPU-accelerated Lattice Boltzmann solver for fast evaluation. To induce realistic ``turbulence,'' we generate stochastic velocity fields that introduce coherent motion and capture multi-scale mixing. 
In the generative process, a neural
network learns to reverse the advection-diffusion operator thus constituting a novel generative model.
We discuss how previous methods emerge as specific cases of our operator, demonstrating that our framework generalizes prior PDE-based corruption techniques.
We illustrate how advection improves the diversity and quality of the
generated images while keeping the overall color palette unaffected. 
This work bridges fluid dynamics, dimensionless PDE theory, and deep generative modeling, 
offering a fresh perspective on physically based inverse problems. 
\end{abstract}

\section{Introduction}
\label{sec:intro}

Denoising probabilistic diffusion-based generative models have made striking strides in recent years, demonstrating high-quality image synthesis through iterative noise addition and subsequent denoising~\citep{SohlDickstein2015DeepUL, Ho2020DenoisingDP, Song2020ScoreBasedGM, Dhariwal2021DiffusionMB}.

A distinct branch of works focuses on introducing different image corruption processes, such as Cold Diffusion~\cite{Bansal_cold_diffusion_23}, Soft Diffusion~\cite{daras2022softdiffusionscorematching}, or more physically inspired processes, such as Inverse Heat Dissipation~\cite{Rissanen2022GenerativeMW} or Blurring Diffusion~\cite{hoogeboom2022blurring}. 
The idea in these works is to replace or augment the pure Gaussian noise with other mechanisms aimed at better preserving color budgets, multi-scale detail, or interpretability. 
A notable subfamily are the PDE-based methods, which  model the image frequencies explicitly and thus deliver a multi-scale perspective with clear frequency-domain interpretation. 
Nevertheless, previous PDE approaches remain purely \emph{isotropic}, ignoring potentially compelling directional flows. 

We introduce the advection--diffusion corruption processes, which allows to unlock \emph{anisotropic} patterns of texture shifts and swirling motions in the forward corruption process. 
This forward operator is physically well-grounded and not covered by earlier works. \\

\textbf{Our main contribution} is hence an \emph{Advection-Diffusion-Reaction Probabilistic Model} which goes \textit{beyond blur}. 
It integrates not only a random reaction terms (Gaussian noise) and blurring (averaging) but also shift (advection) terms, generalizing the previous works into a common framework. 
It is inspired by fluid dynamics and it allows to unlock structured flows. The major advantage over previous methods lies in its physical grounding and its property to enable coherent texture shifts in the forward corruption trajectory that isotropic blur cannot represent. 

To implement the aforementioned advection-diffusion operator on typical datasets efficiently, we propose a scalable GPU-based Lattice Boltzmann Method (LBM) solver, an established fluid simulation technique \cite{krger2017lattice}.
We also introduce a dimensionless formulation of the training process (using similarity numbers) and show that the intensity of the physical process can be easily transferred between images of different resolutions. 

\section{Related Work}
\label{sec:related_work}
The last few years have witnessed tremendous progress in diffusion-based generative methods for image generation and beyond. Several comprehensive surveys \citep{Yang2022DiffusionMA, Cao2022ASO} consolidate the growing literature. 
Building on early variants of \textit{Denoising Diffusion Probabilistic Models} (DDPM) \citep{SohlDickstein2015DeepUL,Ho2020DenoisingDP}, 
the diffusion processes have been generalized to continuous-time stochastic differential equations \citep{Song2020ScoreBasedGM}, 
revealing unifying insights across iterative noising--denoising paradigms. Notably, the notion of learning gradients of the data distribution can be traced back to \citet{Song2019GradientEstimation}, 
which laid the foundation for modern score-based generative modeling.  
Subsequent refinements have focused on \emph{improved sampling} \citep{Song2020DenoisingDI,NicholDhariwal2021Improved,salimans2022progressive,huang2024BlueNoise}, 
\emph{better architectures} \citep{Karras2022ElucidatingTD}, 
\emph{improved training dynamics} \citep{Karras2023TrainingDynamics}, 
and 
\emph{conditional guidance} \citep{Ho2022ClassifierFreeDG}, 
often yielding state-of-the-art image quality \citep{Dhariwal2021DiffusionMB}. 
These advances have extended beyond unconditional image synthesis to tasks such as \emph{super-resolution}~\citep{Saharia2021ImageSV} 
and \emph{time-series imputation}~\citep{Tashiro2021CSDICS}, showcasing the versatility of diffusion approaches. 
A variational perspective on diffusion complements score-based methods~\citep{kingma2021variational}, and  
\emph{Latent Diffusion Models}~\citep{rombach2022high} further scale to high-resolution generation by introducing a compressed latent space, improving computational efficiency and memory usage.

Recently, methods which introduce alternative processes for the corruption of the input image arised. Particularly, \emph{Inverse Heat Dissipation} (IHD) introduced by Rissanen et al. \cite{Rissanen2022GenerativeMW} replaces the conventional Gaussian-based forward corruption with a heat-equation blur with minor additive noise, thereby offering a physically grounded coarse-to-fine scheme. 
Follow-up work has been done by \citet{hoogeboom2022blurring}, who combined blurring with a growing amount of Gaussian noise controlled by an appropriate scheduler~\citep{Chen2023NoiseScheduling}. 
On the other hand, \citet{huang2024BlueNoise} introduced a blue noise characterized by a prescribed energy spectrum for the corruption process. 
Our approach generalizes this PDE-driven philosophy by infusing a velocity-driven \emph{advection} term into the forward process. 

In a so-called \textit{cold diffusion}, \citet{Bansal_cold_diffusion_23} proposed a sampling method that is able to invert a fully deterministic (without noise) degradation process.
They introduced arbitrary deterministic operators (e.g.\ blur, masking) without noise, and learned to invert them without imposing a strict physical PDE. 
\citet{daras2022softdiffusionscorematching} presented a \textit{soft score-matching} loss function and a momentum sampler with application to blurring with a limited amount of noise.
Their model was trained to predict a clean image that would resemble the one after corruption. 

In terms of computational fluid dynamics, the \emph{Lattice Boltzmann Method} (LBM)  
have garnered growing attention \citep{Bedrunka2021LettucePL, Ataei2023XLBAD, lehmann2022accuracy, LANIEWSKIWOLLK2016833} due to its highly local computations suitable for GPU-based acceleration. 
Because LBM naturally solves advection-diffusion type equations, it has been chosen for our goal of embedding a velocity-dependent PDE in the forward corruption pass. 
Finally, the data-driven PDE solvers have exploited deep learning to accelerate numerical simulations 
\citep{Li2020FourierNO, raonic2023convolutionalneuraloperatorsrobust, Kochkov2021MachineLC}.

\section{Background on Diffusion Models}
\label{sec:background_pde}

In diffusion-based generative models (e.g., DDPM), the pixel intensities are progressively corrupted by noise, which can be viewed in physical terms as a \emph{random reaction} process. 
In contrast, recent works highlight that image intensities may also be \emph{transformed} via partial differential equations (PDEs) before adding only marginal noise \citep{Rissanen2022GenerativeMW, daras2022softdiffusionscorematching}. 
Below, we first revisit the classic noise-driven diffusion formulation and then generalize to the PDE-based diffusion process. 

\subsection{Probabilistic Diffusion Models}
\label{sec:gen_models}

Classical diffusion models~\citep{SohlDickstein2015DeepUL,Ho2020DenoisingDP} treat the forward noising as a \emph{Markov chain} \(q(u_k \mid u_{k-1})\) that progressively corrupts an initial data sample \(u_0\). 
Interpreted in continuous time, the forward process can be seen as a \emph{stochastic differential equation} (SDE),
\begin{equation}
  \mathrm{d}u(t)
  \;=\;
  f\bigl(u(t),t\bigr)\,\mathrm{d}t
  \;+\;
  g(t)\,\mathrm{d}w(t),
  \label{eq:SDE_ref}
\end{equation}
where \(w(t)\) is Brownian motion and \(g(t)\) regulates the noise injection. 
The drift term is denoted with \(f\bigl(u(t),t\bigr)\)
The distribution of \(u(t)\) then follows the \emph{Fokker--Planck equation}: 
\begin{equation}
  \frac{\partial p(u,t)}{\partial t}
  \,=\,
  -\nabla_{u} \cdot
  \Bigl(f(u,t)\,p(u,t)\Bigr)
  \,+\,
  \tfrac12 \,\nabla_{u}^2 
  \Bigl(g^2(t)\,p(u,t)\Bigr).
  \label{eq:FokkerPlanck_ref}
\end{equation}
In more intuitive terms, as $t$ increases, ${u}(t)$ becomes increasingly randomized (e.g., converging to an isotropic Gaussian), and a learned \emph{reverse process} iteratively ``denoises'' from that noise-prior back to the data manifold. 
Such purely noise-driven approaches can be viewed physically as a \emph{local ``reaction''} term affecting each pixel independently. 
However, this local scalar reaction may overlook spatial coherence.
 
\begin{figure*}
    \centering
    \setlength{\tabcolsep}{0.1em} %
    \renewcommand{\arraystretch}{1} %
    \begin{tabular}{c c c c c c c c c c c} %
        \adjustimage{width=0.085\textwidth}{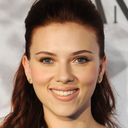} &
        \adjustimage{width=0.085\textwidth}{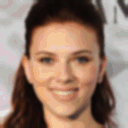} &
        \adjustimage{width=0.085\textwidth}{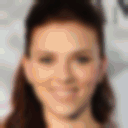} &
        \adjustimage{width=0.085\textwidth}{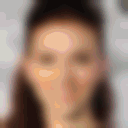} &
        \adjustimage{width=0.085\textwidth}{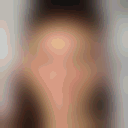} &
        \adjustimage{width=0.085\textwidth}{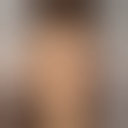} &
        \adjustimage{width=0.085\textwidth}{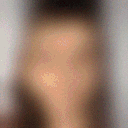} &
        \adjustimage{width=0.085\textwidth}{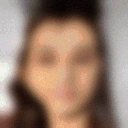} &
        \adjustimage{width=0.085\textwidth}{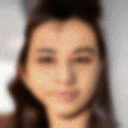} &
        \adjustimage{width=0.085\textwidth}{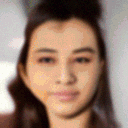} &
        \adjustimage{width=0.085\textwidth}{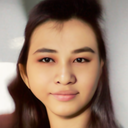} \\
        GT & &  &  &  & Prior &  &  &  &  & Gen \\
    \end{tabular}
    \caption{Example of the corruption and generative process of our method, illustrated over 11 sequential frames in a chain. $\sigma = 20$, $Pe = 0.6$. Please refer to Section~\ref{sec:ADE_forward_description} for details. }
\end{figure*}

\subsection{PDE-Based Diffusion Models}
\label{sec:pde_models_general}
According to terminology coined in \cite{Song2020ScoreBasedGM}, the evolution of moments of the corrupted prior distribution can be described as either being Variance Preserving, $\mathrm{Var}[p(u_t)]=I$, or Variance Exploding, $\mathrm{Var}[p(u_t)] \to \infty $. 
At the same time, the mean of the prior distribution will be either shrinking, $\mathbb{E}[p(u_t)] \to 0 $, or constant $ \mathbb{E}[p(u_t)] = \mathbb{E}[p(u_0)] $.
These properties arise from the interplay between the drift term and Gaussian noise injection in  \cref{eq:SDE_ref}.
When using a conservative PDE, like \emph{blurring}, the forward corruption process \emph{by design} preserves color ``intensity'' .  
Therefore, it can be classified as having a constant mean and exploding variance.
For example, starting from any smoothed ``blue canvas,'' one can generate an image of a sky or an ocean but not an autumn forest. 

\begin{figure}[h]
    \centering    \includegraphics[width=\linewidth]{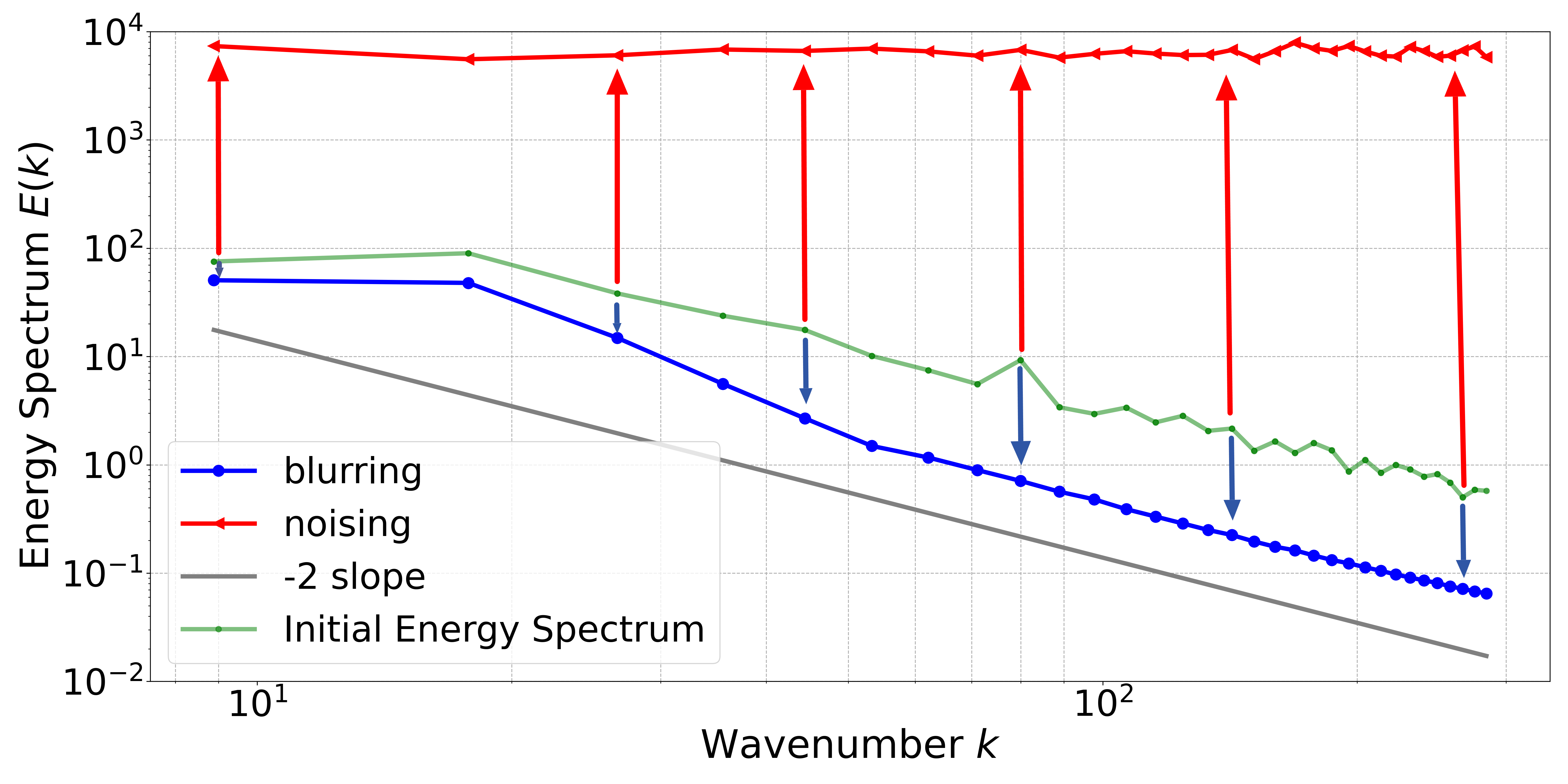}
     \caption{Comparison of the Energy Spectrum (ES) of an image subjected to different corruption processes.
     In the blurring process, the amplitude of energy components decays in a log-linear manner with a constant slope.
     The ES of a blurred image is represented by a blue line with circular markers.
     It is shifted downwards compared to the ES of the clean image denoted with green dotted line.
     On the other hand, the ES of the image in classical DDPM can be viewed as being gradually flooded with noise (red) starting from the higher frequencies..
    } 
     \label{fig:rho_spectrum}
 \end{figure}

Apart of that, the energy spectrum of an image behaves differently depending on the type of corruption process, as shown in \cref{fig:rho_spectrum}. 

\vspace{-10pt}
\paragraph{Examples of PDE corruption operators.}
Recently, models based on that approach have been proposed in literature: 

\begin{itemize}
\setlength{\topsep}{4pt}
\setlength{\itemsep}{4pt} 
  \item \textbf{Heat dissipation (Isotropic Diffusion).}
    The simplest operator sets 
    $\frac{\partial u}{\partial t} = \alpha \,\nabla^2 u$.
    The Inverse Heat Dissipation (IHD)~\cite{Rissanen2022GenerativeMW} utilizes this PDE, generating a ``coarse-to-fine'' representation that naturally preserves color bias.
    To avoid the accumulation of numerical errors, a minor Gaussian perturbation is added during both training and sampling procedures.
  \item \textbf{Blurring Diffusion}. Here, the heat equation is recasted as a full Markov chain in the frequency domain, by injecting dimension-wise noise~\cite{hoogeboom2022blurring}. 
  Although the multi-scale blurring is emphasized, the directional flow is not present.
  \item \textbf{Advection (Ours) or More Complex Dynamics.}
    One can further enrich 
    the physical process 
    with a velocity field to transport pixel intensities spatially. 
\end{itemize}

\noindent
In contrast to local Gaussian noising, the PDE-based corruptions \emph{rearrange} intensities in a spatially coherent way. 
With a purely conservative, PDE-based corruption operator, one may still add small perturbations outside the PDE solution at each step to keep the chain stochastic. 
This framework ensures that color conservation arises from the PDE itself, which can be beneficial for preserving global color palettes.

\section{Advection--Diffusion--Reaction Process}
\label{sec:ADE_forward_description}
Our goal here is to perform the corruption process using a physically grounded mass transport equation, such as the \emph{advection--diffusion--reaction} equation. 
Standard Gaussian-based noising can be seen as a special
case of this equation in which both the advective transport and the averaging (blurring) terms are omitted.
Let us discuss the \emph{advection--diffusion--reaction} equation in the general form,
\begin{equation}
\frac{\partial u}{\partial t} 
\;+\;
\underbrace{\nabla \cdot\! (\mathbf{v} \! \,u) }_{\text{advection}}
\;=\;
\underbrace{\nabla \cdot ( \alpha \, \nabla  u)}_{\text{diffusion}}
+ \underbrace{\dot{Q}(t) }_{\text{reaction}}.
\label{eq:advection_diffusion_pde} 
\end{equation}
The velocity field, $\mathbf{v}=\mathbf{v}(x,y)$, displaces the image intensities, $u=u(x,y,t)$. 
We omit the vector notation here, but the same approach is applied for each color channel.
Over time, the diffusion term continues to reduce high-frequency features with a time-varying coefficient $\alpha=\alpha(t)$. 
Simultaneously, the reaction term, $\dot{Q}(t)$, can modify the amount of the quantity $u$ over time. 
In the context of DDPM, the reaction term would correspond to the injection of the Gaussian noise, $\dot{Q} \sim \mathcal{N}(\bm0, \sigma \bm I)$, while the advection and diffusion terms may be interpreted as the drift term, $f(u(t),t)$, in Eq.\ref{eq:SDE_ref}.
This kind of PDEs naturally arises in physical processes that describe the transport of quantities such as mass or energy and are known as conservation laws.

\begin{figure}[t]
  \centering  
  \includegraphics[width=0.99\linewidth]{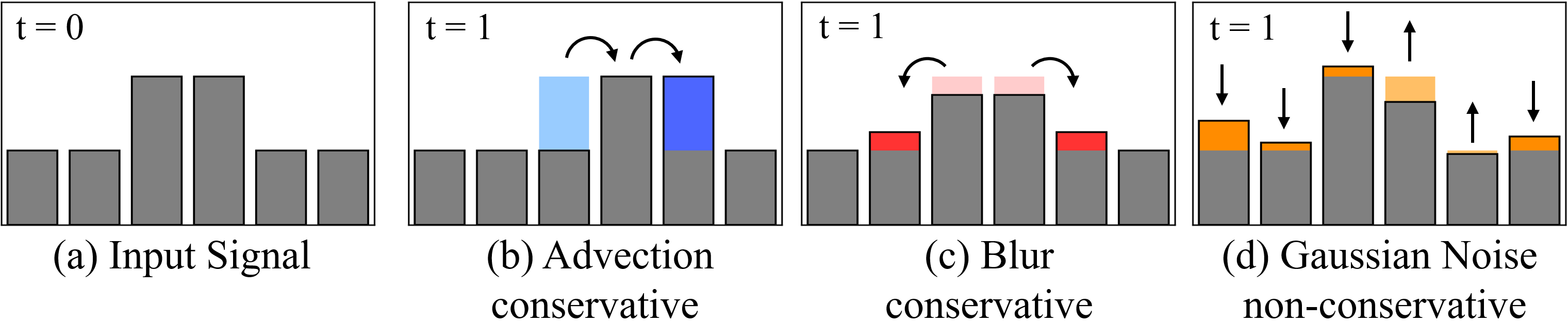}
   \caption{Corruption process: (a) input image, (b) advection  and (c) blur ``redistribute'' the intensities but preserve the total ``mass'', i.e., pixel-intensity sum (conservative). (d) Gaussian noise adds or subtracts ``mass'' (non-conservative). }
   \label{fig:ade_terms_visualized}
   \vspace{-8pt}
\end{figure}

\begin{figure}[h]
    \centering
    \includegraphics[width=\linewidth]{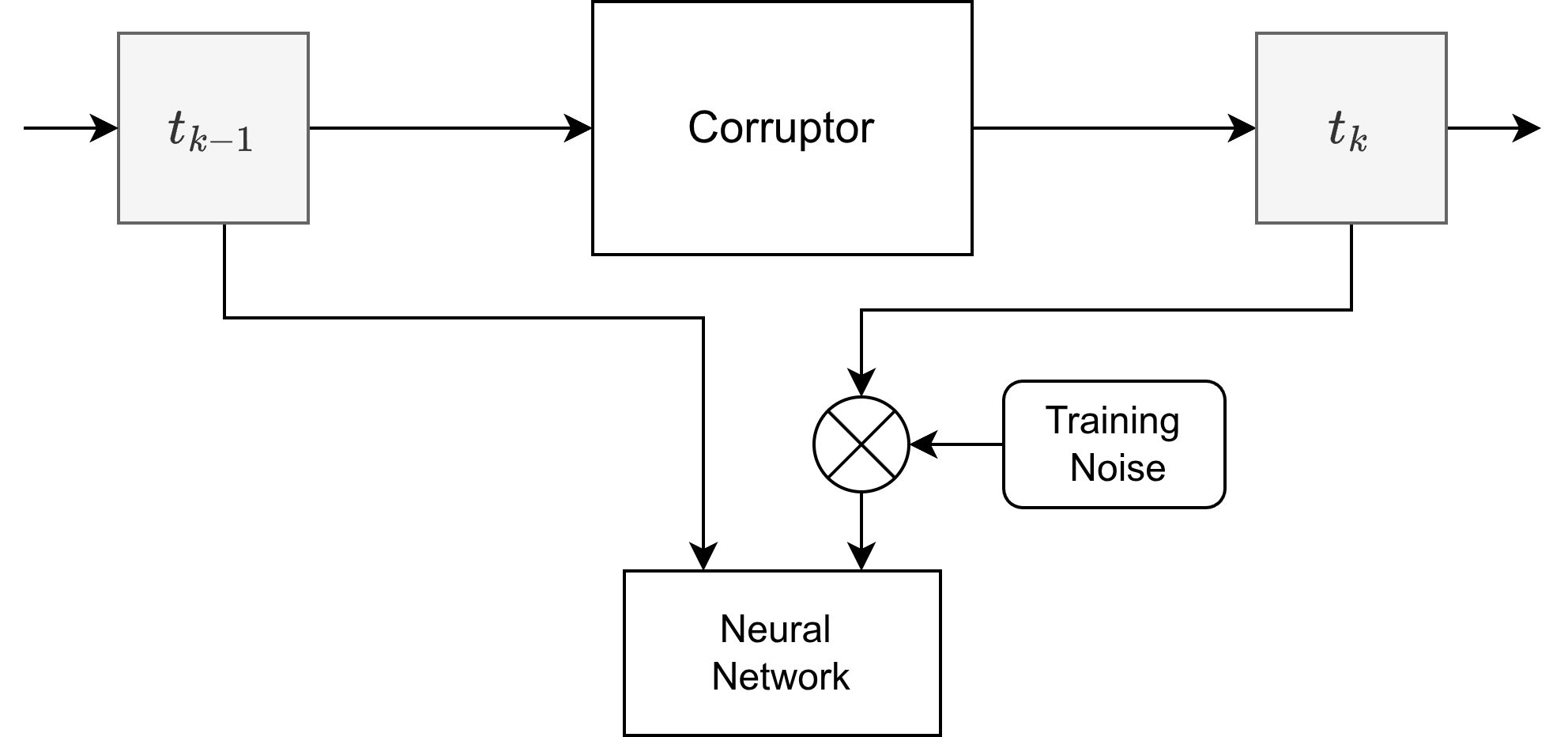}
    \caption{An overview of the NN training pipeline.
    The image corruptor applies advection--diffusion operator $\mathcal{A}(t_k)$ during each of the discrete time steps. 
    The NN is trained on pairs of images destroyed up to the prescribed time, as dictated by the scheduler.}
    \label{fig:trainingflowchart}
\end{figure}

\cref{fig:ade_terms_visualized} shows that the redistribution of pixels' intensities can be achieved solely by the reaction term (addition/subtraction) or via the advection-diffusion operator which would shift the intensity (being a vector flux rather than a scalar reaction term).
Interestingly, from the point of view of a standalone observer who sees only the effective change in intensity, the underlying processes may be indistinguishable.
Our research aims to explore the advection effect while keeping the stochastic \textit{scalar} noise marginal.
Consequently, in our method we do \emph{not} inject noise directly into the PDE itself in order to keep the process conservative.
Let as define the $\mathcal{A}(t_k)\bigl[u \bigr]$ as shorthand for the advection-diffusion terms given in \cref{eq:advection_diffusion_pde}. 
This equation is solved numerically for $k$ steps using a numerical LBM solver as described in \cref{sec:gpu-lbm-solver}.
We add a small Gaussian perturbation just before passing the data to the neural network,
\begin{equation}
u_{k}(x,y)
\;=\;
\underbrace{\mathcal{A}(t_k)\bigl[u_{0}\bigr]}_{\substack{\text{advection-diffusion} \\ \text{forward chain}}}
\;+\;
\underbrace{\epsilon_t \sim \mathcal{N}(\bm0, \sigma_T \bm I)}_{\substack{\text{Gaussian} \\ \text{training noise}}} \,.
\label{eq:ADE_fwd_eq}
\end{equation}
Here, $\sigma_T = 0.01$ is a small constant during all steps. 
In contrast to classical DDPM, the noise is not accumulated along the corruption process.
As illustrated in Fig.\ref{fig:trainingflowchart}, this separation keeps the corruption process physically consistent.
The primary purpose of these small Gaussian perturbations is to alleviate the accumulation of numerical errors (c.f. \citet{daras2022softdiffusionscorematching}). 
Readers interested in work with high noise-to-blur ratios are referred to \cite{hoogeboom2022blurring}.

\subsection{Turbulent Velocity Field Generator}
\label{sec:velocity-field-generator}
The presented advection--diffusion process relies on a velocity field
\(\mathbf{v} = \mathbf{v}(\mathbf{x},t)\) to transport image intensities during the forward process. 
In real-world phenomena, the time varying, turbulent velocity fields exhibit both large and small-scale coherent flows \cite{wilcox2006turbulence}. 
To capture these multi-scale effects we propose a \emph{Spectral Turbulence Generator} that synthesizes a \(\mathbf{v}\) with controllable spectral energy. %
 
The turbulent velocity field can swirl, convect, or otherwise \emph{mix} image content over time, along the forward corruption trajectory. 
Our turbulence generator achieves this by parameterizing \(\mathbf{v}\) in the frequency domain and drawing random phases to yield diverse realizations. 
Moreover, the spectral parametrization allows to control energy at specific wavenumbers, thereby tailoring how features are displaced over different spatial scales.

\vspace{-10pt}
\paragraph{Spectral Parameterization.}
The synthetic turbulent velocity field is inspired by the Random Fourier Modes (RFM) approach \cite{Fung_Hunt_Malik_Perkins_1992}. 
The core idea is to construct a velocity field in the Fourier space by assigning random phases to Fourier modes whose amplitudes are determined by a prescribed energy spectrum.  
Consider the two-dimensional domain \(\Omega \subset \mathbb{R}^2\) with coordinates
\(\mathbf{x} = (x,y)\).
We form discrete wavevectors \(\boldsymbol{\kappa}=(\kappa_x,\kappa_y)\) on a grid 
determined by the spatial resolution \(N \times N\).
Let us denote the 2D velocity field $\mathbf{v} = [v_x, v_y]$ and its Fourier transform as $\hat{\mathbf{v}}(\mathbf{k}) = \mathcal{F} (\mathbf{v})$.
In the RFM approach, the time-varying velocity components are generated in the spectral space as,
\begin{align}
\nonumber 
\hat{v_x}(\boldsymbol{\kappa}) &= A(\|\boldsymbol{\kappa}\|) e^{i (\phi(\boldsymbol{\kappa})  + \omega t)} \nonumber \\ 
\nonumber 
\hat{v_y}(\boldsymbol{\kappa}) &= A(\|\boldsymbol{\kappa}\|) e^{i (\phi(\boldsymbol{\kappa}) + \omega t)},
\end{align}
where $\|\boldsymbol{\kappa}\| =  \sqrt{k_x^2 + k_y^2}$ is the wavenumber magnitude.
The angular frequency $\omega$ is calculated as $\omega = \|\boldsymbol{\kappa}\| \, dt $ where we set $dt = 10^{-4}$ as a small increment so that phase evolution remains slow and stable in the spectral domain. 
The random phases are denoted as $\phi(\boldsymbol{\kappa})$ and are uniformly distributed in $[0,2\pi)$.
The amplitude $A(\|\boldsymbol{\kappa}\|)$ is determined by the prescribed energy spectrum  $E(\|\boldsymbol{\kappa}\|) \propto \|\boldsymbol{\kappa}\| ^{-2}$. 

\begin{figure}[t]
    \centering
    \begin{subfigure}[t]{0.443\linewidth}
        \centering
        \includegraphics[width=\linewidth]{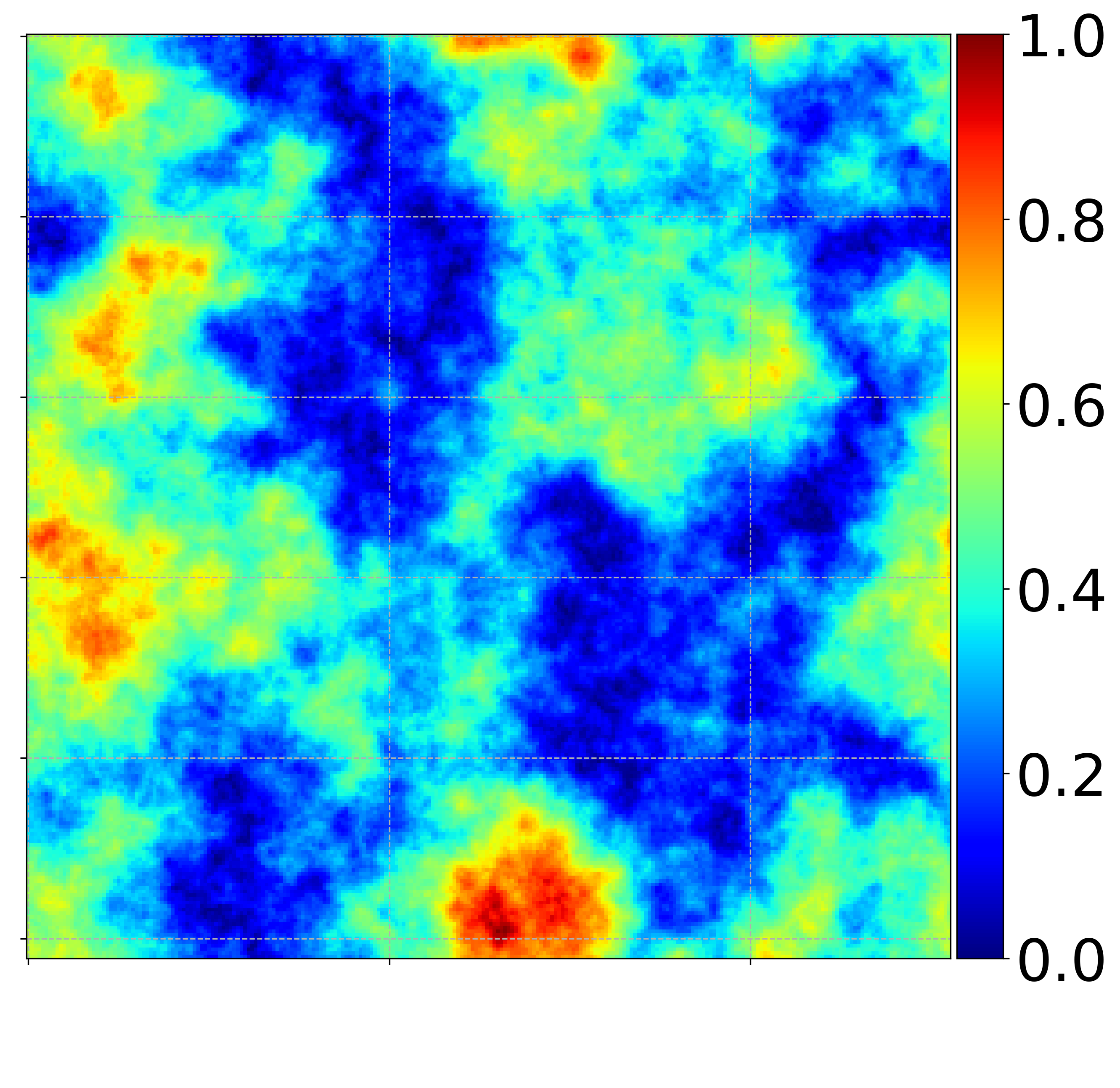}
        \caption{Generated velocity field. Colored by the velocity field magnitude (normalized).} %
        \label{fig:force_plt_mag}
    \end{subfigure}\hfill\hfill
    \begin{subfigure}[t]{0.435\linewidth}
        \centering
        \includegraphics[width=\linewidth]{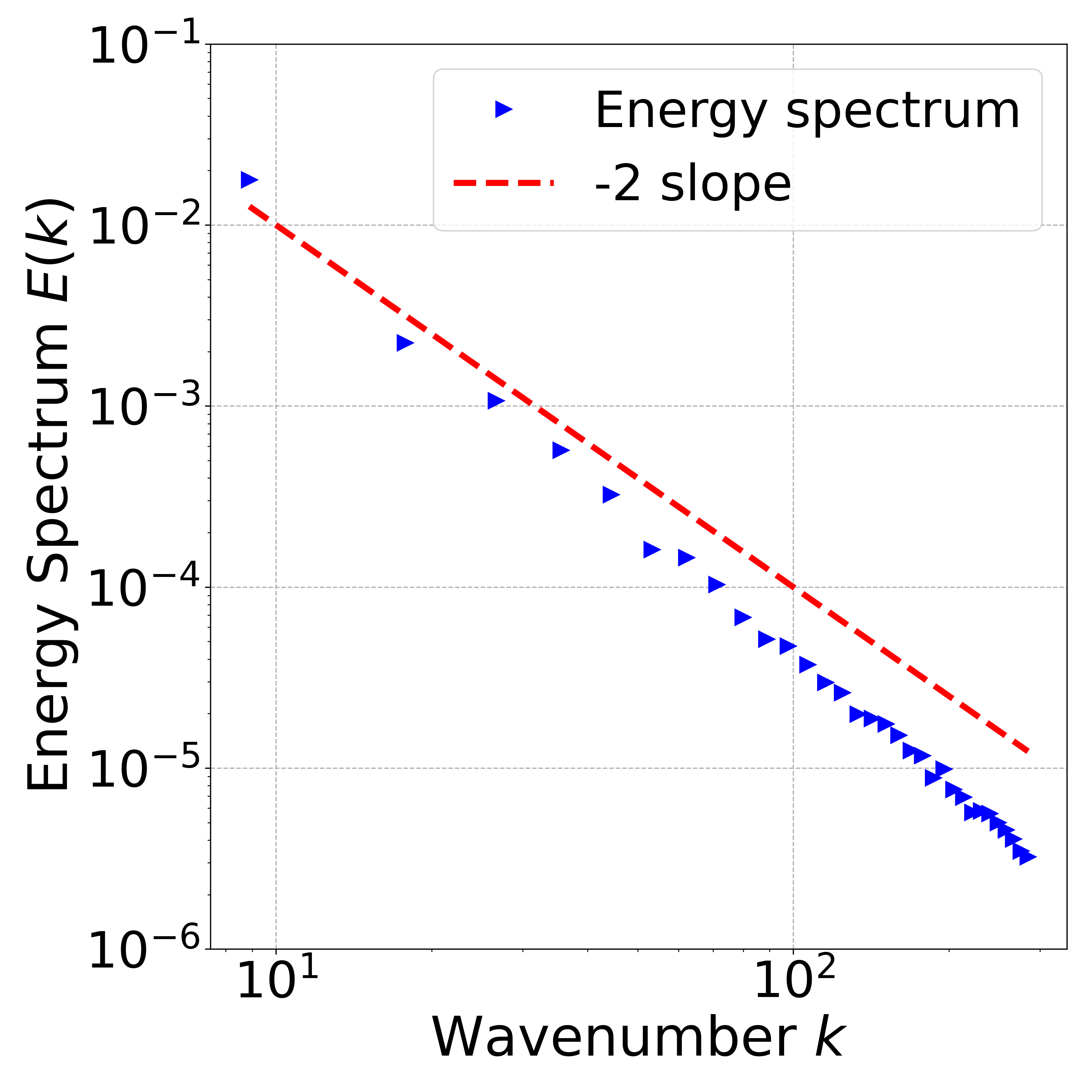}
        \caption{Spectrum of the turbulent velocity field obeys the imposed decay slope.} %
        \label{fig:force_spectrum}
    \end{subfigure}
    \caption{Generated turbulent velocity field and its corresponding spectrum. Please refer to Section~\ref{sec:velocity-field-generator} for details. }
    \label{fig:force_combined}
\end{figure}

We restrict \(\boldsymbol{\kappa}\) to a chosen band to shape the flow’s dominant frequencies; in our experiments we choose \([\kappa_{\min}, \kappa_{\max}] = [2\pi / N,  2 \pi / (1024N) ] \).
By inverse Fourier transform we obtain the spatial velocity components \(\mathbf{v}(\mathbf{x},t)\)  
updated across time steps \(t=1,\dots,t_N\). 
This approach naturally handles swirling,
drifting, and other turbulent-like motions when \(\kappa_{\max}\) is large enough to include fine-scale eddies.
Fig.~\ref{fig:force_combined} depicts an example of the generated velocity field and its corresponding spectrum.

\vspace{-10pt}
\paragraph{Velocity Magnitude Limiting.}
Our generator can also cap the velocity magnitude to control numerical stability by maintaining realistic motion speeds. 
Specifically, after transforming to the spatial domain 
we apply a soft-limiter \(\tanh(\cdot)\)-based function to ensure that the generated field does not exceed a maximum threshold, $\|\mathbf{v}(\mathbf{x},t)\|<10^{-3}$. One can vary this threshold to tune the swirl intensity or keep flows numerically stable. 

\vspace{-10pt}
\paragraph{Integration into the PDE.}
At forward step \(k\) the generator outputs \(\mathbf{v}_k(\mathbf{x})\). Substituting into
\eqref{eq:advection_diffusion_pde}, we solve for the advected and diffused image field. Over multiple
timesteps, this results in a gradual flow-driven smoothing that captures both large-scale drift and
fine-grained turbulence.

\subsection{Lattice Boltzmann Solver}
\label{sec:gpu-lbm-solver}

While one can solve the heat equation in the frequency domain, the presence of advection makes an efficient real-space discretization necessary. 
We adopt the \emph{Lattice Boltzmann Method} (LBM), which models the macroscopic field \(u(x,y,t)\) via particle-like \emph{distribution functions} at each grid node:
\(\{u_i(x,y,t)\}_{i=0,\dots,8}\) in a D2Q9 arrangement, where D is the number of dimensions and Q is the number of discrete distributions.
Conceptually, \(u_i(x,y,t)\) denotes distributions moving in one of nine discrete velocity directions, so that \(u(x,y,t)=\sum_i u_i(x,y,t)\) recovers the
physical density (or intensity) at \((x,y)\). It can be shown \citep{krger2017lattice}, that the LBM algorithm naturally handles advection and diffusion,
and easily scales to large images on GPUs.

\vspace{-10pt}
\paragraph{The LBM Routine.}
At each time step, the LBM applies the following fundamental stages:
\begin{enumerate}
\item \emph{Compute equilibrium distribution}:
The advection is driven by the externally generated velocity field $\mathbf{v}$ which influences the equilibrium distribution: 
\begin{align}
\nonumber 
 u_i ^{eq}(\mathbf{x},t, \mathbf{v}) =  w_i u 
 \left[ 1 + \frac{\mathbf{c}_i \mathbf{v}}{c_s^2 } + \frac{ (\mathbf{c}_i \mathbf{v})^2}{2 c_s^4} - \frac{\mathbf{v}^2 } {2c_s^2 } \right] \,,
\end{align}
where $c_s^2 = \frac{1}{3}$, is the lattice speed of sound and $ w_i$ is a directional weight.
The relaxation time \(\tau\) is directly correlated with the diffusion coefficient 
$ \alpha = c_s^2 \left(\tau - \frac{1}{2} \right)$.

\item \emph{Collision}: The local distributions $u_i$  at each grid node relax toward an equilibrium
        \(u_i^{\mathrm{eq}}(u,\mathbf{v})\) at a rate \(\tfrac{1}{\tau}\),
       \begin{align} \nonumber 
          \tilde{u}_i(x,y,t)
          \;=\;
          u_i(x,y,t)
          \;-\;
          \frac{1}{\tau}\bigl[u_i(x,y,t)-u_i^{\mathrm{eq}}\bigr].
        \end{align}
        
\item \emph{Streaming}: The post-collision distributions \(\tilde{u}_i\) shift (or``stream'') to their
        neighboring nodes according to the discrete lattice directions \((c_{ix},\,c_{iy})\). On the D2Q9 lattice, each node has eight neighbors and the rest particle 
        \begin{align}
          u_i\bigl(x+c_{ix},\,y+c_{iy},\,t+\Delta t\bigr)
          \;=\;
          \tilde{u}_i(x,y,t).
        \end{align}
\end{enumerate}
After streaming, one sums the updated \(u_i\) to recover the macroscopic field \(u=\sum_i u_i\). 
Physical boundary conditions are implemented using the bounce-back rule which simply reverses the distributions back into the domain and achieves the no-flux condition.

\vspace{-10pt}
\paragraph{Computational Advantages.}
The LBM collision and streaming are purely local memory operations, ideal for GPU acceleration. This enables
\(\mathcal{O}(HW)\) updates on an \(H\times W\) grid with limited communication overhead. 
The result is a stable, easily parallelizable solver that captures both directional flow and diffusion in one coherent framework.
We implement the LBM scheme in CUDA using the Taichi framework \citep{hu2019taichi}. 
The distribution functions are stored in a
structure-of-arrays layout for memory coalescing. 
The explicit LBM stencil in the pseudo-code is available in the Appendix C of our work.

\subsection{Dimensionless PDE Formulation}
\label{sec:dimensionless-advection-diffusion}

A key step to a fair comparison of our advection--diffusion method to others (we compare to the IHD method~\cite{Rissanen2022GenerativeMW})
is to express the forward PDE in a dimensionless form. 
We do this by introducing two dimensionless parameters: 
the \emph{Fourier number} (\(\mathrm{Fo}\)) and the \emph{Peclet number} (\(\mathrm{Pe}\)) which together
control how much diffusion and advection is applied. 
This strips away arbitrary scale factors (like image resolution or absolute blur sizes), letting us match the essential ``diffusion budget'' (Fourier number) and ``flow to diffusion ratio'' (Peclet number) between methods so that any performance differences are not merely due to mismatched scales.  

\vspace{-8pt}
\paragraph{Characteristic Scales.}
Let \(\alpha\) be the diffusion coefficient from \cref{eq:advection_diffusion_pde}, \(L\) a characteristic
length scale (e.g., the image width), and \(V\) the maximum or typical speed for the velocity field
\(\mathbf{v}\). We define the dimensionless spatial/time coordinates as 
\begin{align}
\nonumber
  \mathbf{x}^* \;=\;\frac{\mathbf{x}}{L}, 
  \quad
  t^* \;=\; \frac{\alpha \, t}{\,L^2\,},
\end{align}
with \(\nabla = \frac{1}{L}\nabla^*\) and 
\(\frac{\partial}{\partial t} = \frac{L^2}{\alpha}\frac{\partial}{\partial t^*}\). We then scale the velocity with \(\mathbf{v}^*(x^*,t^*) = \tfrac{\,\mathbf{v}(x,t)\,}{V}\). Reinterpreting the image
\(u(x,t)\) as \(u^*(x^*,t^*)\) completes the non-dimensional setup.

\newcommand{\Fo}{\mathrm{Fo}}
\newcommand{\Pe}{\mathrm{Pe}}

\vspace{4pt}
\noindent
\textbf{Fourier Number (Fo).\quad}
When discretizing the PDE into time increments of length \(\Delta t\), we define
\begin{align}
  \Fo
  \;=\;\frac{\alpha\,\Delta t}{\,L^2\,}
  \;=\;\Delta t^*,
\end{align}
which dictates how much diffusion occurs per time step in dimensionless time. 
Equating \(\Fo\) across
different forward processes ensure that the same corruption schedule and the same physical time apply to data at different resolutions.

\vspace{-8pt}
\paragraph{Peclet Number (Pe).}
To quantify the ratio of advective transport to diffusion rate, we use 
\begin{align}
   \Pe 
   \;=\;\frac{\,V\,L\,}{\alpha}.
\end{align}
Increasing \(\mathrm{Pe}\) intensifies the directional flow.

\vspace{-8pt}
\paragraph{Dimensionless PDE.}
Starting from
\(\displaystyle \frac{\partial u}{\partial t}
\;+\;\alpha\,(\mathbf{v}\!\cdot\nabla u)
  \;=\;\alpha\,\nabla^2 u,\)
and applying the above scalings yields
\begin{equation}
  \label{eq:FoPe-AdvectionDiffusion}
  \frac{\partial u^*}{\partial t^*}
  \;+\;
  \mathrm{Fo}\,\mathrm{Pe}\,
  \bigl(\mathbf{v}^*\!\cdot\nabla^*\bigr)\,u^*
  \;=\;
  \mathrm{Fo}\,\nabla^{*2}u^*,
\end{equation}
where the diffusion term is scaled by \(\mathrm{Fo}\) and the advection term by \(\mathrm{Fo}\,\mathrm{Pe}\).
When \(\mathrm{Pe}=0\),
\cref{eq:FoPe-AdvectionDiffusion} reduces to forward blur only (no advection). 
In practice, we fix the
\(\mathrm{Fo}\) schedule (thereby defining the diffusivity $\alpha$ per solver step) and then we choose \(\mathrm{Pe}\) to modulate how much advective flow is added. 
This dimensionless framing ensures a clean comparison: both processes consume the same ``diffusion budget'', but differ in how strongly they advect.

\subsection{Scheduler}
\label{sec:scheduler}
In the work of Rissanen et al.~\citep{Rissanen2022GenerativeMW}, the amount of blur has been solely defined by a parameter $\sigma$. 
From the physical standpoint, it be can connected with $\Fo$ as $\sigma = \sqrt{2 t \alpha}$ therefore  $\Fo=\frac{\sigma}{L^2}$.
To compare our results to Rissanen et al. \cite{Rissanen2022GenerativeMW}, we found the exponential schedule most suitable. 
The blur schedule, expressed in dimensionless time $\Fo_t$ for $t = 0, 1, ..., T-1$, is generated using an exponential spacing.
The formula reads,
\begin{align*}
    \Fo_t = \left(\frac{\Fo_{\text{max}}}{\Fo_{\text{min}}}\right)^{\frac{t}{T-1}} \Fo_{\text{min}},
\end{align*}
where $\Fo_{\text{min}}$ and $\Fo_{\text{max}}$ is the initial and final dimensionless time. The value $T$ is the total number of denoising steps in the schedule and $t$ is the index of the current step, ranging from 0 to $T-1$.
Notice that the number of denoising steps $T$ does not have to coincide with the number of steps performed by the numerical solver.
Therefore the relaxation coefficient in conductivity in Eq. \ref{eq:advection_diffusion_pde} has to be calculated for each pair of $Fo_{t}$ and $Fo_{t+1}$.
\subsection{Learning the Reverse Process}
\label{sec:reverse_process}
Given an image, let us denote the field of its pixels intensities as $u_k$. 
It evolves in discrete time steps $k=1,\dots,K$.
We define a \emph{forward} chain $q(u_{1{:}K}\mid u_0)$ by applying our PDE-based advection-diffusion operator plus noise. 
Rather than building a variational bound, the neural network $\NN_\theta$ 
learns differences between chain elements $p_\theta(u_{k-1}) - p_\theta(u_{k})$ in a purely in a regression-style manner. 
Specifically, we start from a corrupted prior $p(u_K)$ (e.g.\ kernel density estimates of blurred and advected training images).
The training and sampling noise, denoted as $\epsilon_T$ and $\epsilon_S$ respectively, is injected between pairs \((u_k,u_{k-1})\) for all $k$.
Their ratio is fixed as $\sigma_T/\sigma_S=1.25$ and acts solely as a mild regularization.

Observe, that the MSE loss in \cref{alg:training_alg} corresponds to a finite difference of $\frac{\partial u}{\partial t} \approx \frac{u_{k-1} - u_{k}}{dt}$ where $dt = 1/K$. 
As the number of steps grows to infinity, $K \to \infty$, the loss converges to the denoising score matching objective (which is the gradient of the log-probability density), re-using the derivation presented in IHD App.\ A.4 ~\cite{Rissanen2022GenerativeMW} .

\begin{algorithm}[t]
\caption{Training}\label{alg:training_alg}
\begin{algorithmic}
	\State $ \mathbf{u}_k, \mathbf{u}_{k-1}, k \leftarrow \text{drawn a training batch}$ %
	\State $ \epsilon_T \sim \mathcal{N}\left(\mathbf{0}, \sigma_T^2\right)$ \Comment{Training noise}
	\State $ \hat{\mathbf{u}}_k \leftarrow \mathbf{u}_k + \epsilon_T$
	\State $ \Delta^{\mathbf{u}} = \NN_\theta(\hat{\mathbf{u}}_k, k)$
	\State $\mathcal{L}_{\text{MSE}}  = \| \Delta^u + (\mathbf{u}_{k-1} - \hat{\mathbf{u}}_k) \|_2^2$
\end{algorithmic}
\end{algorithm}
To speed up the \cref{alg:training_alg}, the $\mathbf{u}_k$ and $\mathbf{u}_{k-1}$ are actually stored in a precomputed training data tensor from which the batch is drawn. 
The precomputation procedure follows \cref{eq:ADE_fwd_eq},  $\mathbf{u}_k = \mathcal{A}(t_k)\bigl[u_{0}\bigr]$. The neural network is an U-Net~\cite{Ronneberger2015UNet} with attention layers, please refer to the supplemental material for further details.

\begin{algorithm}[t]
\caption{Sampling}\label{alg:sampling_alg}
 \begin{algorithmic}
 \State $k \leftarrow K$ \Comment{Start from terminal state}
 \State $\mathbf{u} \sim p\left(\mathbf{u}_K\right)$ \Comment{Sample from the blurry prior}
 \While{$k > 0$}
     \State $ \epsilon_S \sim \mathcal{N}\left(\mathbf{0}, \sigma_S^2\right)$ \Comment{Sampling noise}
     \State $ \hat{\mathbf{u}}_k \leftarrow \mathbf{u}_k + \epsilon_S $
     \State $ \Delta^{\mathbf{u}} \leftarrow \NN_\theta(\hat{\mathbf{u}}_k, k) $
     \State $ \mathbf{u}_{k-1} \leftarrow \hat{\mathbf{u}}_k + \Delta^{\mathbf{u}} $ \Comment{Reverse step}
     \State $ k \leftarrow k-1$
 \EndWhile
 \end{algorithmic}
 \end{algorithm}

\FloatBarrier
\section{Experiments}
\label{sec:experiments}

We evaluate our approach on datasets commonly used in generative modeling: \textrm{FFHQ-128} ($128\times128$ resolution, $70{,}000$ training samples), 
\textrm{MNIST} ($28\times28$ pixels, $60{,}000$ training images, $10{,}000$ testing). We present additional qualitative results on \textrm{LSUN Church} dataset in the Appendix.
To obtain the initial state (blurry prior), we corrupt the clean images with according to the PDE, as described in  \cref{sec:ADE_forward_description}.

\vspace{-10pt}
\paragraph{Impact of Peclet Number on Generated Samples. } %
We first provide qualitative demonstrations of our model’s ability to generate high-fidelity images. 
In all datasets we observe that a directional flow (via \(\mathrm{Pe}\neq 0\)) yields visually richer details. 
\Cref{fig:generative_noise_sig16_variation_subfig} shows side-by-side generated images on the FFHQ-128 dataset, illustrating that even mild Peclet numbers can enhance the local structure.

\renewcommand{\genfacelength}{0.195} %
\renewcommand{\gencolPelength}{0.15} %
\renewcommand{\gencolimglength}{0.85} %

\begin{figure}[t]
  \centering
  \begin{subfigure}[c]{\columnwidth}
    \begin{minipage}[c]{0.15\columnwidth} %
      \centering
      \textbf{Pe}
    \end{minipage}%
    \begin{minipage}[c]{0.9\columnwidth} %
      \begin{minipage}[c]{0.2\columnwidth}
        \centering \textbf{Prior}
      \end{minipage}%
      \hfill
      \begin{minipage}[c]{0.725\columnwidth}
        \centering 
        \textbf{Generated samples}
      \end{minipage}%
    \end{minipage}
  \end{subfigure}

  \begin{subfigure}[c]{\columnwidth}
    \begin{minipage}[c]{\gencolPelength \columnwidth}
      \centering
       (IHD) \\ 0.00
    \end{minipage}%
    \begin{minipage}[c]{\gencolimglength \columnwidth}
      \centering
      \includegraphics[width=\genfacelength \columnwidth]{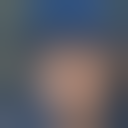}%
      \hfill
      \includegraphics[width=\genfacelength \columnwidth]{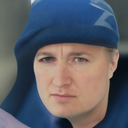}%
      \hfill
      \includegraphics[width=\genfacelength \columnwidth]{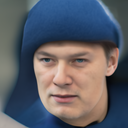}%
      \hfill
      \includegraphics[width=\genfacelength \columnwidth]{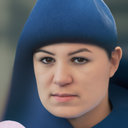}%
      \hfill
      \includegraphics[width=\genfacelength \columnwidth]{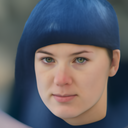}%
    \end{minipage}
  \end{subfigure}

  \begin{subfigure}[c]{\columnwidth}
    \begin{minipage}[c]{\gencolPelength \columnwidth}
      \centering
        0.02
    \end{minipage}%
    \begin{minipage}[c]{\gencolimglength \columnwidth}
      \centering
      \includegraphics[width=\genfacelength \columnwidth]{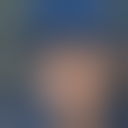}%
      \hfill
      \includegraphics[width=\genfacelength \columnwidth]{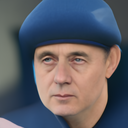}%
      \hfill
      \includegraphics[width=\genfacelength \columnwidth]{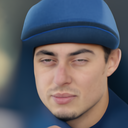}%
      \hfill
      \includegraphics[width=\genfacelength \columnwidth]{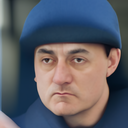}%
      \hfill
      \includegraphics[width=\genfacelength \columnwidth]{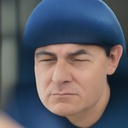}%
    \end{minipage}
  \end{subfigure}

  \begin{subfigure}[c]{\columnwidth}
    \begin{minipage}[c]{\gencolPelength \columnwidth}
      \centering
        0.04
    \end{minipage}%
    \begin{minipage}[c]{\gencolimglength \columnwidth}
      \centering
      \includegraphics[width=\genfacelength \columnwidth]{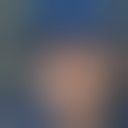}%
      \hfill
      \includegraphics[width=\genfacelength \columnwidth]{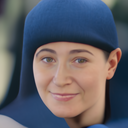}%
      \hfill
      \includegraphics[width=\genfacelength \columnwidth]{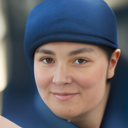}%
      \hfill
      \includegraphics[width=\genfacelength \columnwidth]{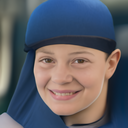}%
      \hfill
      \includegraphics[width=\genfacelength \columnwidth]{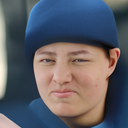}%
    \end{minipage}
  \end{subfigure}

  \begin{subfigure}[c]{\columnwidth}
    \begin{minipage}[c]{\gencolPelength \columnwidth}
      \centering
        0.06
    \end{minipage}%
    \begin{minipage}[c]{\gencolimglength \columnwidth}
      \centering
      \includegraphics[width=\genfacelength \columnwidth]{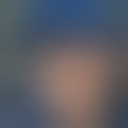}%
      \hfill
      \includegraphics[width=\genfacelength \columnwidth]{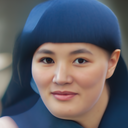}%
      \hfill
      \includegraphics[width=\genfacelength \columnwidth]{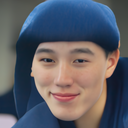}%
      \hfill
      \includegraphics[width=\genfacelength \columnwidth]{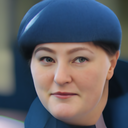}%
      \hfill
      \includegraphics[width=\genfacelength \columnwidth]{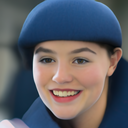}%
    \end{minipage}
  \end{subfigure}
  
  \begin{subfigure}[c]{\columnwidth}
    \begin{minipage}[c]{\gencolPelength \columnwidth}
      \centering
        0.08
    \end{minipage}%
    \begin{minipage}[c]{\gencolimglength \columnwidth}
      \centering
      \includegraphics[width=\genfacelength \columnwidth]{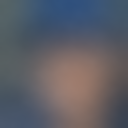}%
      \hfill
      \includegraphics[width=\genfacelength \columnwidth]{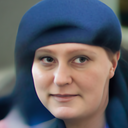}%
      \hfill
      \includegraphics[width=\genfacelength \columnwidth]{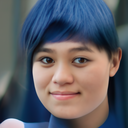}%
      \hfill
      \includegraphics[width=\genfacelength \columnwidth]{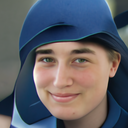}%
      \hfill
      \includegraphics[width=\genfacelength \columnwidth]{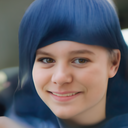}%
    \end{minipage}
  \end{subfigure}

  \begin{subfigure}[c]{\columnwidth}
    \begin{minipage}[c]{\gencolPelength \columnwidth}
      \centering
        0.10
    \end{minipage}%
    \begin{minipage}[c]{\gencolimglength \columnwidth}
      \centering
      \includegraphics[width=\genfacelength \columnwidth]{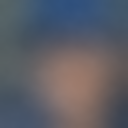}%
      \hfill
      \includegraphics[width=\genfacelength \columnwidth]{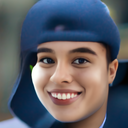}%
      \hfill
      \includegraphics[width=\genfacelength \columnwidth]{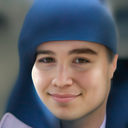}%
      \hfill
      \includegraphics[width=\genfacelength \columnwidth]{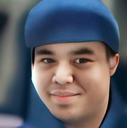}%
      \hfill
      \includegraphics[width=\genfacelength \columnwidth]{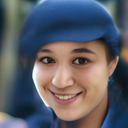}%
    \end{minipage}
  \end{subfigure}

  \begin{subfigure}[c]{\columnwidth}
    \begin{minipage}[c]{\gencolPelength \columnwidth}
      \centering
        0.12
    \end{minipage}%
    \begin{minipage}[c]{\gencolimglength \columnwidth}
      \centering
      \includegraphics[width=\genfacelength \columnwidth]{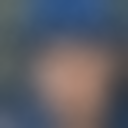}%
      \hfill
      \includegraphics[width=\genfacelength \columnwidth]{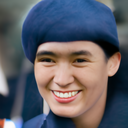}%
      \hfill
      \includegraphics[width=\genfacelength \columnwidth]{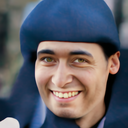}%
      \hfill
      \includegraphics[width=\genfacelength \columnwidth]{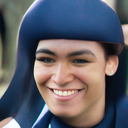}%
      \hfill
      \includegraphics[width=\genfacelength \columnwidth]{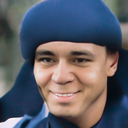}%
    \end{minipage}
  \end{subfigure}

  \begin{subfigure}[c]{\columnwidth}
    \begin{minipage}[c]{\gencolPelength \columnwidth}
      \centering
        0.14
    \end{minipage}%
    \begin{minipage}[c]{\gencolimglength \columnwidth}
      \centering
      \includegraphics[width=\genfacelength \columnwidth]{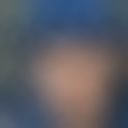}%
      \hfill
      \includegraphics[width=\genfacelength \columnwidth]{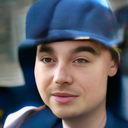}%
      \hfill
      \includegraphics[width=\genfacelength \columnwidth]{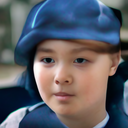}%
      \hfill
      \includegraphics[width=\genfacelength \columnwidth]{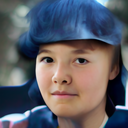}%
      \hfill
      \includegraphics[width=\genfacelength \columnwidth]{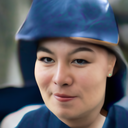}%
    \end{minipage}
  \end{subfigure}
  \caption{Samples for $\sigma =16$, comparing different Peclet numbers. 
  }
  \label{fig:generative_noise_sig16_variation_subfig}
\end{figure}

To quantify the impact of our advection-diffusion model on image quality and generative performance, we report evaluation metrics in \cref{tab:Pe_mega_table} for various Peclet numbers ($Pe$), which control the strength of the advective term. The experiments are conducted on the FFHQ dataset at $128{\times}128$ resolution using 200 sampling steps, and repeated under two final blur conditions, parameterized by spatial diffusion scales $\sigma{=}16$ and $\sigma{=}20$.  
We report Fréchet Inception Distance (FID)\cite{Heusel2017} to measure distributional alignment and the LPIPS metric\cite{zhang2018LPIPS}—to capture perceptual diversity with respect to ground truth images. 
In addition, we evaluate generation fidelity and sample diversity using the Precision, Recall, Density, and Coverage (PRDC) metrics~\cite{naeem2020reliable}. 
As expected, increasing the Peclet number enhances sample diversity (higher $\sigma_{\text{LPIPS}}$ and PRDC scores) but generally degrades fidelity (higher FID), consistent across both $\sigma$ regimes. The arrow annotations in the column headers indicate the preferred direction for each metric.

Hoogeboom et al.~\cite{hoogeboom2022blurring} noticed that better FID metric can be achieved for higher final blur $\sigma$ at the price of a more difficult (longer) training. 
Eventually, the optimal $\Pe$ depends on the final amount of blur.

\begin{table}[t]
\small
\centering
\caption{Evaluation metrics for Peclet numbers on FFHQ $128{\times}128$. Top: $\sigma{=}16$, bottom: $\sigma{=}20$. Lower FID is better. Higher $\mu_{\text{LPIPS}}$ and PRDC values indicate greater diversity and coverage.
When $Pe=0$, the model achieves a baseline values corresponding to the purely blurring approach.
}
\label{tab:Pe_mega_table}
\setlength{\tabcolsep}{5pt} %
\renewcommand{\arraystretch}{1.1} %
\begin{tabularx}{\linewidth}{c *{7}{>{\centering\arraybackslash}X}}
\toprule
Pe & FID\,$\downarrow$ 
   & $\mu_{\text{LPIPS}}$\,$\uparrow$ 
   & $\sigma_{\text{LPIPS}}$\,$\uparrow$ 
   & P\,$\uparrow$ 
   & R\,$\uparrow$ 
   & D\,$\uparrow$ 
   & C\,$\uparrow$ \\
  \midrule
\multicolumn{8}{c}{$\sigma = 16$} \\
\midrule
0 (IHD) & 52.10 & 0.238 & 0.011 & 0.789 & 0.118 & 0.763 & 0.507 \\
0.02  & 52.33 & 0.255 & 0.013 & 0.788 & 0.112 & 0.770 & 0.506 \\
0.04  & 53.54 & 0.260 & 0.014 & 0.797 & 0.119 & 0.803 & 0.499 \\
0.06  & 46.93 & 0.265 & 0.012 & \cellcolor{LimeGreen}0.812 & 0.139 & 0.858 & 0.539 \\
0.08  & 46.24 & 0.248 & 0.017 & 0.806 & 0.153 & 0.870 & 0.565 \\
0.10  & 47.57 & \cellcolor{GreenYellow}0.297 & 0.017 & 0.800 & 0.171 & 0.892 & 0.550 \\
0.12  & \cellcolor{GreenYellow}45.36 & 0.296 & 0.018 & \cellcolor{GreenYellow}0.808 & \cellcolor{GreenYellow}0.173 & \cellcolor{GreenYellow}0.956 & \cellcolor{GreenYellow}0.587 \\
0.14  & \cellcolor{LimeGreen}38.10 & \cellcolor{LimeGreen}0.302 & 0.023 & 0.798 & \cellcolor{LimeGreen}0.223 & \cellcolor{LimeGreen}0.958 & \cellcolor{LimeGreen}0.627 \\
\midrule
\multicolumn{8}{c}{$\sigma = 20$} \\
\midrule
0 (IHD)    & 55.87 & 0.265 & 0.016 & 0.798 & 0.109 & 0.762 & 0.482 \\
0.02  & 56.57 & 0.293 & 0.013 & 0.797 & 0.102 & 0.806 & 0.491 \\
0.04  & 51.44 & 0.286 & 0.022 & 0.815 & 0.115 & 0.921 & 0.539 \\
0.06  & \cellcolor{LimeGreen}36.64 & 0.315 & 0.019 & \cellcolor{LimeGreen}0.826 & \cellcolor{GreenYellow}0.243 & \cellcolor{GreenYellow}1.040 & \cellcolor{LimeGreen}0.665 \\
0.08  & \cellcolor{GreenYellow}37.41 & 0.305 & 0.019 & \cellcolor{GreenYellow}0.817 & \cellcolor{LimeGreen}0.247 & \cellcolor{LimeGreen}1.043 & \cellcolor{GreenYellow}0.662 \\
0.10  & 42.88 & 0.311 & 0.018 & 0.764 & 0.187 & 0.854 & 0.556 \\
0.12  & 48.62 & \cellcolor{GreenYellow}0.344 & 0.022 & 0.688 & 0.183 & 0.683 & 0.510 \\
0.14  & 54.56 & \cellcolor{LimeGreen}0.348 & 0.019 & 0.632 & 0.150 & 0.565 & 0.429 \\
\bottomrule
\end{tabularx}
\end{table}

\vspace{-10pt}
\paragraph{Image Interpolation. }
We also evaluate our method on interpolated initial states and compare it with a baseline approach (IHD). 
Two samples from FFHQ dataset undergo the forward process, followed by linear interpolation in the latent space. 
The interpolated states are then denoised with generative noise and interpolated with SLERP~\citet{SLERP85}.
The analysis provides an insight into the smoothness and consistency of the learned latent space. 
By denoising the linearly interpolated noisy inputs we assess the model's ability to generate coherent transitions. 
\Cref{fig:faces_interpolation} depicts the qualitative results.

\vspace{-10pt}
\paragraph{Discussion and Limitations. }

An immediate drawback of introducing a nonlinear advection term into the diffusion process is the necessity of step-by-step computation up to the given time. 
In practice, we solve the PDE which governs the corruption process for the dataset and store the result to avoid repeated runs.
This procedure consumes $\sim10\%$ of total training time.
In our examples, the corruption process lead to the blurry prior from which new samples can be generated. 
As a consequence, we have not found a distribution (like a standard Gaussian) from which samples could be easily drawn.
However, a sufficiently long process with properly tuned balance between advection and diffusion terms shall converge to 
a well defined stationary turbulent field analogous to the noising Markov process.

\begin{figure*}

\begin{minipage}{\textwidth}
  \centering
  
\begin{minipage}{0.1\textwidth}
   \centering IHD\\Pe=$0.0$
  \end{minipage}%
  \begin{minipage}{0.9\textwidth}
    \centering
    \includegraphics[width=\textwidth]{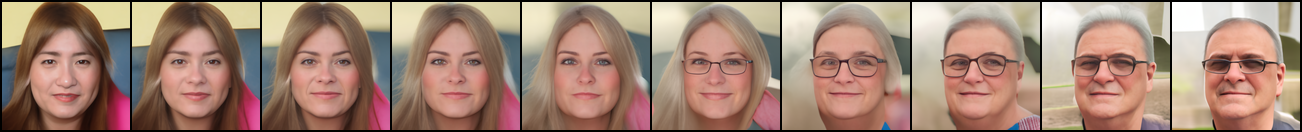}
  \end{minipage}
  \vspace{0.05cm} %

  \begin{minipage}{0.1\linewidth}
	\centering Our\\Pe=$0.02$
  \end{minipage}%
  \begin{minipage}{0.9\linewidth}
    \centering
    \includegraphics[width=\linewidth]{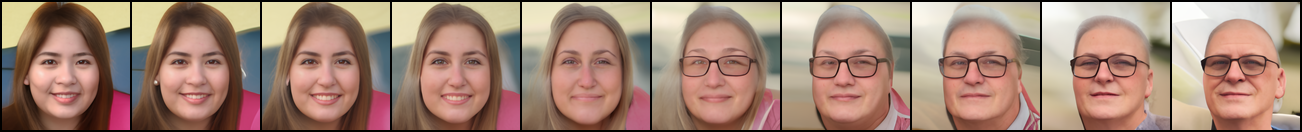}
  \end{minipage}
  \vspace{0.05cm} %
  
  \begin{minipage}{0.1\linewidth}
	\centering Our\\Pe=$0.04$
  \end{minipage}%
  \begin{minipage}{0.9\linewidth}
    \centering
    \includegraphics[width=\linewidth]{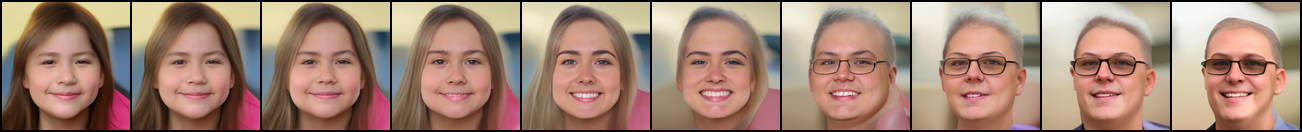}
  \end{minipage}
  \vspace{0.05cm} %
  
\begin{minipage}{0.1\linewidth}
	\centering Our\\Pe=$0.06$
  \end{minipage}%
  \begin{minipage}{0.9\linewidth}
    \centering
    \includegraphics[width=\linewidth]{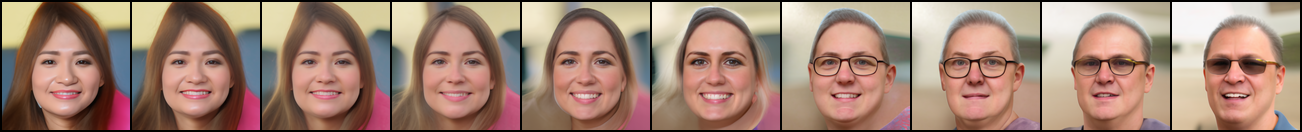}
  \end{minipage}
  \vspace{0.05cm} %
  
  \begin{minipage}{0.1\linewidth}
	\centering Our\\Pe=$0.08$
  \end{minipage}%
  \begin{minipage}{0.9\linewidth}
    \centering
    \includegraphics[width=\linewidth]{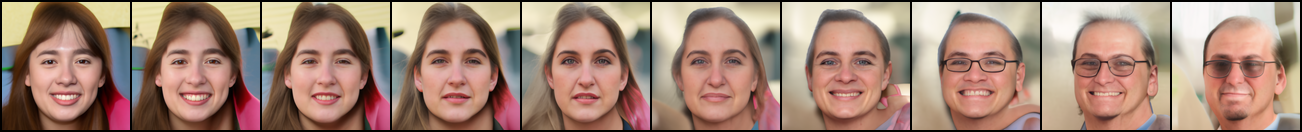}
\end{minipage}
  \vspace{0.05cm} %
  
\begin{minipage}{0.1\linewidth}
	\centering Our\\Pe=$0.10$
  \end{minipage}%
  \begin{minipage}{0.9\linewidth}
    \centering
    \includegraphics[width=\linewidth]{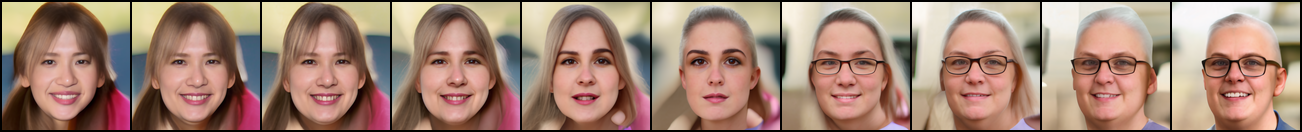}
  \end{minipage}
  \vspace{0.05cm} %

\begin{minipage}{0.1\linewidth}
	\centering Our\\Pe=$0.12$
  \end{minipage}%
  \begin{minipage}{0.9\linewidth}
    \centering
    \includegraphics[width=\linewidth]{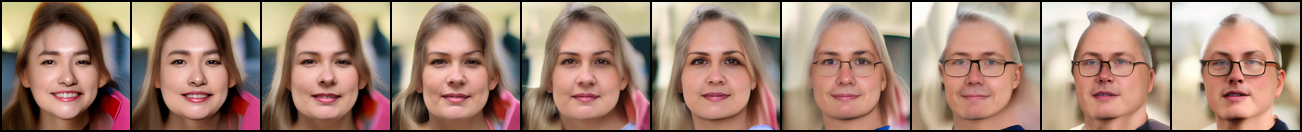}
  \end{minipage}
  \vspace{0.05cm} %

  \begin{minipage}{0.1\linewidth}
	\centering Our\\Pe=$0.14$
  \end{minipage}%
  \begin{minipage}{0.9\linewidth}
    \centering
    \includegraphics[width=\linewidth]{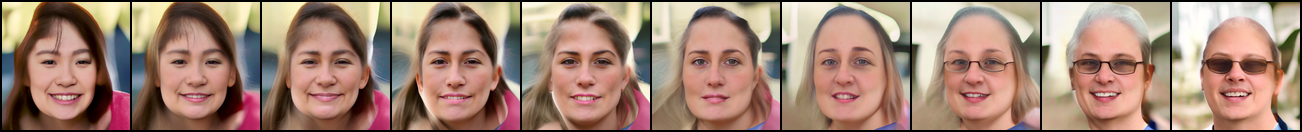}
  \end{minipage}
  \vspace{0.05cm} %
  
  \captionof{figure}{Visual comparison of interpolations between two FFHQ samples.
  Each undergoes the forward process up to $ \sigma =16$, followed by linear interpolation and denoising with SLERP-interpolated generative noise added at each step. } %
  \label{fig:faces_interpolation} %
\end{minipage}
    
\end{figure*}

\section{Future Outlook and Conclusions.}
\label{sec:conclusion}
We proposed a novel PDE-based diffusion model based on the advection-diffusion equation. The model introduces a turbulent mixing term to the forward diffusion process, which, to our knowledge, has not been attempted before. 
In our experiments, we showed that adding the advective term besides the diffusion one improve the quality (FID) of generated images compared to the baseline approach (IHD)~\cite{Rissanen2022GenerativeMW}.
The external velocity field, which can transfer pixels' intensities in a spatially coherent way, offers a new way of introducing corruption into the forward process.  
In the future, the influence of different turbulence generators and the slope of turbulent kinetic energy can be further investigated. 
The interplay with a Gaussian noise corruption as proposed in \citet{hoogeboom2022blurring} also seems a natural direction of research.
Finally, it would be interesting to evaluate the model using different training objective, 
for instance the \textit{flow matching} approach~\cite{lipman2023flowmatchinggenerativemodeling}. 
Then, the effect of the ``overall'' destruction of the image (final state) could be investigated separately from the forward corruption trajectory dictated by a particular of PDE.  

\section*{Acknowledgments}
\label{app:acknowledgement}
We gratefully acknowledge Polish high-performance computing infrastructure PLGrid (HPC Center: ACK Cyfronet AGH) for providing computer facilities and support within computational grant no. PLG/2025/017969. We would also like to thank Kamil Deja and Przemyslaw Spurek for aiding us with in-house reviews.

{
\balance
\small    
\bibliographystyle{ieeenat_fullname}
\bibliography{bibliography}
}

\clearpage
\onecolumn

\appendix

\section*{Appendix}
In this appendix, 
\S\ref{app:nn_architecture} contains hyperparameters and experiment setup.
\S\ref{app:additional_samples} contains additional samples and interpolations for FFHQ-128, MNIST and LSUN Church datasets.
\S\ref{app:pseudocode} contains solver implementation details.

\section{Hyperparameter settings}
\label{app:nn_architecture}
We present the complete experimental configuration for our model architecture, including network topology details and optimization parameters.
Our implementation utilizes a modified U-Net architecture with residual blocks and multi-head self-attention layers. Table \ref{tab:app:hyperparameters} summarizes the dataset-specific configurations. The spatial resolution and batch size were selected to maximize GPU memory utilization while maintaining stable training dynamics.

\begin{table*}[h]
    \normalsize
    \centering
    \caption{Neural network hyperparameters used during experiments on different datasets.}
    \begin{tabular}{c c c c c c c c}
        \hline
        Dataset & Network param. & Layer multipliers & Base Channels & Learning rate & Resolution & Batch & Attention lvls   \\
        \hline
        \hline
        FFHQ & 210904835 & (1, 2, 3, 4, 5) & 128 & 2e-05 & 128$\times$128 & 32 & (2, 3, 4)   \\
        LSUN & 261828227 & (1, 2, 3, 4, 5) & 128 & 2e-05 & 128$\times$128 & 32 & (2, 3, 4)   \\
        MNIST & 42082049 & (1, 2, 2) & 128  & 2e-04 & 28$\times$28 & 128 & (2,)         \\
        \hline

    \end{tabular}
    \label{tab:app:hyperparameters}
\end{table*}
All experiments were conducted on NVIDIA A100 GPUs using PyTorch 2.6.0. The FFHQ model trained for 1M iterations ($\approx$146 hours), MNIST converged within 500k iterations ($\approx$32 hours), we additionally train a model for LSUN Church dataset for 500k iterations ($\approx$ 80 hours). We employed random horizontal flipping (p=0.5) for FFHQ augmentation, with no augmentation applied to MNIST and LSUN Church.

\section{Additional samples}
\label{app:additional_samples}

\begin{figure*}[p]
    \centering
    \adjustboxset{scale=0.8, valign=c}
    \begin{tabular}{w{c}{40pt} c c}
        & \textbf{$\sigma=16$ } & \textbf{$\sigma=20$} \\
        \addlinespace[10pt] %
        
        \begin{subfigure}{\linewidth}
        \centering
        (IHD) \\ Pe=$0.00$
        \end{subfigure} &
        \begin{subfigure}{0.4\linewidth}
        \centering
        \includegraphics[trim={0 522pt 0 0}, clip, width=\linewidth, valign=c]{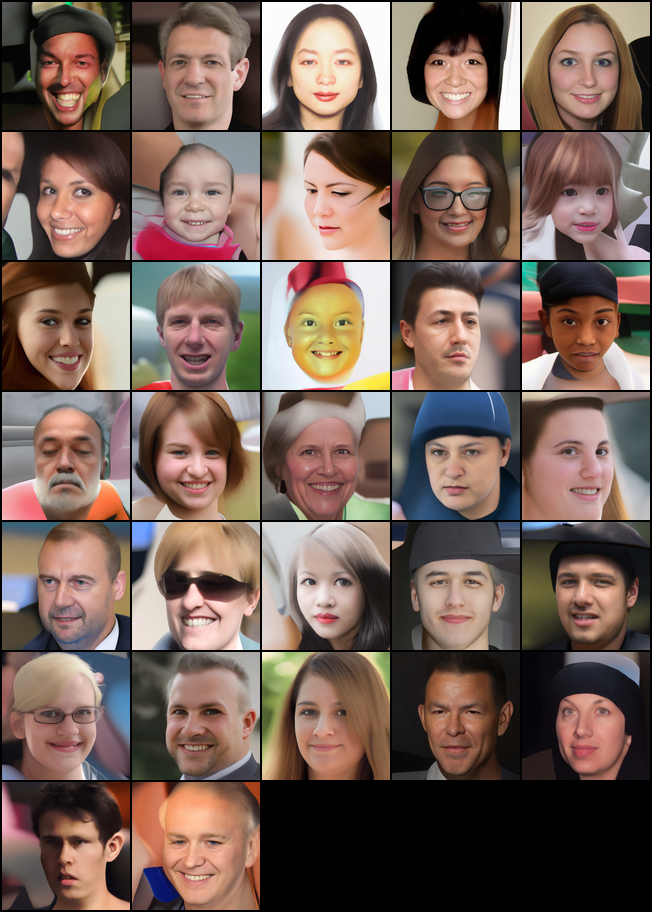}
        \end{subfigure} &
        \begin{subfigure}{0.4\linewidth}
        \centering
        \includegraphics[trim={0 522pt 0 0}, clip, width=\linewidth, valign=c]{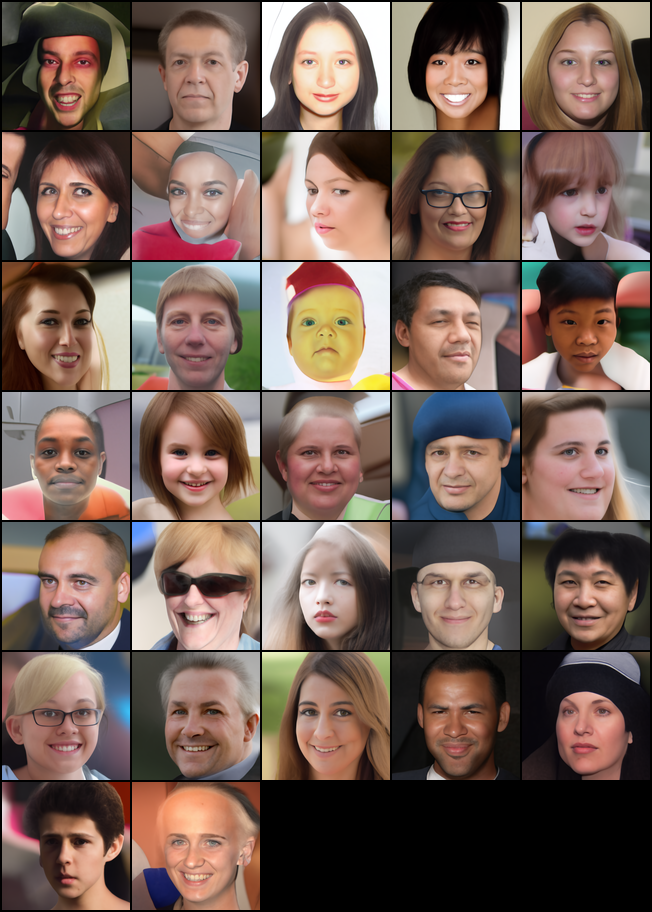}
        \end{subfigure} \\
        \addlinespace[5pt] %

        \begin{subfigure}{\linewidth}
        \centering
        Pe=$0.02$
        \end{subfigure} &
        \begin{subfigure}{0.4\linewidth}
        \centering
        \includegraphics[trim={0 522pt 0 0}, clip, width=\linewidth, valign=c]{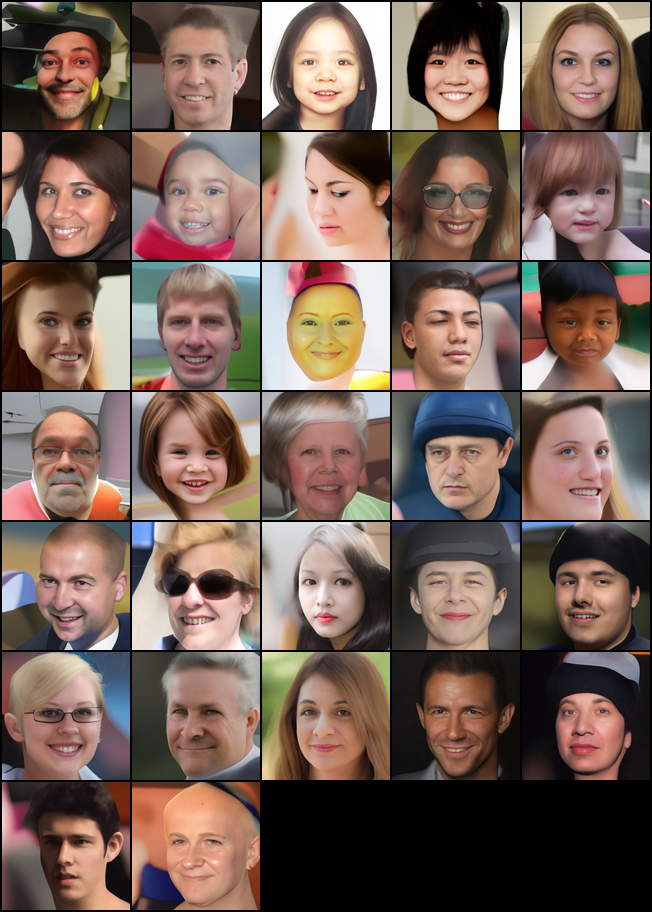}
        \end{subfigure} &
        \begin{subfigure}{0.4\linewidth}
        \centering
        \includegraphics[trim={0 522pt 0 0}, clip, width=\linewidth, valign=c]{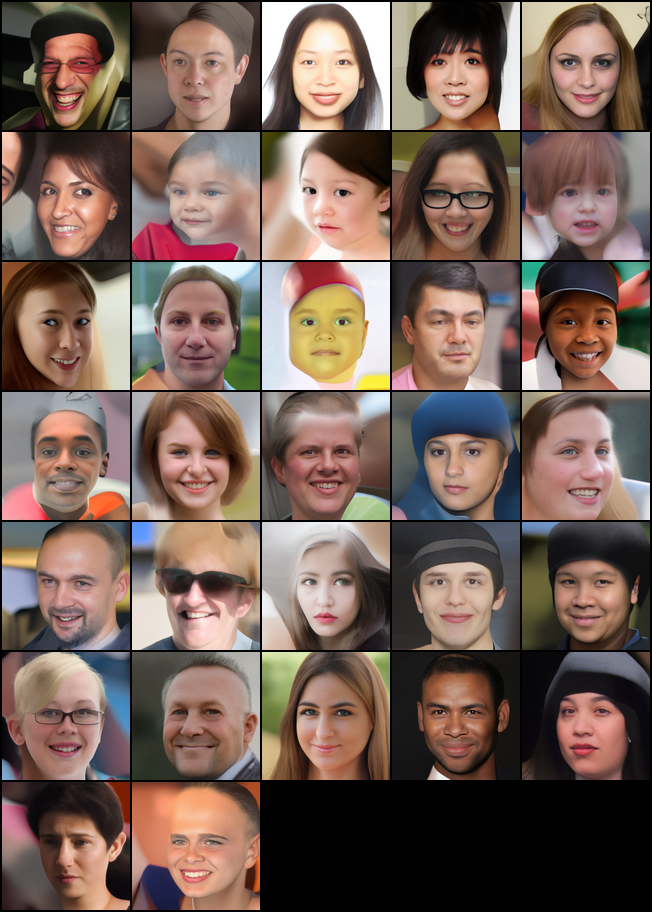}
        \end{subfigure} \\
        \addlinespace[5pt] %

        \begin{subfigure}{\linewidth}
        \centering
        Pe=$0.04$
        \end{subfigure} &
        \begin{subfigure}{0.4\linewidth}
        \centering
        \includegraphics[trim={0 522pt 0 0}, clip, width=\linewidth, valign=c]{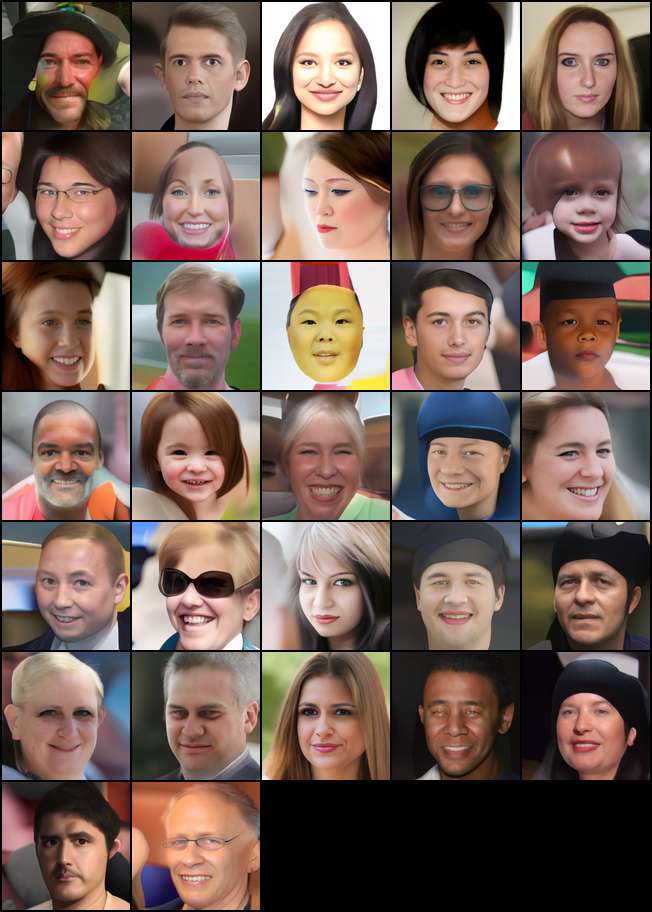}
        \end{subfigure} &
        \begin{subfigure}{0.4\linewidth}
        \centering
        \includegraphics[trim={0 522pt 0 0}, clip, width=\linewidth, valign=c]{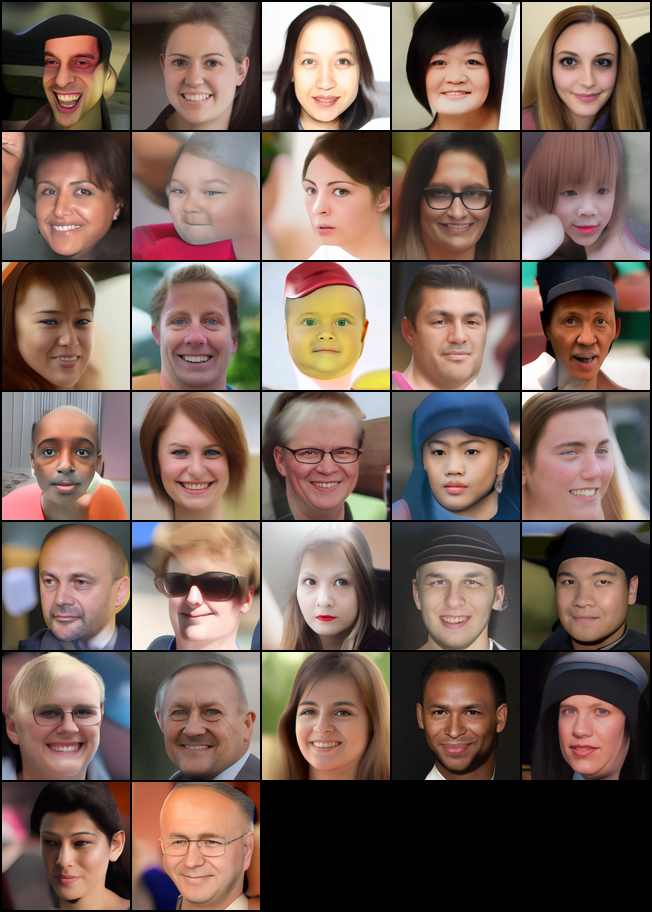}
        \end{subfigure} \\
        \addlinespace[5pt] %

        \begin{subfigure}{\linewidth}
        \centering
        Pe=$0.06$
        \end{subfigure} &
        \begin{subfigure}{0.4\linewidth}
        \centering
        \includegraphics[trim={0 522pt 0 0}, clip, width=\linewidth, valign=c]{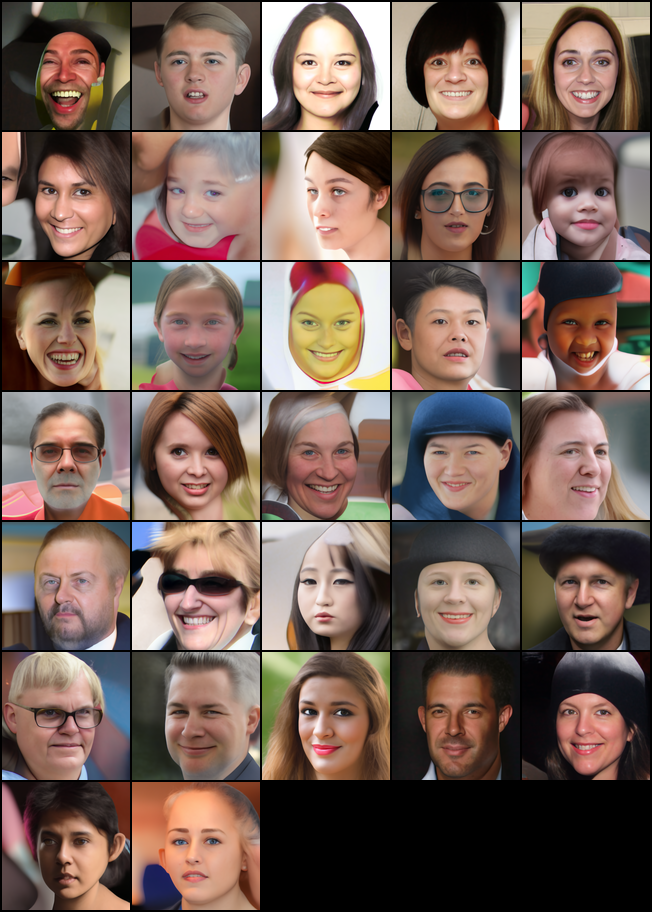}
        \end{subfigure} &
        \begin{subfigure}{0.4\linewidth}
        \centering
        \includegraphics[trim={0 522pt 0 0}, clip, width=\linewidth, valign=c]{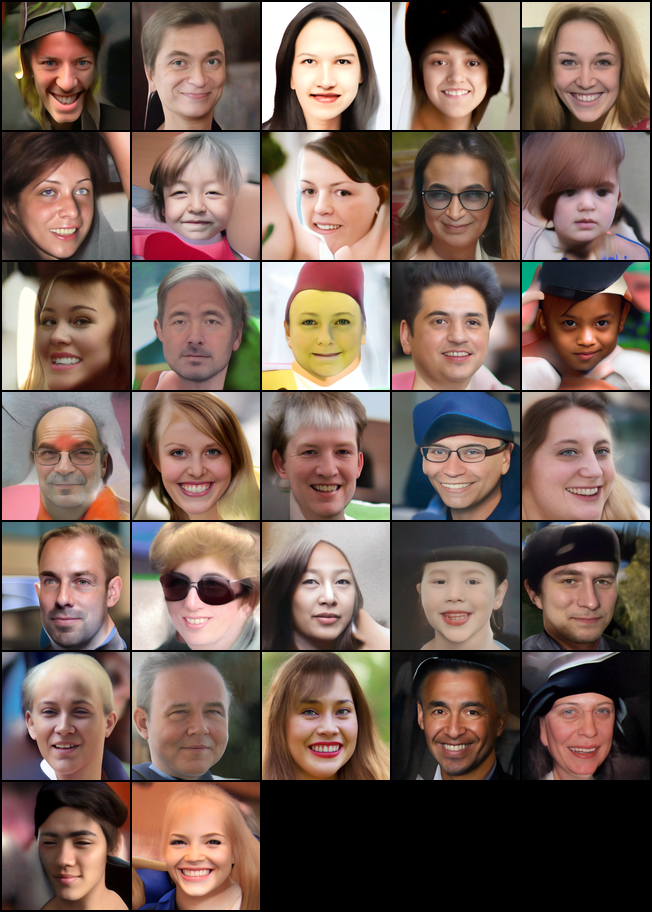}
        \end{subfigure} \\
        \addlinespace[5pt] %
        
    \end{tabular}
    \caption{Visual comparison of the results of our method and the IHD method on the FFHQ dataset.}
    \label{fig:visual_comp_ffhq_part1}
\end{figure*}

\begin{figure*}[p]
    \centering
    \adjustboxset{scale=0.8, valign=c}
    \begin{tabular}{w{c}{40pt} c c}
        & \textbf{$\sigma=16$ } & \textbf{$\sigma=20$} \\
        \addlinespace[10pt] %
        
        \begin{subfigure}{\linewidth}
        \centering
        Pe=$0.08$
        \end{subfigure} &
        \begin{subfigure}{0.4\linewidth}
        \centering
        \includegraphics[trim={0 522pt 0 0}, clip, width=\linewidth, valign=c]{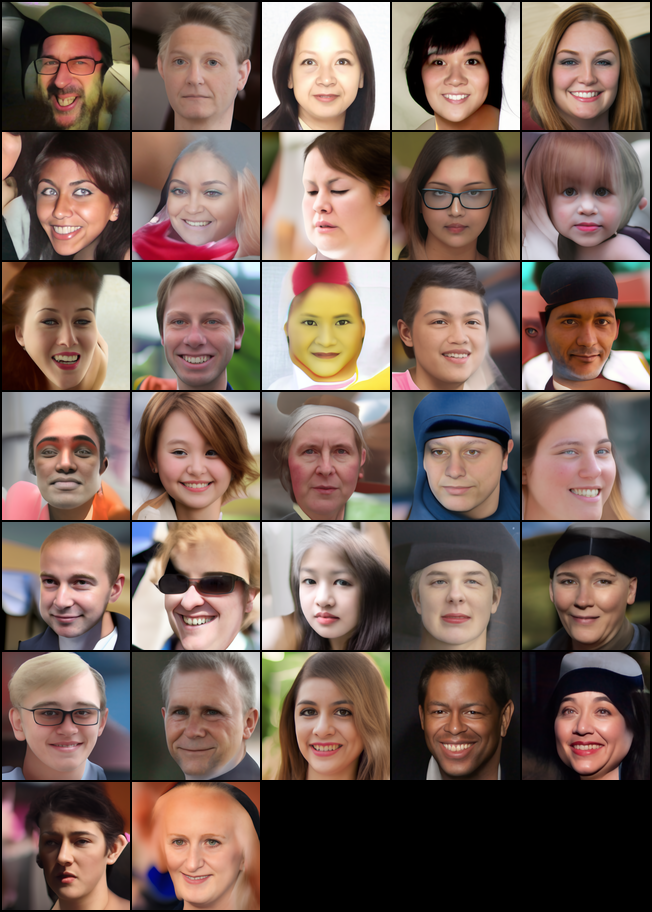}
        \end{subfigure} &
        \begin{subfigure}{0.4\linewidth}
        \centering
        \includegraphics[trim={0 522pt 0 0}, clip, width=\linewidth, valign=c]{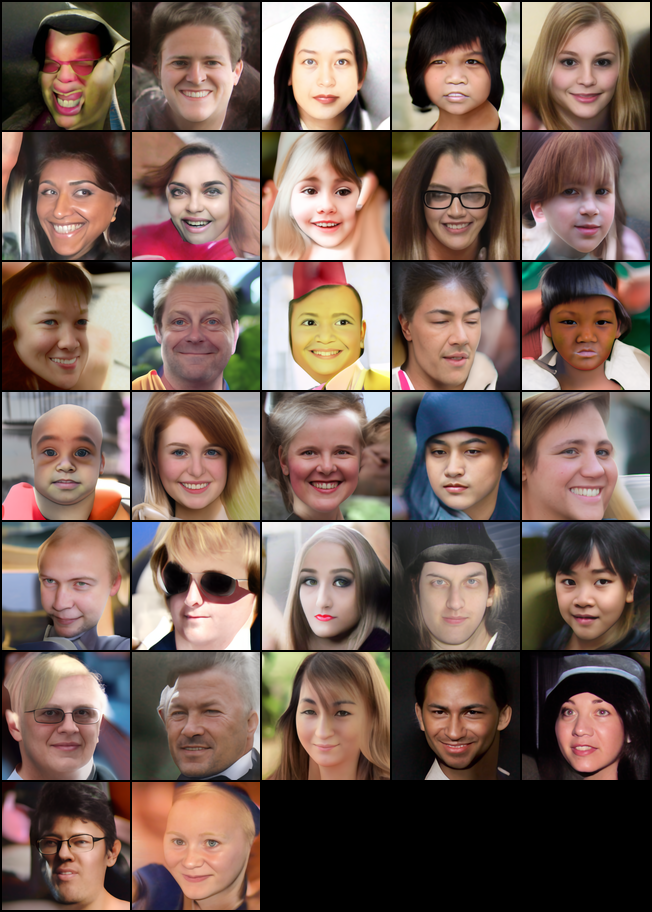}
        \end{subfigure} \\
        \addlinespace[5pt] %

        \begin{subfigure}{\linewidth}
        \centering
        Pe=$0.10$
        \end{subfigure} &
        \begin{subfigure}{0.4\linewidth}
        \centering
        \includegraphics[trim={0 522pt 0 0}, clip, width=\linewidth, valign=c]{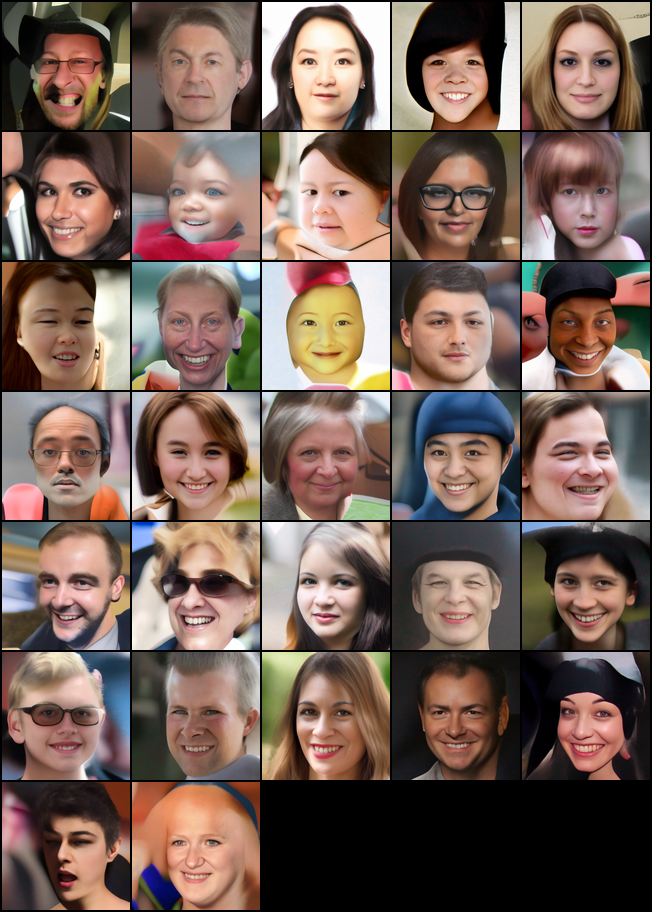}
        \end{subfigure} &
        \begin{subfigure}{0.4\linewidth}
        \centering
        \includegraphics[trim={0 522pt 0 0}, clip, width=\linewidth, valign=c]{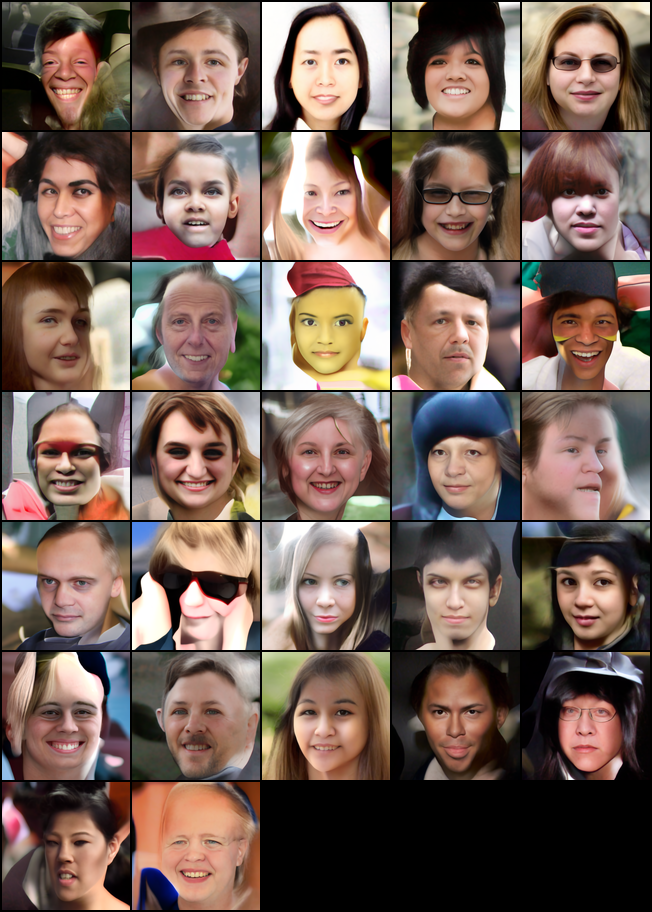}
        \end{subfigure} \\
        \addlinespace[5pt] %

        \begin{subfigure}{\linewidth}
        \centering
        Pe=$0.12$
        \end{subfigure} &
        \begin{subfigure}{0.4\linewidth}
        \centering
        \includegraphics[trim={0 522pt 0 0}, clip, width=\linewidth, valign=c]{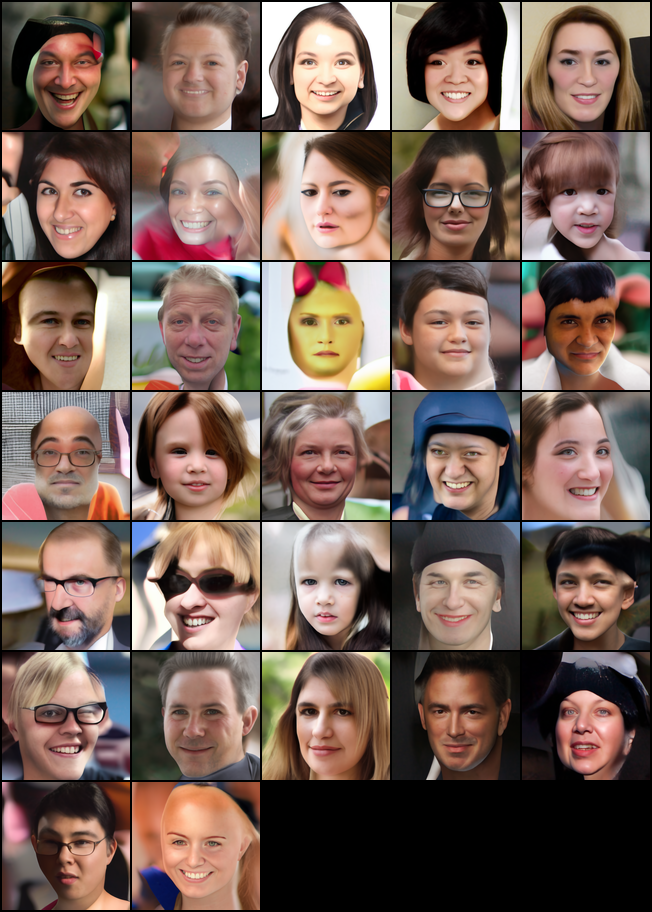}
        \end{subfigure} &
        \begin{subfigure}{0.4\linewidth}
        \centering
        \includegraphics[trim={0 522pt 0 0}, clip, width=\linewidth, valign=c]{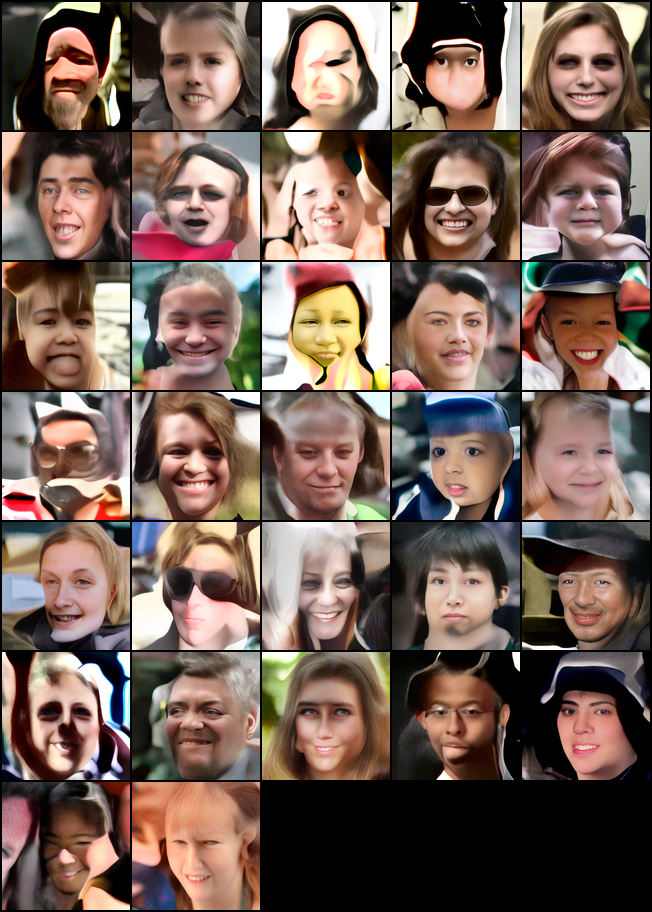}
        \end{subfigure} \\
        \addlinespace[5pt] %

        \begin{subfigure}{\linewidth}
        \centering
        Pe=$0.14$
        \end{subfigure} &
        \begin{subfigure}{0.4\linewidth}
        \centering
        \includegraphics[trim={0 522pt 0 0}, clip, width=\linewidth, valign=c]{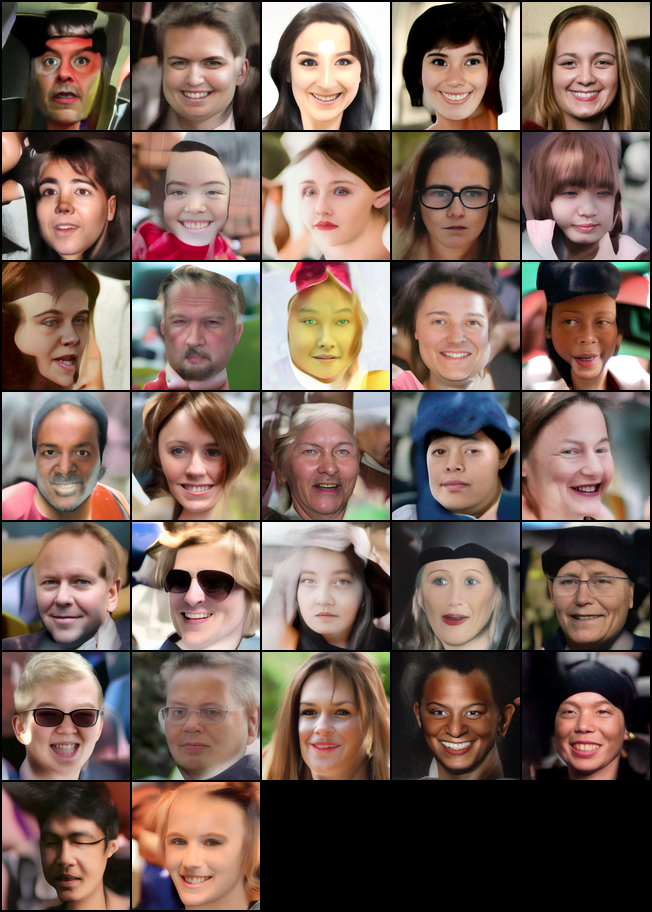}
        \end{subfigure} &
        \begin{subfigure}{0.4\linewidth}
        \centering
        \includegraphics[trim={0 522pt 0 0}, clip, width=\linewidth, valign=c]{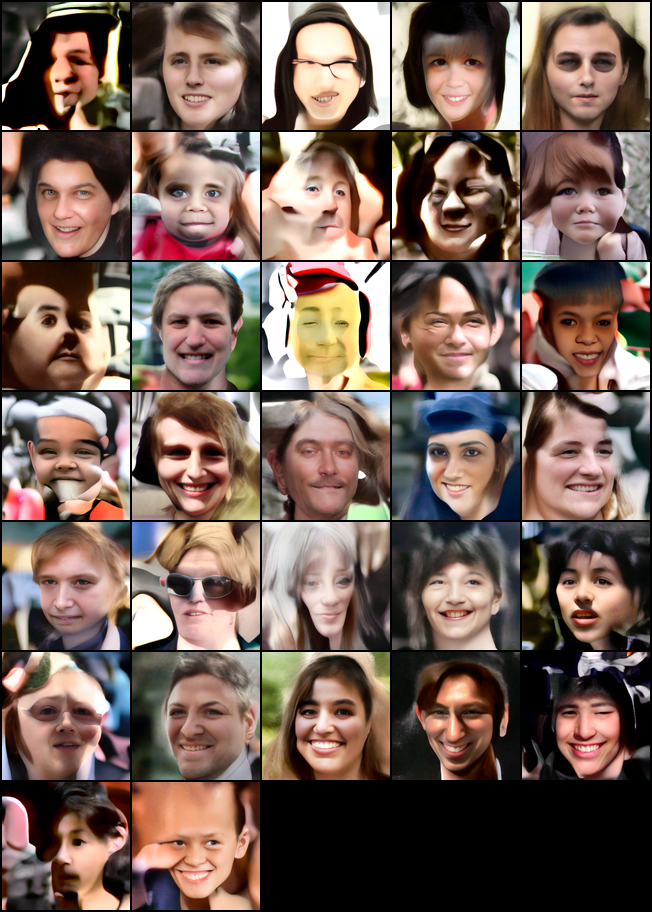}
        \end{subfigure} \\
        \addlinespace[5pt] %
        
    \end{tabular}
    \caption{Visual comparison of the results of our method and the IHD method on the FFHQ dataset.}
    \label{fig:visual_comp_ffhq_part2}
\end{figure*}

\begin{figure*}[h!]
    \centering
    \includegraphics[width=\linewidth]{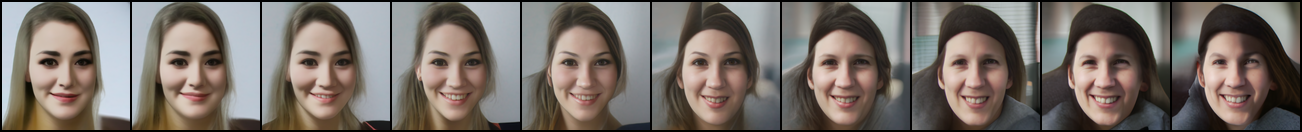}
    \includegraphics[width=\linewidth]{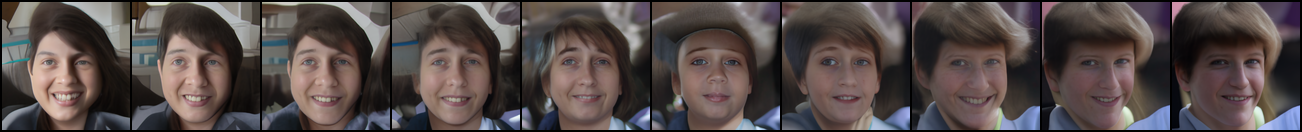}
    \includegraphics[width=\linewidth]{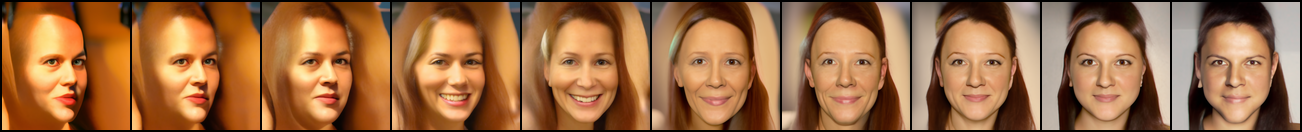}
    \includegraphics[width=\linewidth]{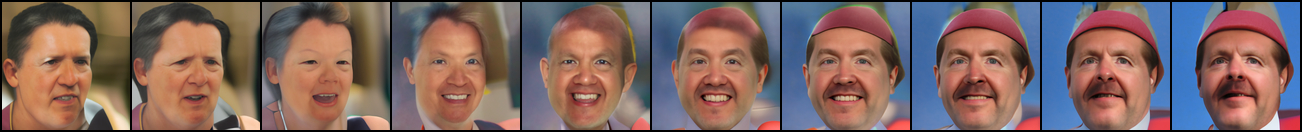}
    \includegraphics[width=\linewidth]{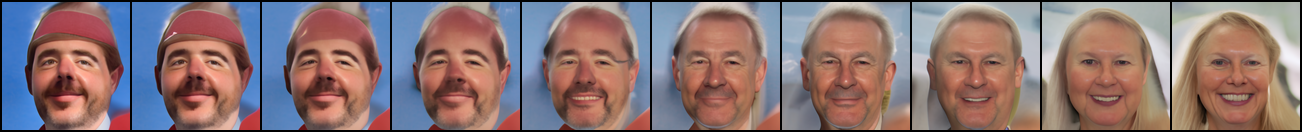}
    \includegraphics[width=\linewidth]{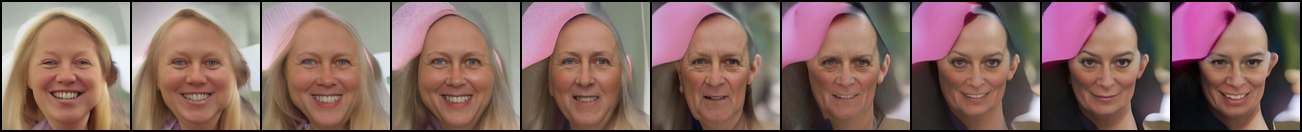}
    \includegraphics[width=\linewidth]{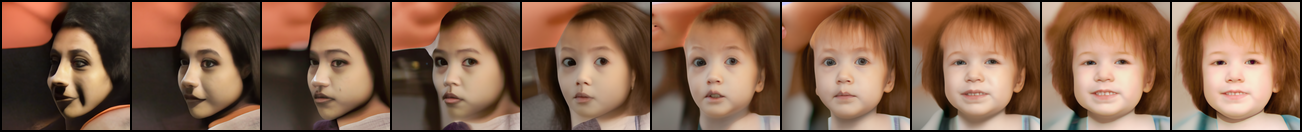}
    \includegraphics[width=\linewidth]{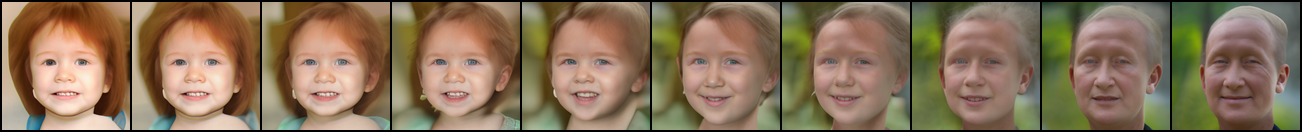}
    \includegraphics[width=\linewidth]{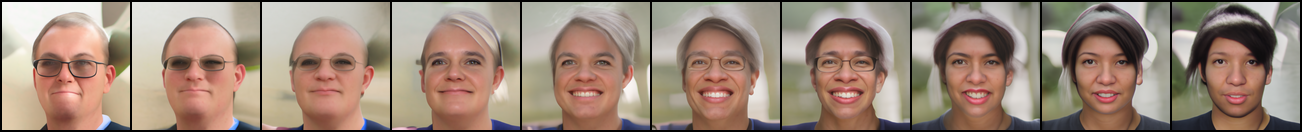}
    \caption{Interpolations between two random images on FFHQ 128 $\times$ 128. $\sigma$ = 16, Pe=0.8}
    \label{fig:faces_interpolation_16_0.8}
\end{figure*}

\begin{figure*}[h!]
    \centering
    \includegraphics[width=\linewidth]{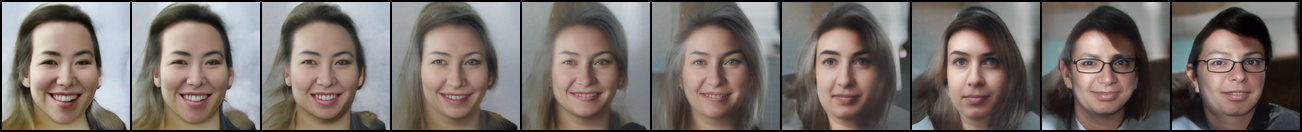}
    \includegraphics[width=\linewidth]{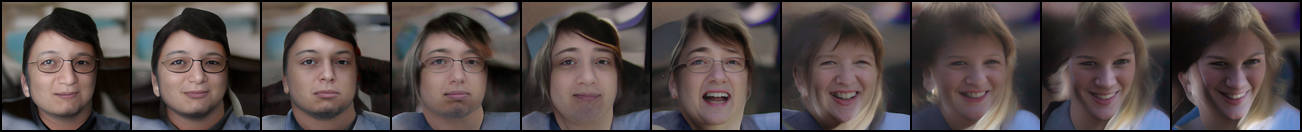}
    \includegraphics[width=\linewidth]{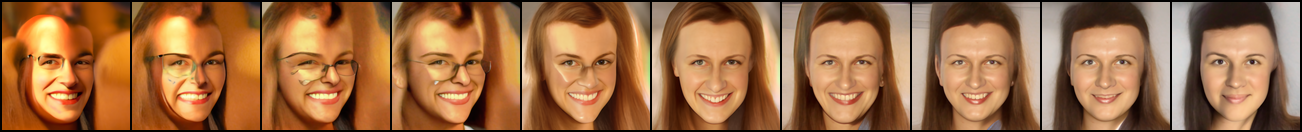}
    \includegraphics[width=\linewidth]{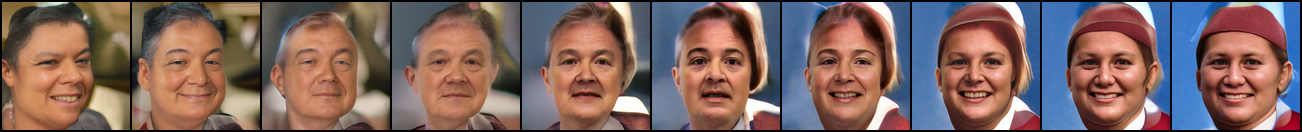}
    \includegraphics[width=\linewidth]{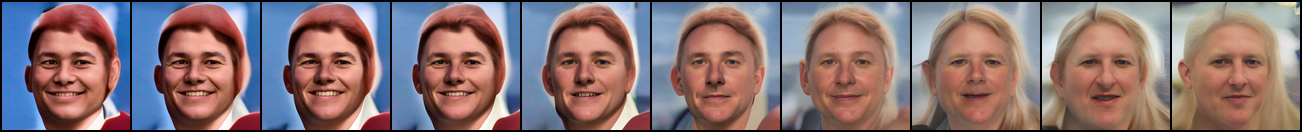}
    \includegraphics[width=\linewidth]{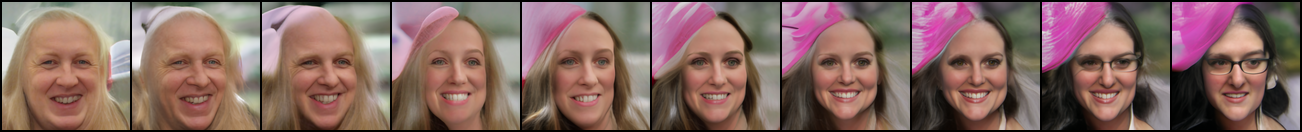}
    \includegraphics[width=\linewidth]{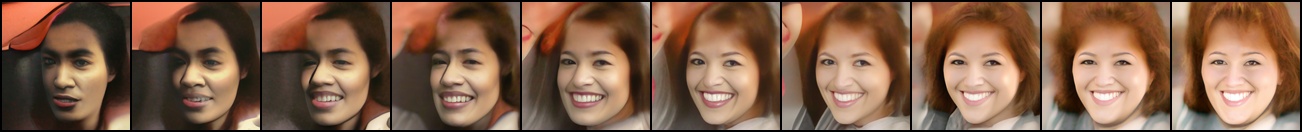}
    \includegraphics[width=\linewidth]{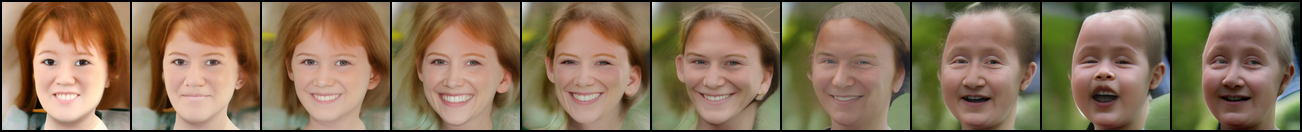}
    \includegraphics[width=\linewidth]{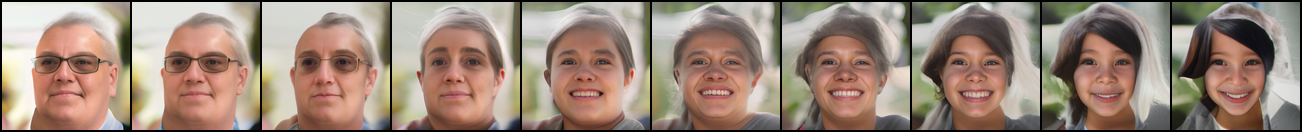}
    \caption{Interpolations between two random images on FFHQ 128 $\times$ 128. $\sigma$ = 20, Pe=0.8}
    \label{fig:faces_interpolation_20_0.8}
\end{figure*}

\begin{figure*}[h]

\begin{minipage}{0.9\textwidth}
  \centering
  \offinterlineskip %
\begin{minipage}{0.1\textwidth}
   \centering IHD\\Pe=$0.0$
  \end{minipage}%
  \begin{minipage}{0.9\textwidth}
    \centering
    \includegraphics[width=\textwidth]{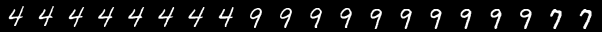}
    \includegraphics[width=\textwidth]{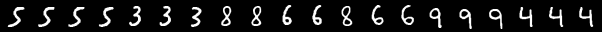}
    \includegraphics[width=\textwidth]{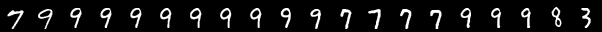}
  \end{minipage}
  \\
  \vspace{10pt}
  \begin{minipage}{0.1\linewidth}
	\centering Our\\Pe=$0.02$
  \end{minipage}%
  \begin{minipage}{0.9\linewidth}
    \centering
    \includegraphics[width=\textwidth]{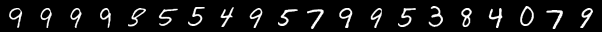}
    \includegraphics[width=\textwidth]{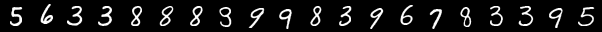}
    \includegraphics[width=\textwidth]{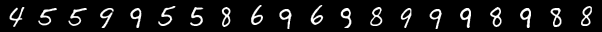}
  \end{minipage}
  \\
  \vspace{10pt}
    
  \begin{minipage}{0.1\linewidth}
	\centering Our\\Pe=$0.04$
  \end{minipage}%
  \begin{minipage}{0.9\linewidth}
    \centering
    \includegraphics[width=\textwidth]{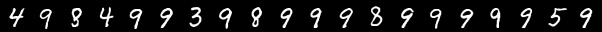}
    \includegraphics[width=\textwidth]{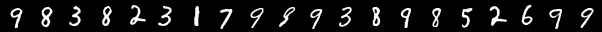}
    \includegraphics[width=\textwidth]{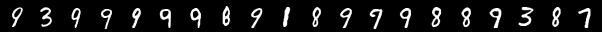}
  \end{minipage}
  \\
  \vspace{10pt}
  
\begin{minipage}{0.1\linewidth}
	\centering Our\\Pe=$0.06$
  \end{minipage}%
  \begin{minipage}{0.9\linewidth}
    \centering
    \includegraphics[width=\textwidth]{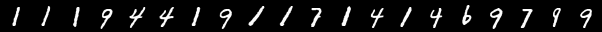}
    \includegraphics[width=\textwidth]{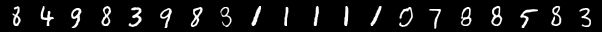}
    \includegraphics[width=\textwidth]{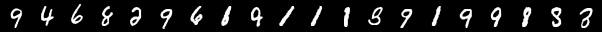}
  \end{minipage}
  \\
  \vspace{10pt}
  \begin{minipage}{0.1\linewidth}
	\centering Our\\Pe=$0.08$
  \end{minipage}%
  \begin{minipage}{0.9\linewidth}
    \centering
    \includegraphics[width=\textwidth]{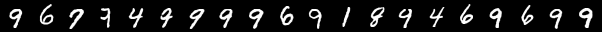}
    \includegraphics[width=\textwidth]{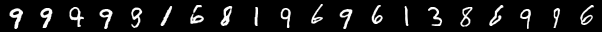}
    \includegraphics[width=\textwidth]{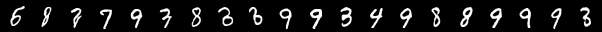}
  \end{minipage}
  \\
  \vspace{10pt}

  \begin{minipage}{0.1\linewidth}
	\centering Our\\Pe=$0.10$
  \end{minipage}%
  \begin{minipage}{0.9\linewidth}
    \centering`
    \includegraphics[width=\textwidth]{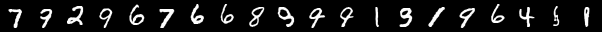}
    \includegraphics[width=\textwidth]{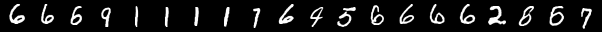}
    \includegraphics[width=\textwidth]{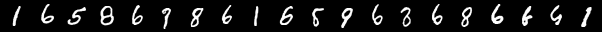}
  \end{minipage}

  \captionof{figure}{Visual comparison of interpolations between two MNIST samples.
  Each undergoes the forward process up to $ \sigma = 20$, followed by linear interpolation and denoising with SLERP-interpolated generative noise added at each step. } %
  \label{fig:mnist_interpolation} %
\end{minipage}
    
\end{figure*}

\begin{figure*}[h]
    \centering
    \begin{subfigure}{0.9\linewidth}
        \centering
        \includegraphics[trim={0 30pt 0pt 0}, clip, width=\linewidth]{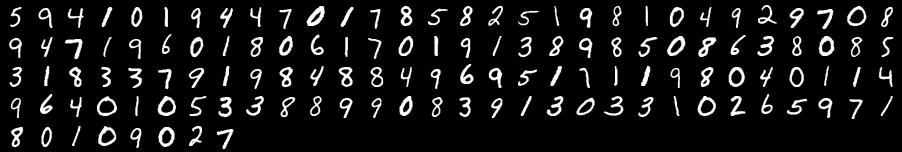}
        \caption{\normalsize (IHD) $Pe=0.00$}
    \end{subfigure}
    
    \vspace{0.3cm}

    \begin{subfigure}{0.9\linewidth}
        \centering
        \includegraphics[trim={0 30pt 0pt 0}, clip, width=\linewidth]{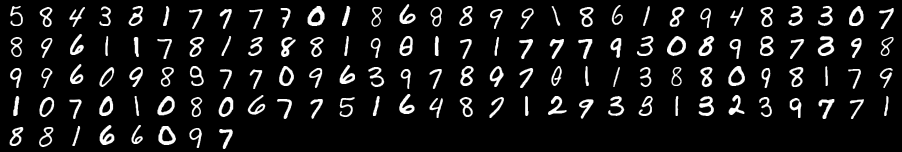}
        \caption{\normalsize $Pe=0.02$}
    \end{subfigure}

    \vspace{0.3cm}

    \begin{subfigure}{0.9\linewidth}
        \centering
        \includegraphics[trim={0 30pt 0pt 0}, clip, width=\linewidth]{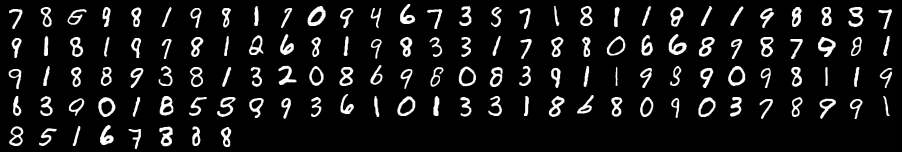}
        \caption{\normalsize $Pe=0.04$}
    \end{subfigure}

    \vspace{0.3cm}

    \begin{subfigure}{0.9\linewidth}
        \centering
        \includegraphics[trim={0 30pt 0pt 0}, clip, width=\linewidth]{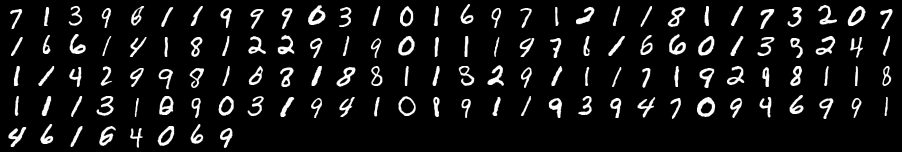}
        \caption{\normalsize $Pe=0.06$}
    \end{subfigure}

    \vspace{0.3cm}

    \begin{subfigure}{0.9\linewidth}
        \centering
        \includegraphics[trim={0 30pt 0pt 0}, clip, width=\linewidth]{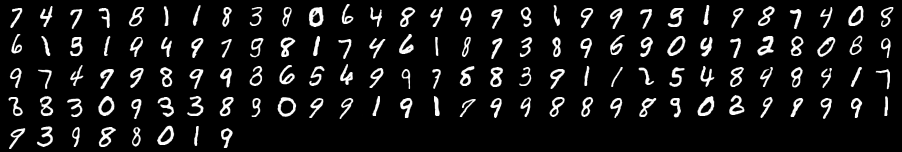}
        \caption{\normalsize $Pe=0.08$}
    \end{subfigure}

    \vspace{0.3cm}

    \begin{subfigure}{0.9\linewidth}
        \centering
        \includegraphics[trim={0 30pt 0pt 0}, clip, width=\linewidth]{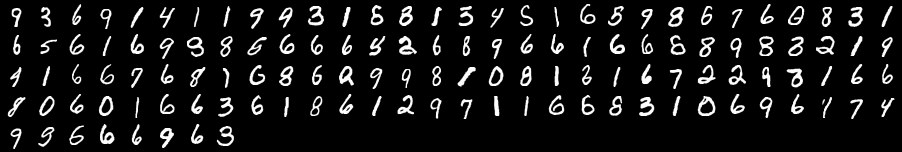}
        \caption{\normalsize $Pe=0.10$}
    \end{subfigure}

    \caption{Visual comparison of the results of our method and the IHD method on the MNIST dataset.}
    \label{fig:visual_comp_mnist}
\end{figure*}

\begin{figure*}[h!]
    \centering
    \includegraphics[width=0.95\linewidth]{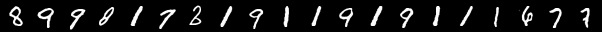}
    \includegraphics[width=0.95\linewidth]{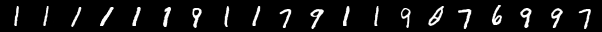}
    \includegraphics[width=0.95\linewidth]{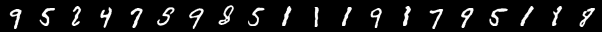}
    \includegraphics[width=0.95\linewidth]{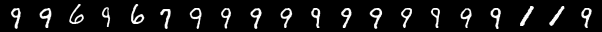}
    \includegraphics[width=0.95\linewidth]{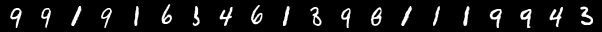}
    \includegraphics[width=0.95\linewidth]{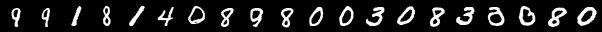}
    \includegraphics[width=0.95\linewidth]{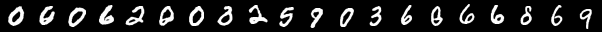}
    \includegraphics[width=0.95\linewidth]{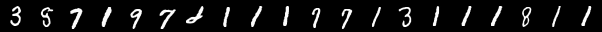}
    \includegraphics[width=0.95\linewidth]{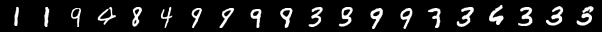}
    \includegraphics[width=0.95\linewidth]{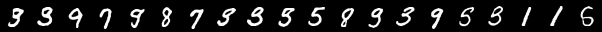}
    \includegraphics[width=0.95\linewidth]{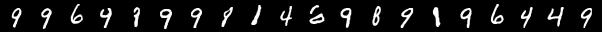}
    \includegraphics[width=0.95\linewidth]{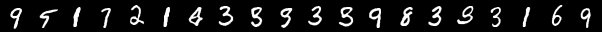}
    \includegraphics[width=0.95\linewidth]{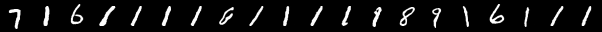}
    \includegraphics[width=0.95\linewidth]{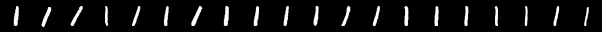}
    \includegraphics[width=0.95\linewidth]{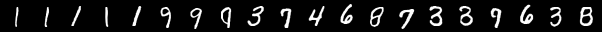}
    \includegraphics[width=0.95\linewidth]{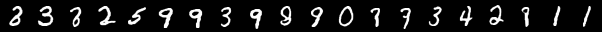}
    \includegraphics[width=0.95\linewidth]{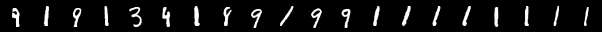}
    \includegraphics[width=0.95\linewidth]{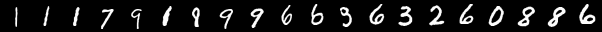}
    \includegraphics[width=0.95\linewidth]{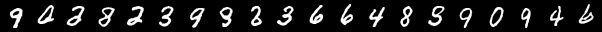}
    \includegraphics[width=0.95\linewidth]{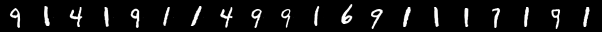}
    \includegraphics[width=0.95\linewidth]{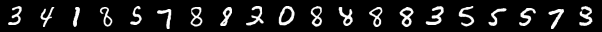}
    \includegraphics[width=0.95\linewidth]{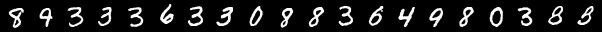}
    \includegraphics[width=0.95\linewidth]{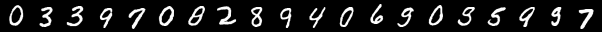}
    \caption{Interpolations between two random images on MNIST, $\sigma=20$, $Pe=0.6$}
    \label{fig:mnist_interpolation_20_0.6}
\end{figure*}

\begin{figure*}[h!]
    \centering
    \includegraphics[width=0.95\linewidth]{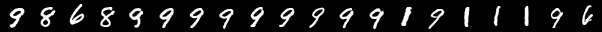}
    \includegraphics[width=0.95\linewidth]{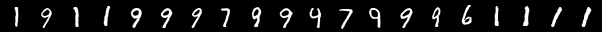}
    \includegraphics[width=0.95\linewidth]{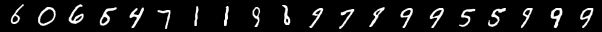}
    \includegraphics[width=0.95\linewidth]{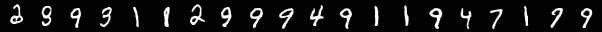}
    \includegraphics[width=0.95\linewidth]{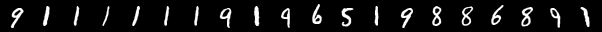}
    \includegraphics[width=0.95\linewidth]{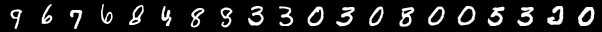}
    \includegraphics[width=0.95\linewidth]{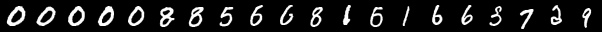}
    \includegraphics[width=0.95\linewidth]{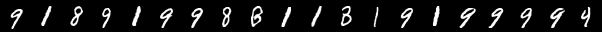}
    \includegraphics[width=0.95\linewidth]{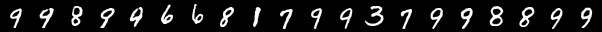}
    \includegraphics[width=0.95\linewidth]{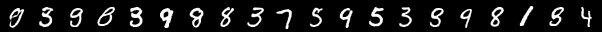}
    \includegraphics[width=0.95\linewidth]{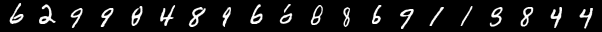}
    \includegraphics[width=0.95\linewidth]{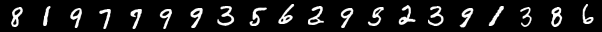}
    \includegraphics[width=0.95\linewidth]{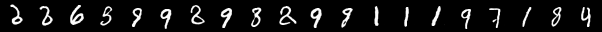}
    \includegraphics[width=0.95\linewidth]{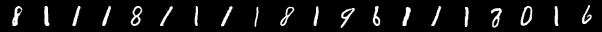}
    \includegraphics[width=0.95\linewidth]{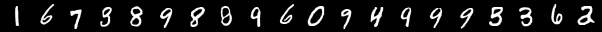}
    \includegraphics[width=0.95\linewidth]{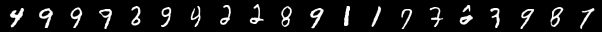}
    \includegraphics[width=0.95\linewidth]{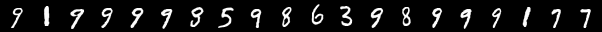}
    \includegraphics[width=0.95\linewidth]{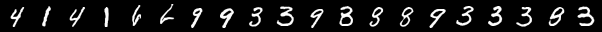}
    \includegraphics[width=0.95\linewidth]{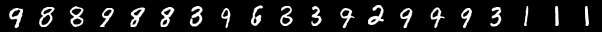}
    \includegraphics[width=0.95\linewidth]{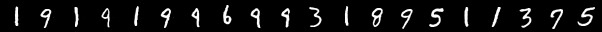}
    \includegraphics[width=0.95\linewidth]{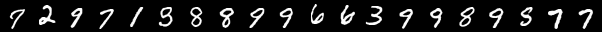}
    \includegraphics[width=0.95\linewidth]{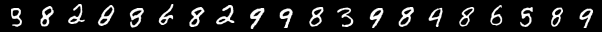}
    \includegraphics[width=0.95\linewidth]{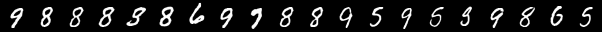}
    \caption{Interpolations between two random images on MNIST, $\sigma=20$, $Pe=0.8$}
    \label{fig:mnist_interpolation_20_0.8}
\end{figure*}

\begin{figure*}[b]
\small
\begin{minipage}{\textwidth}
  \centering
  
\begin{minipage}{0.1\textwidth}
   \centering IHD\\Pe=$0.0$
  \end{minipage}%
  \begin{minipage}{0.9\textwidth}
    \centering
    \includegraphics[width=\textwidth]{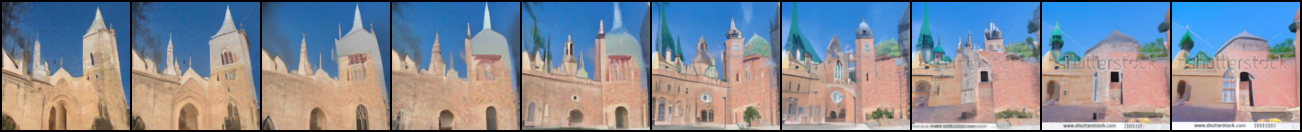}
    \\[1pt]
    \includegraphics[width=\textwidth]{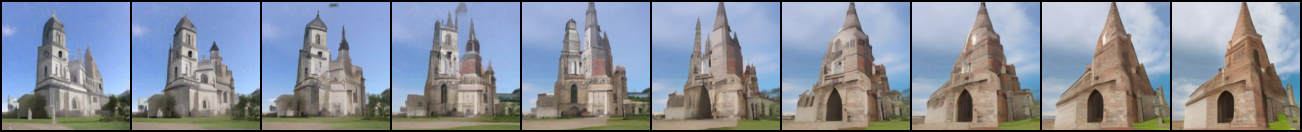}
  \end{minipage}\\
  \vspace{5pt} %

  \begin{minipage}{0.1\linewidth}
	\centering Our\\Pe=$0.02$
  \end{minipage}%
  \begin{minipage}{0.9\linewidth}
    \centering
    \includegraphics[width=\linewidth]{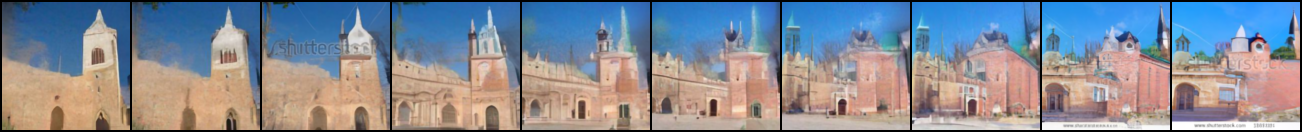}
    \\[1pt]
    \includegraphics[width=\textwidth]{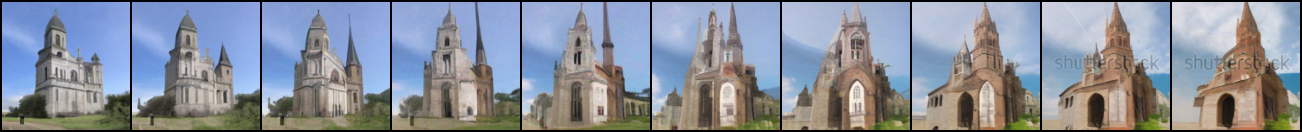}
  \end{minipage}
  \vspace{5pt}\\ %
  
  \begin{minipage}{0.1\linewidth}
	\centering Our\\Pe=$0.04$
  \end{minipage}%
  \begin{minipage}{0.9\linewidth}
    \centering
    \includegraphics[width=\linewidth]{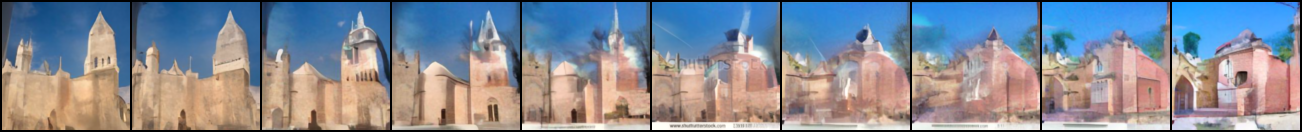}
    \\[1pt]
    \includegraphics[width=\textwidth]{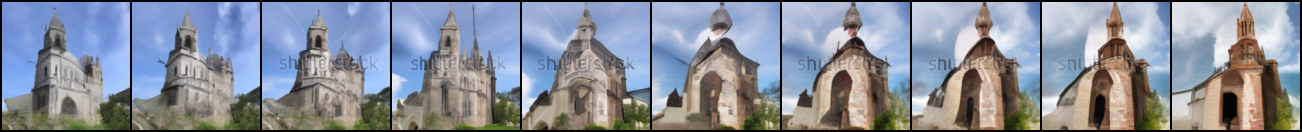}
  \end{minipage}\\
  \vspace{5pt} %
  
\begin{minipage}{0.1\linewidth}
	\centering Our\\Pe=$0.06$
  \end{minipage}%
  \begin{minipage}{0.9\linewidth}
    \centering
    \includegraphics[width=\linewidth]{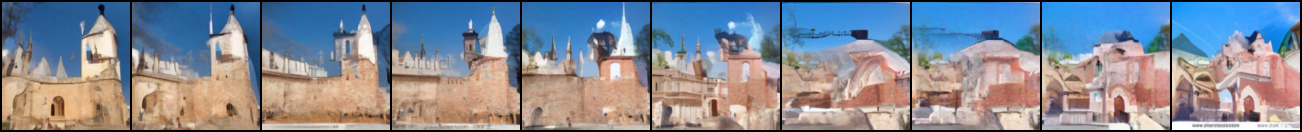}
    \\[1pt]
    \includegraphics[width=\textwidth]{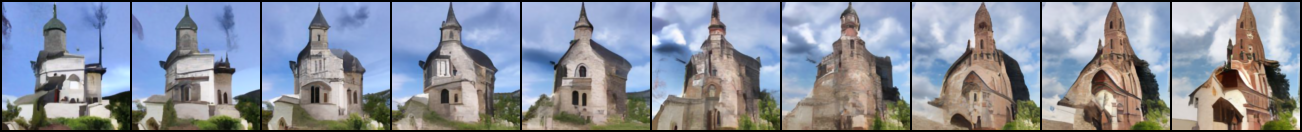}
  \end{minipage}\\
  \vspace{5pt} %
  
  \begin{minipage}{0.1\linewidth}
	\centering Our\\Pe=$0.08$
  \end{minipage}%
  \begin{minipage}{0.9\linewidth}
    \centering
    \includegraphics[width=\linewidth]{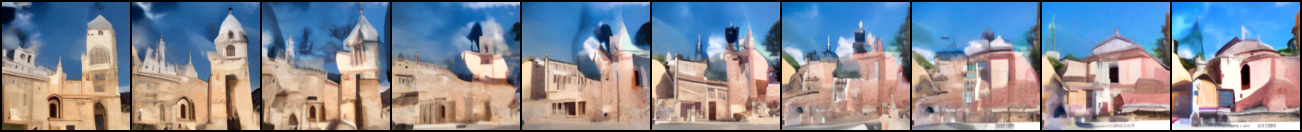}
    \\[1pt]
    \includegraphics[width=\textwidth]{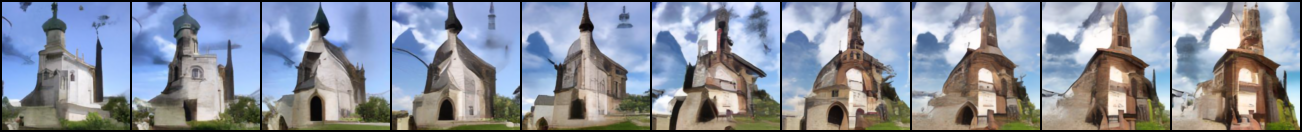}
\end{minipage}\\
  \vspace{5pt} %

  \captionof{figure}{Visual comparison of interpolations between two LSUN Church samples.
  Each undergoes the forward process up to $ \sigma =20$, followed by linear interpolation and denoising with SLERP-interpolated generative noise added at each step. 
  } %
  \label{fig:lsun_faces_interpolation} %
\end{minipage}
    
\end{figure*}

\begin{figure*}[p]
    \centering
    \adjustboxset{scale=0.8, valign=c}
    \begin{tabular}{w{c}{40pt} c}
        & \textbf{$\sigma=20$} \\
        \addlinespace[10pt] %
        
        \begin{subfigure}{\linewidth}
        \centering
        (IHD) \\ Pe=$0.00$
        \end{subfigure} &
        \begin{subfigure}{0.87\linewidth}
        \centering
        \includegraphics[trim={0 1174.5 0 0}, clip, width=\linewidth, valign=c]{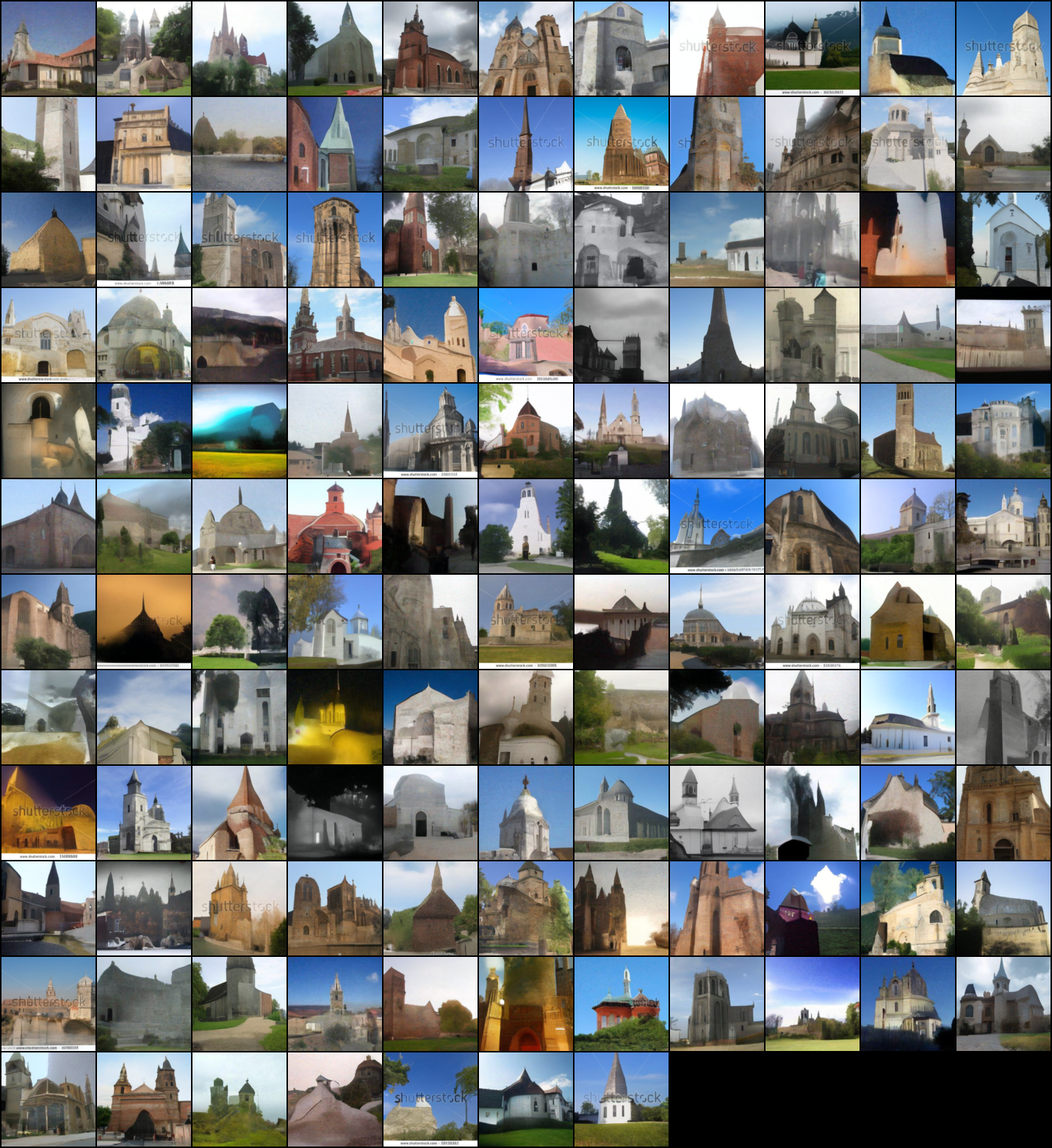}
        \end{subfigure} \\
        \addlinespace[5pt] %

        \begin{subfigure}{\linewidth}
        \centering
        Pe=$0.02$
        \end{subfigure} &
        \begin{subfigure}{0.87\linewidth}
        \centering
        \includegraphics[trim={0 1174.5pt 0 0}, clip, width=\linewidth, valign=c]{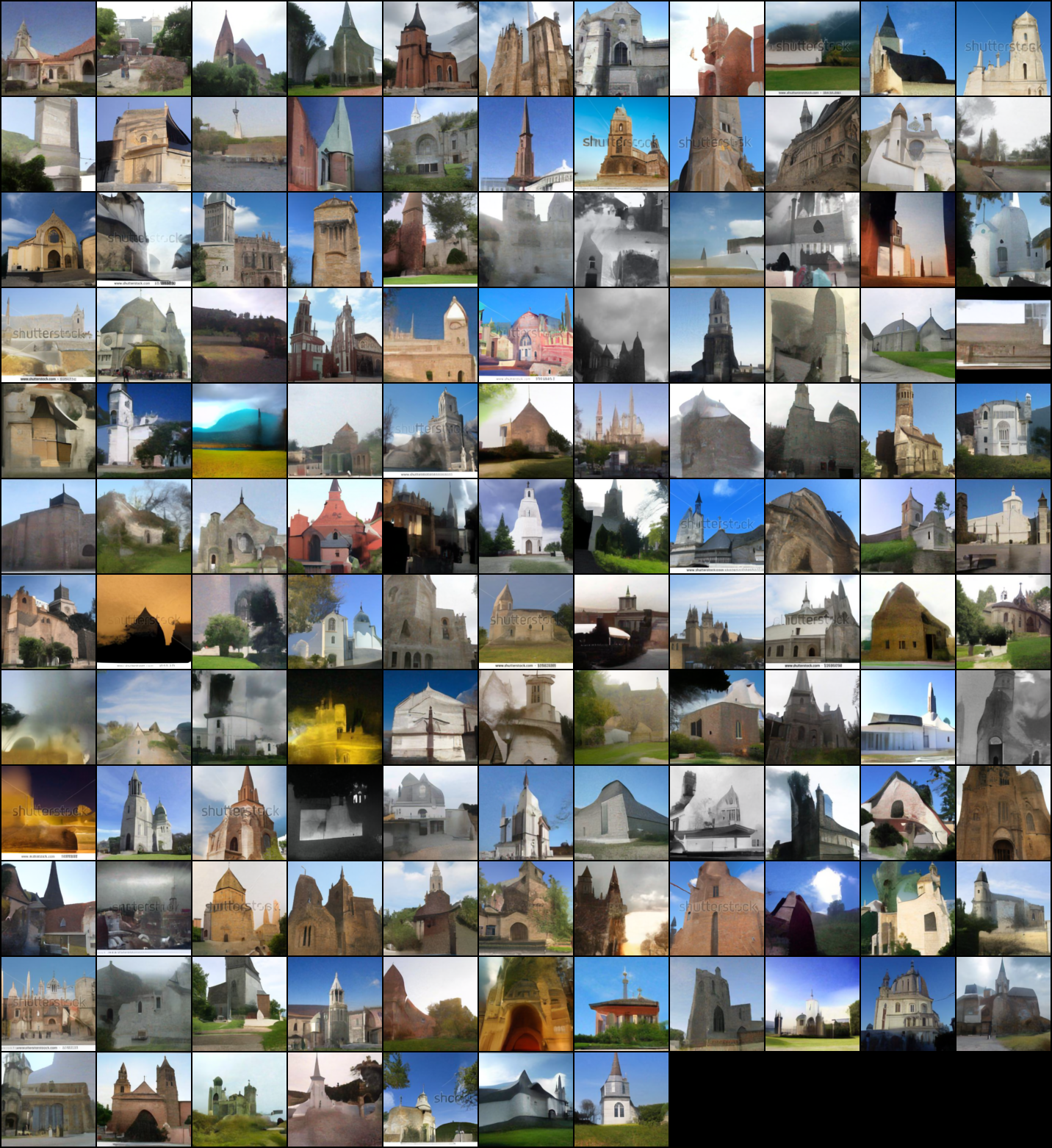}
        \end{subfigure} \\
        \addlinespace[5pt] %

        \begin{subfigure}{\linewidth}
        \centering
        Pe=$0.04$
        \end{subfigure} &
        \begin{subfigure}{0.87\linewidth}
        \centering
        \includegraphics[trim={0 1174.5pt 0 0}, clip, width=\linewidth, valign=c]{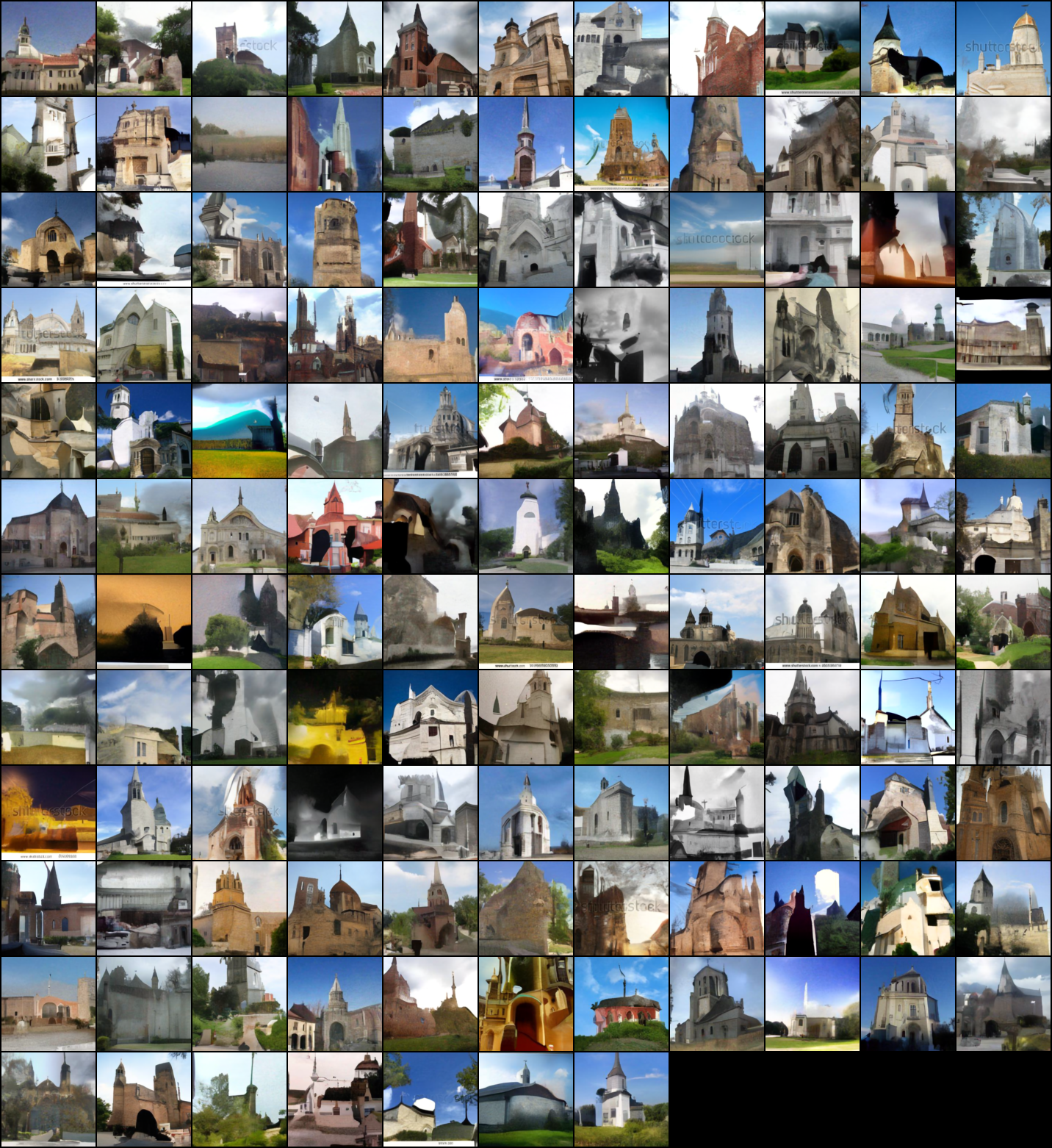}
        \end{subfigure} \\
        \addlinespace[5pt] %

        \begin{subfigure}{\linewidth}
        \centering
        Pe=$0.06$
        \end{subfigure} &
        \begin{subfigure}{0.87\linewidth}
        \centering
        \includegraphics[trim={0 1174.5pt 0 0}, clip, width=\linewidth, valign=c]{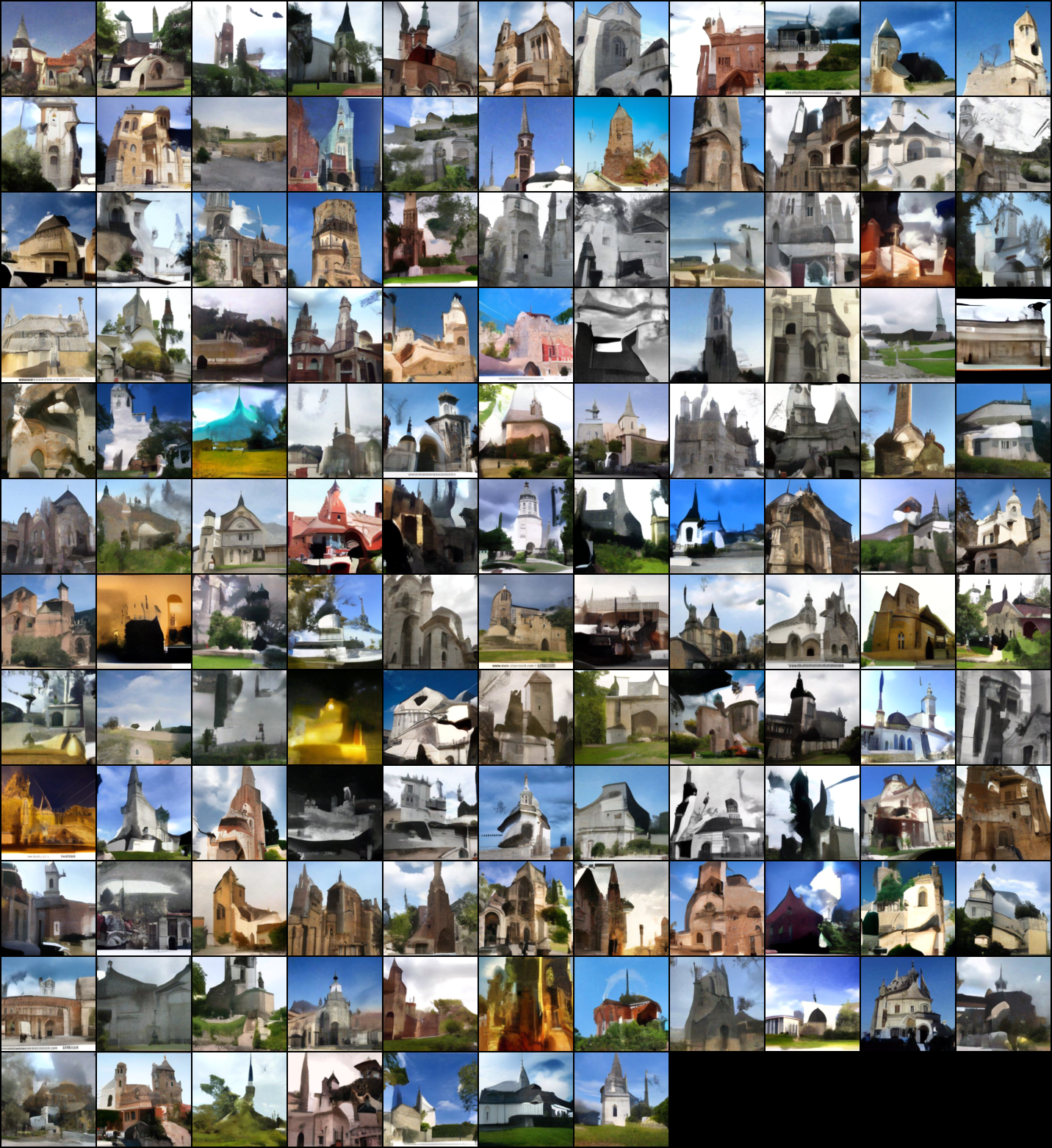}
        \end{subfigure} \\
        \addlinespace[5pt] %

        \begin{subfigure}{\linewidth}
        \centering
        Pe=$0.08$
        \end{subfigure} &
        
        \begin{subfigure}{0.87\linewidth}
        \centering
        \includegraphics[trim={0 1174.5pt 0 0}, clip, width=\linewidth, valign=c]{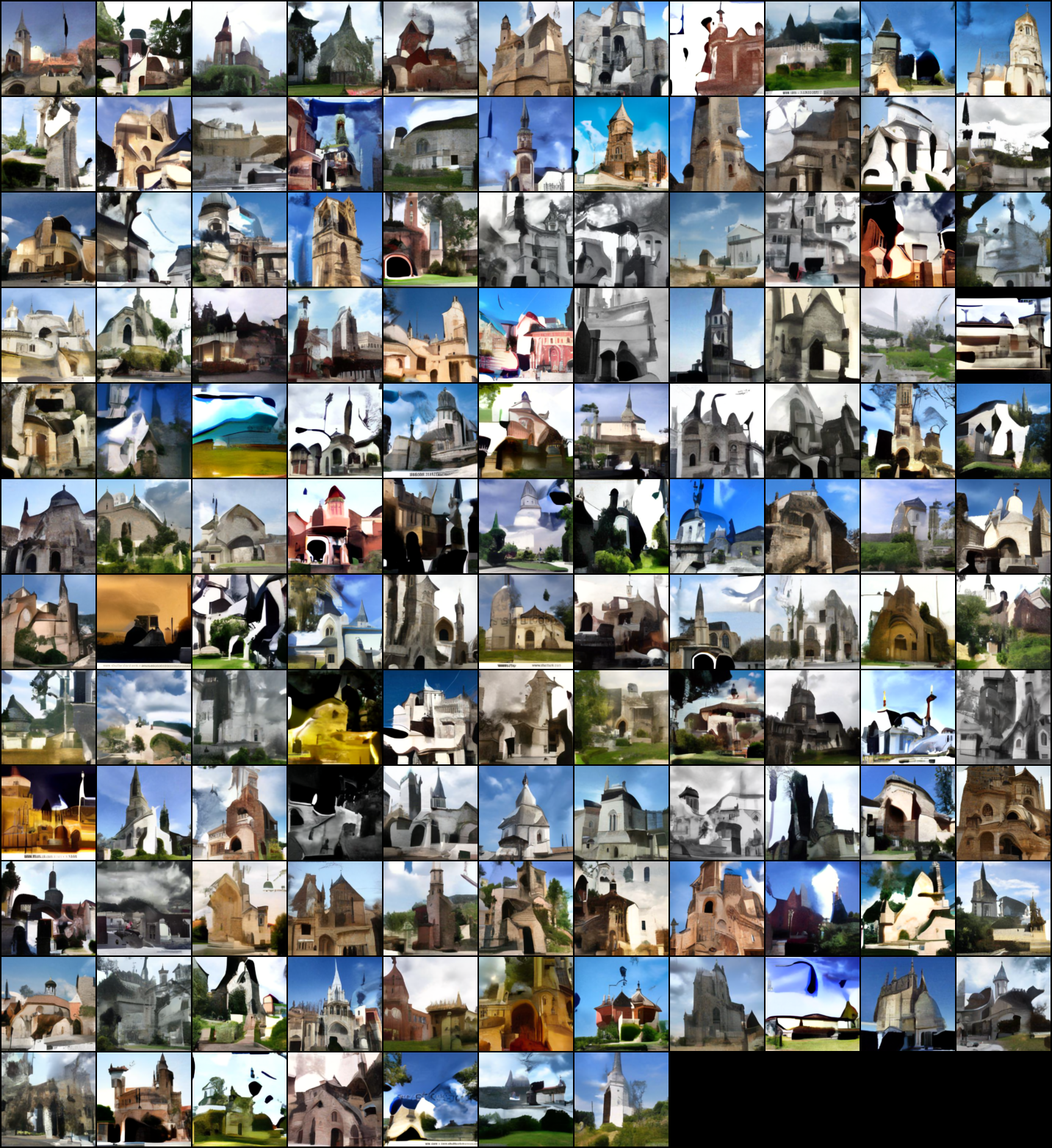}
        \end{subfigure} \\
        \addlinespace[5pt] %
        
    \end{tabular}
    \caption{Visual comparison of the results of our method and the IHD method on the LSUN Church dataset.}
    \label{fig:lsun_visual_comp}
\end{figure*}

\begin{figure*}[h!]
    \centering
    \includegraphics[width=\linewidth]{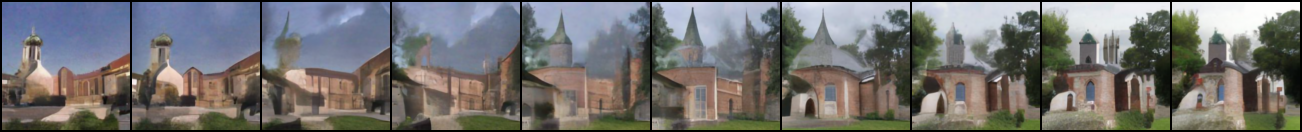}
    \includegraphics[width=\linewidth]{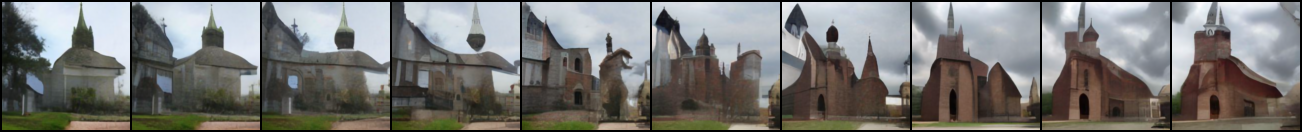}
    \includegraphics[width=\linewidth]{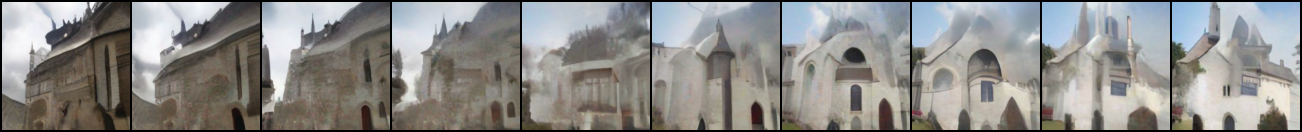}
    \includegraphics[width=\linewidth]{figures/results/LSUN/interpolations/Pe_4e-2/interpolation_37.png}
    \includegraphics[width=\linewidth]{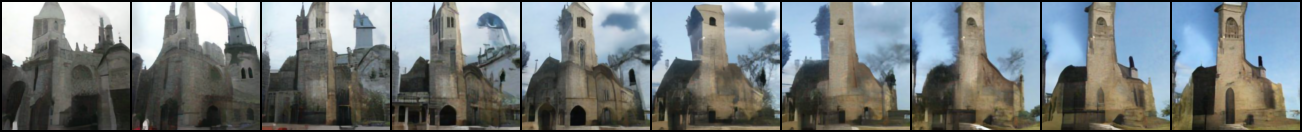}
    \includegraphics[width=\linewidth]{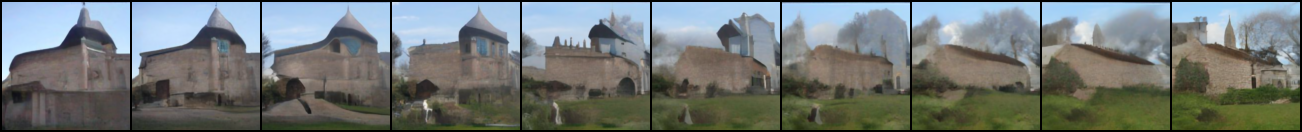}
    \includegraphics[width=\linewidth]{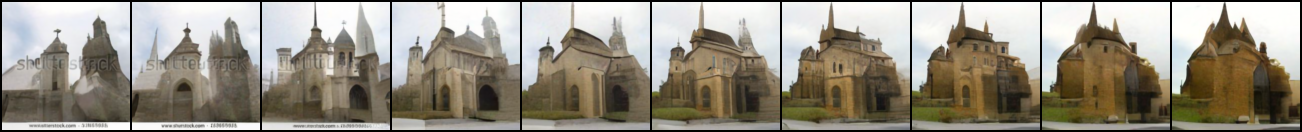}
    \includegraphics[width=\linewidth]{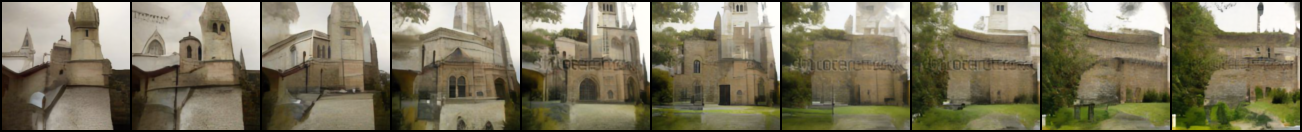}
    \includegraphics[width=\linewidth]{figures/results/LSUN/interpolations/Pe_4e-2/interpolation_89.png}
    \includegraphics[width=\linewidth]{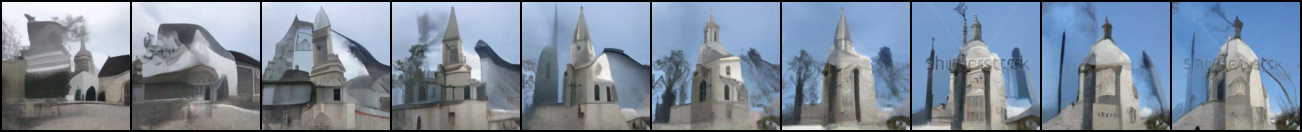}
    \includegraphics[width=\linewidth]{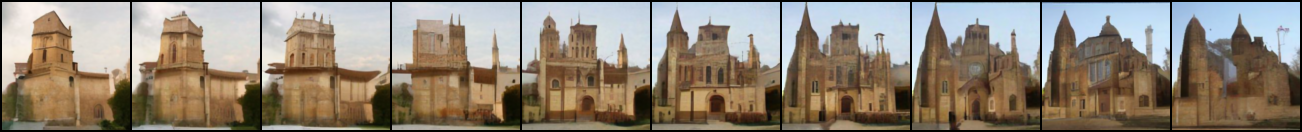}
    \includegraphics[width=\linewidth]{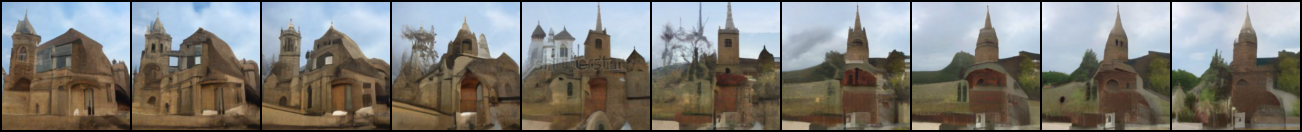}
    \caption{Interpolations between two random images on LSUN Church 128 $\times$ 128. $\sigma$ = 20, Pe=0.4}
    \label{fig:lsun_faces_interpolation_0.4}
\end{figure*}

\begin{figure*}[h!]
    \centering
    \includegraphics[width=\linewidth]{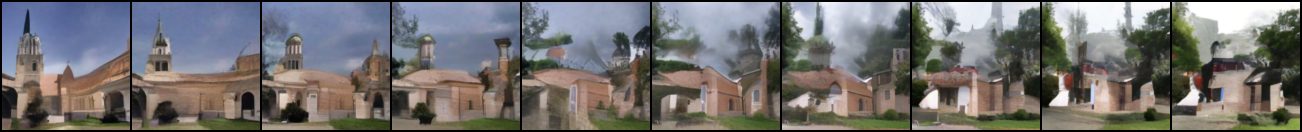}
    \includegraphics[width=\linewidth]{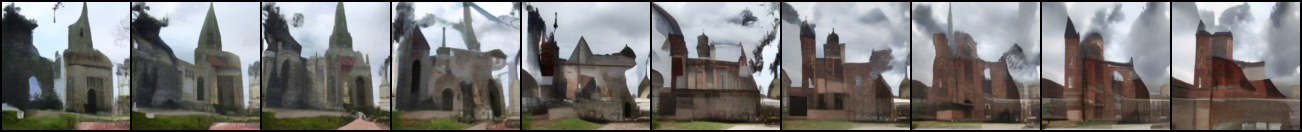}
    \includegraphics[width=\linewidth]{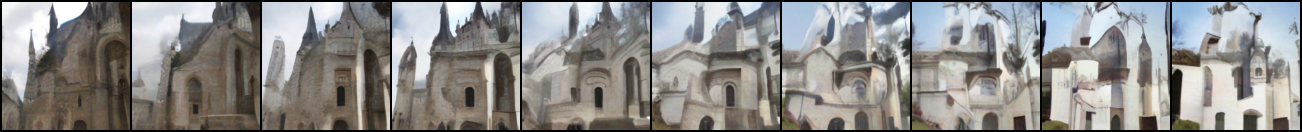}
    \includegraphics[width=\linewidth]{figures/results/LSUN/interpolations/Pe_6e-2/interpolation_37.png}
    \includegraphics[width=\linewidth]{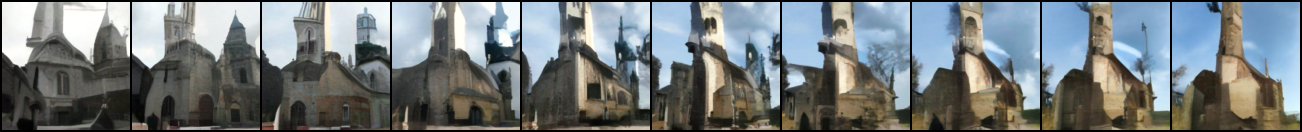}
    \includegraphics[width=\linewidth]{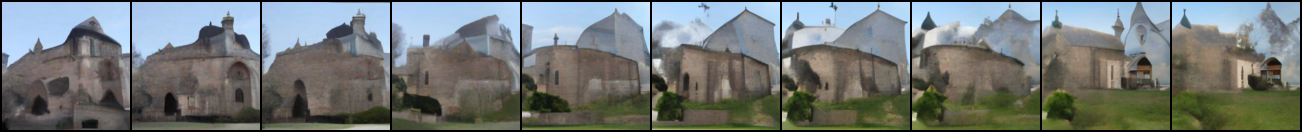}
    \includegraphics[width=\linewidth]{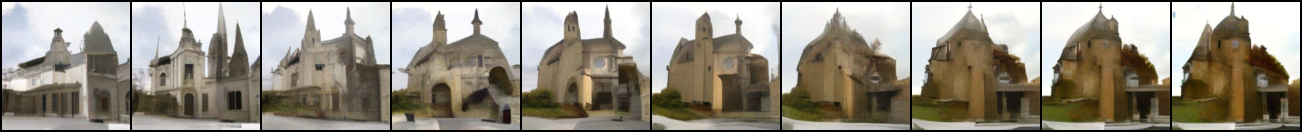}
    \includegraphics[width=\linewidth]{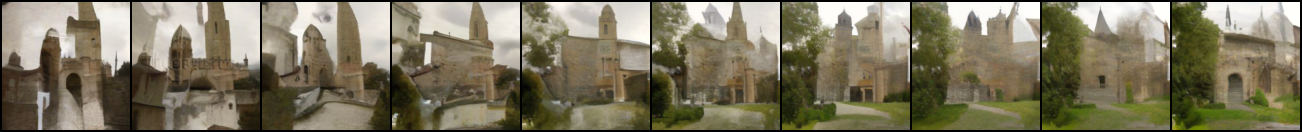}
    \includegraphics[width=\linewidth]{figures/results/LSUN/interpolations/Pe_6e-2/interpolation_89.png}
    \includegraphics[width=\linewidth]{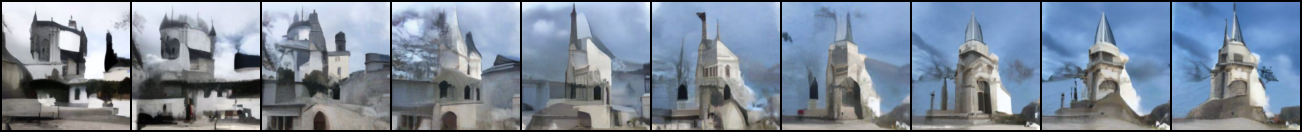}
    \includegraphics[width=\linewidth]{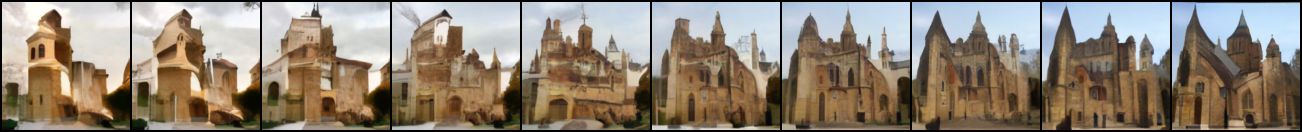}
    \includegraphics[width=\linewidth]{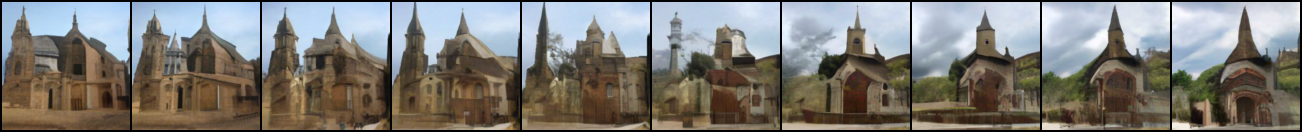}
    \caption{Interpolations between two random images on LSUN Church 128 $\times$ 128. $\sigma$ = 20, Pe=0.6}
    \label{fig:lsun_faces_interpolation_0.6}
\end{figure*}

We present supplementary experimental results from parameter ablations examining the influence of the Peclet number (Pe) on image synthesis quality and sample distribution diversity. 
Experiments examine Peclet number (Pe) impacts on FFHQ-128 ($\text{Pe} \in \{0.0, 0.02, 0.04, 0.06, 0.08, 0.1, 0.12, 0.14\}$), MNIST ($\text{Pe} \in \{0.0, 0.02, 0.04, 0.06, 0.08, 0.1\}$) and LSUN Church ($\text{Pe} \in \{0.0, 0.02, 0.04, 0.06, 0.08\}$). This appendix documents comparative studies conducted on the FFHQ-128, MNIST and LSUN Church datasets, organized as follows:
(i) single-initial-state sampling through generation, where ground truth (GT) images are propagated through the forward diffusion process with Pe-specific dynamics and reconstructed through model inference (Figs.~\ref{fig:initial_state_sampling_ffhq_sigma20}, \ref{fig:initial_state_sampling_mnist_sigma20}, \ref{fig:initial_state_sampling_lsun_sigma20});
(ii) interpolation trajectories with Peclet ablations, illustrating transition dynamics under varying diffusion constraints (Figs.~\ref{fig:faces_interpolation}, \ref{fig:mnist_interpolation}, \ref{fig:lsun_faces_interpolation});
(iii) multiple initial-state sampling via stochastic generation from diversified initial states, emphasizing Pe's role in governing output variability across distinct trajectory initializations (Figs.~\ref{fig:visual_comp_ffhq_part1}, \ref{fig:visual_comp_ffhq_part2}, \ref{fig:visual_comp_mnist}, \ref{fig:lsun_visual_comp});
and (iv) uncurated interpolation demonstrating raw model behavior (Figs.~\ref{fig:faces_interpolation_16_0.8}, \ref{fig:faces_interpolation_20_0.8}, \ref{fig:mnist_interpolation_20_0.6}, \ref{fig:mnist_interpolation_20_0.8}, \ref{fig:lsun_faces_interpolation_0.4}, \ref{fig:lsun_faces_interpolation_0.6}). 

\renewcommand{\genfacelength}{0.19} %
\renewcommand{\gencolPelength}{0.13} %
\renewcommand{\gencolimglength}{0.78} %

\begin{figure*}[h]
  \centering
  \small
  \begin{subfigure}[c]{\textwidth}
    \begin{minipage}[c]{0.12\textwidth} %
      \centering
      \textbf{Pe}
    \end{minipage}%
    \begin{minipage}[c]{0.8\textwidth} %
      \begin{minipage}[c]{0.21\textwidth}
        \centering \textbf{Prior}
      \end{minipage}%
      \hfill
      \begin{minipage}[c]{0.74\textwidth}
        \centering 
        \hspace{-34pt}\textbf{Generated samples}
      \end{minipage}%
    \end{minipage}
  \end{subfigure}
  
  \begin{subfigure}[c]{\textwidth}
    \begin{minipage}[c]{\gencolPelength \textwidth}
      \centering
        0.00
    \end{minipage}%
    \begin{minipage}[c]{\gencolimglength \textwidth}
      \centering
      \includegraphics[width=\genfacelength \textwidth]{figures/results/FFHQ/FID_experiment_grapics/GRID_FOLDER_TURBAN/i_000.png}%
      \hfill
      \includegraphics[width=\genfacelength \textwidth]{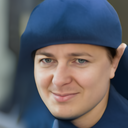}%
      \hfill
      \includegraphics[width=\genfacelength \textwidth]{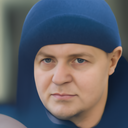}%
      \hfill
      \includegraphics[width=\genfacelength \textwidth]{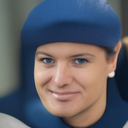}%
      \hfill
      \includegraphics[width=\genfacelength \textwidth]{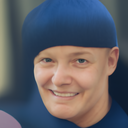}%
    \end{minipage}
  \end{subfigure}

  \begin{subfigure}[c]{\textwidth}
    \begin{minipage}[c]{\gencolPelength \textwidth}
      \centering
        0.02
    \end{minipage}%
    \begin{minipage}[c]{\gencolimglength \textwidth}
      \centering
      \includegraphics[width=\genfacelength \textwidth]{figures/results/FFHQ/FID_experiment_grapics/GRID_FOLDER_TURBAN/i_002.png}%
      \hfill
      \includegraphics[width=\genfacelength \textwidth]{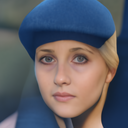}%
      \hfill
      \includegraphics[width=\genfacelength \textwidth]{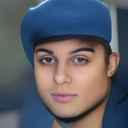}%
      \hfill
      \includegraphics[width=\genfacelength \textwidth]{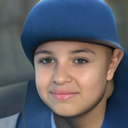}%
      \hfill
      \includegraphics[width=\genfacelength \textwidth]{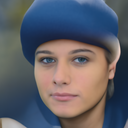}%
    \end{minipage}
  \end{subfigure}

  \begin{subfigure}[c]{\textwidth}
    \begin{minipage}[c]{\gencolPelength \textwidth}
      \centering
        0.04
    \end{minipage}%
    \begin{minipage}[c]{\gencolimglength \textwidth}
      \centering
      \includegraphics[width=\genfacelength \textwidth]{figures/results/FFHQ/FID_experiment_grapics/GRID_FOLDER_TURBAN/i_004.png}%
      \hfill
      \includegraphics[width=\genfacelength \textwidth]{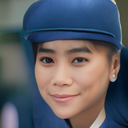}%
      \hfill
      \includegraphics[width=\genfacelength \textwidth]{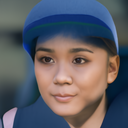}%
      \hfill
      \includegraphics[width=\genfacelength \textwidth]{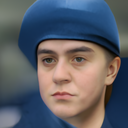}%
      \hfill
      \includegraphics[width=\genfacelength \textwidth]{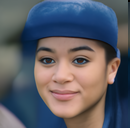}%
    \end{minipage}
  \end{subfigure}

  \begin{subfigure}[c]{\textwidth}
    \begin{minipage}[c]{\gencolPelength \textwidth}
      \centering
        0.06
    \end{minipage}%
    \begin{minipage}[c]{\gencolimglength \textwidth}
      \centering
      \includegraphics[width=\genfacelength \textwidth]{figures/results/FFHQ/FID_experiment_grapics/GRID_FOLDER_TURBAN/i_006.png}%
      \hfill
      \includegraphics[width=\genfacelength \textwidth]{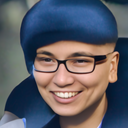}%
      \hfill
      \includegraphics[width=\genfacelength \textwidth]{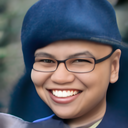}%
      \hfill
      \includegraphics[width=\genfacelength \textwidth]{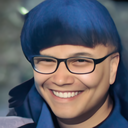}%
      \hfill
      \includegraphics[width=\genfacelength \textwidth]{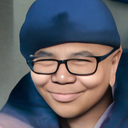}%
    \end{minipage}
  \end{subfigure}
  
  \begin{subfigure}[c]{\textwidth}
    \begin{minipage}[c]{\gencolPelength \textwidth}
      \centering
        0.08
    \end{minipage}%
    \begin{minipage}[c]{\gencolimglength \textwidth}
      \centering
      \includegraphics[width=\genfacelength \textwidth]{figures/results/FFHQ/FID_experiment_grapics/GRID_FOLDER_TURBAN/i_008.png}%
      \hfill
      \includegraphics[width=\genfacelength \textwidth]{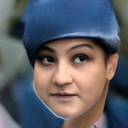}%
      \hfill
      \includegraphics[width=\genfacelength \textwidth]{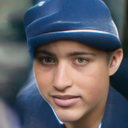}%
      \hfill
      \includegraphics[width=\genfacelength \textwidth]{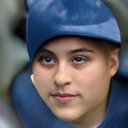}%
      \hfill
      \includegraphics[width=\genfacelength \textwidth]{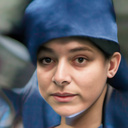}%
    \end{minipage}
  \end{subfigure}

  \begin{subfigure}[c]{\textwidth}
    \begin{minipage}[c]{\gencolPelength \textwidth}
      \centering
        0.10
    \end{minipage}%
    \begin{minipage}[c]{\gencolimglength \textwidth}
      \centering
      \includegraphics[width=\genfacelength \textwidth]{figures/results/FFHQ/FID_experiment_grapics/GRID_FOLDER_TURBAN/i_010.png}%
      \hfill
      \includegraphics[width=\genfacelength \textwidth]{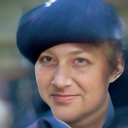}%
      \hfill
      \includegraphics[width=\genfacelength \textwidth]{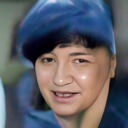}%
      \hfill
      \includegraphics[width=\genfacelength \textwidth]{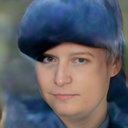}%
      \hfill
      \includegraphics[width=\genfacelength \textwidth]{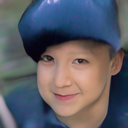}
    \end{minipage}
  \end{subfigure}

  \begin{subfigure}[c]{\textwidth}
    \begin{minipage}[c]{\gencolPelength \textwidth}
      \centering
        0.12
    \end{minipage}%
    \begin{minipage}[c]{\gencolimglength \textwidth}
      \centering
      \includegraphics[width=\genfacelength \textwidth]{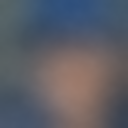}%
      \hfill
      \includegraphics[width=\genfacelength \textwidth]{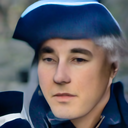}%
      \hfill
      \includegraphics[width=\genfacelength \textwidth]{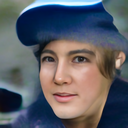}%
      \hfill
      \includegraphics[width=\genfacelength \textwidth]{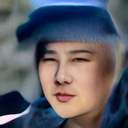}%
      \hfill
      \includegraphics[width=\genfacelength \textwidth]{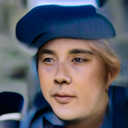}
    \end{minipage}
  \end{subfigure}

  \begin{subfigure}[c]{\textwidth}
    \begin{minipage}[c]{\gencolPelength \textwidth}
      \centering
        0.14
    \end{minipage}%
    \begin{minipage}[c]{\gencolimglength \textwidth}
      \centering
      \includegraphics[width=\genfacelength \textwidth]{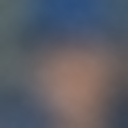}%
      \hfill
      \includegraphics[width=\genfacelength \textwidth]{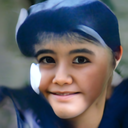}%
      \hfill
      \includegraphics[width=\genfacelength \textwidth]{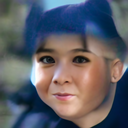}%
      \hfill
      \includegraphics[width=\genfacelength \textwidth]{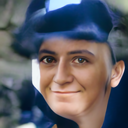}%
      \hfill
      \includegraphics[width=\genfacelength \textwidth]{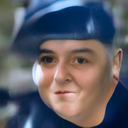}
    \end{minipage}
  \end{subfigure}
          
  \caption{Results for $\sigma=20$, showing inverse processes with varying Pe numbers. The image prior is consistent across rows for visual comparison, preserving the color palette. }
  \label{fig:initial_state_sampling_ffhq_sigma20}
\end{figure*}

\begin{figure*}[b]
\small
\begin{minipage}{\textwidth}
  \centering
  
\begin{minipage}{0.1\textwidth}
   \centering IHD\\Pe=$0.0$
  \end{minipage}%
  \begin{minipage}{0.9\textwidth}
    \centering
    \includegraphics[width=\textwidth]{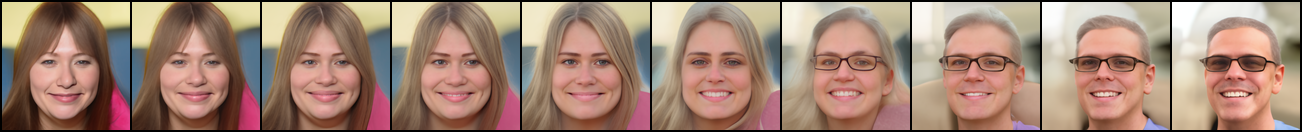}
  \end{minipage}
  \vspace{0.05cm} %

  \begin{minipage}{0.1\linewidth}
	\centering Our\\Pe=$0.02$
  \end{minipage}%
  \begin{minipage}{0.9\linewidth}
    \centering
    \includegraphics[width=\linewidth]{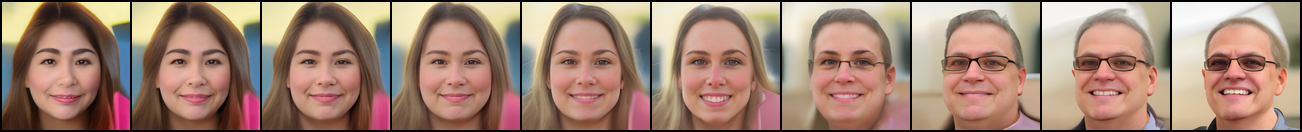}
  \end{minipage}
  \vspace{0.05cm} %
  
  \begin{minipage}{0.1\linewidth}
	\centering Our\\Pe=$0.04$
  \end{minipage}%
  \begin{minipage}{0.9\linewidth}
    \centering
    \includegraphics[width=\linewidth]{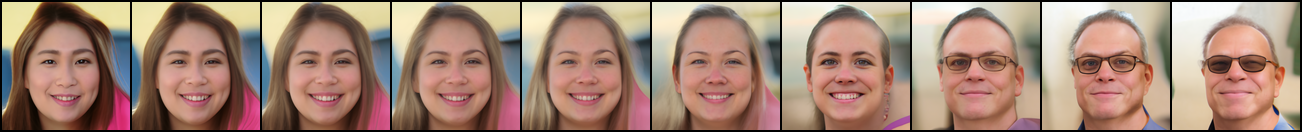}
  \end{minipage}
  \vspace{0.05cm} %
  
\begin{minipage}{0.1\linewidth}
	\centering Our\\Pe=$0.06$
  \end{minipage}%
  \begin{minipage}{0.9\linewidth}
    \centering
    \includegraphics[width=\linewidth]{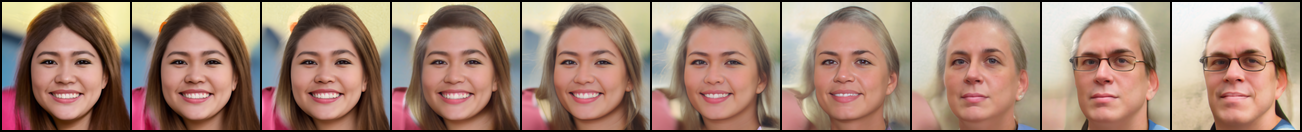}
  \end{minipage}
  \vspace{0.05cm} %
  
  \begin{minipage}{0.1\linewidth}
	\centering Our\\Pe=$0.08$
  \end{minipage}%
  \begin{minipage}{0.9\linewidth}
    \centering
    \includegraphics[width=\linewidth]{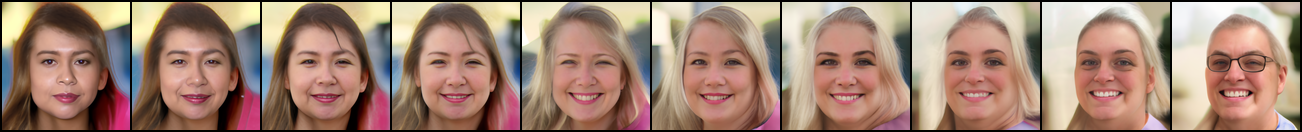}
\end{minipage}
  \vspace{0.05cm} %
  
\begin{minipage}{0.1\linewidth}
	\centering Our\\Pe=$0.10$
  \end{minipage}%
  \begin{minipage}{0.9\linewidth}
    \centering
    \includegraphics[width=\linewidth]{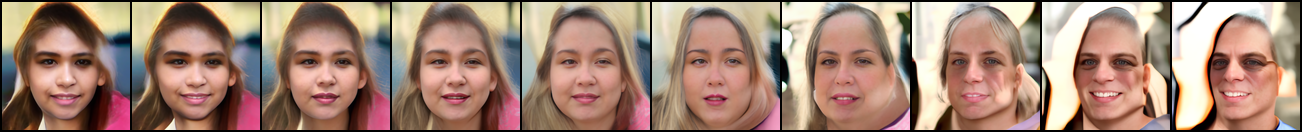}
  \end{minipage}
  \vspace{0.05cm} %

\begin{minipage}{0.1\linewidth}
	\centering Our\\Pe=$0.12$
  \end{minipage}%
  \begin{minipage}{0.9\linewidth}
    \centering
    \includegraphics[width=\linewidth]{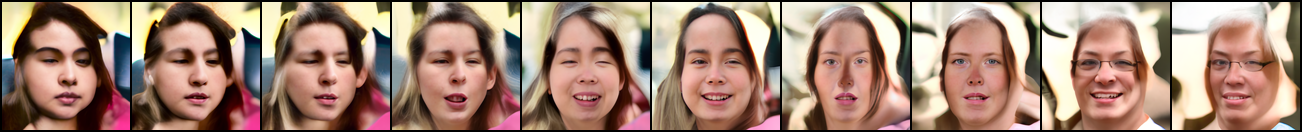}
  \end{minipage}
  \vspace{0.05cm} %

  \begin{minipage}{0.1\linewidth}
	\centering Our\\Pe=$0.14$
  \end{minipage}%
  \begin{minipage}{0.9\linewidth}
    \centering
    \includegraphics[width=\linewidth]{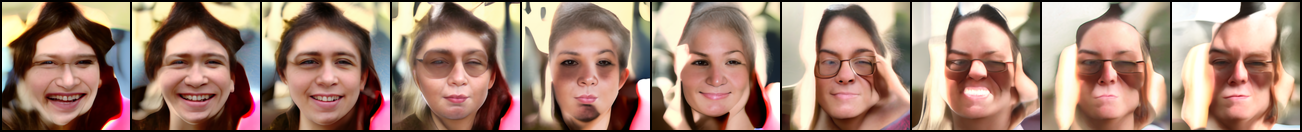}
  \end{minipage}
  \vspace{0.05cm} %
  
  \captionof{figure}{Visual comparison of interpolations between two FFHQ samples.
  Each undergoes the forward process up to $ \sigma =20$, followed by linear interpolation and denoising with SLERP-interpolated generative noise added at each step. 
  } %
  \label{fig:faces_interpolation} %
\end{minipage}
    
\end{figure*}

\setlength{\cellspacetoplimit}{0.5\tabcolsep}
\setlength{\cellspacebottomlimit}{\cellspacetoplimit}

\setlength{\cellspacetoplimit}{0.5\tabcolsep}
\setlength{\cellspacebottomlimit}{\cellspacetoplimit}

\clearpage

\renewcommand{\genfacelength}{0.2} %
\renewcommand{\gencolPelength}{0.135} %
\renewcommand{\gencolimglength}{0.82} %
\begin{figure*}[h]
  \centering
  \begin{subfigure}[c]{\textwidth}
    \begin{minipage}[c]{0.135\textwidth} %
      \centering
      \textbf{Pe}
    \end{minipage}%
    \begin{minipage}[c]{0.81\textwidth} %
      \begin{minipage}[c]{0.15\textwidth}
      \hspace{18pt}  \centering \textbf{Prior}
      \end{minipage}%
      \hfill
      \begin{minipage}[c]{0.41625\textwidth}
        \centering 
        \hspace{-150pt} \textbf{Generated samples}
      \end{minipage}%
    \end{minipage}
  \end{subfigure}
  
  \begin{subfigure}[c]{\textwidth}
    \begin{minipage}[c]{\gencolPelength \textwidth}
      \centering
       (IHD) \\ 0.00
    \end{minipage}%
    \begin{minipage}[c]{\gencolimglength \textwidth}
      \centering
      \includegraphics[trim={0 240pt 210pt 0}, clip, width=\genfacelength \textwidth]{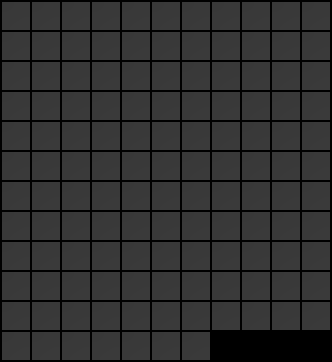}%
      \hfill
      \includegraphics[trim={0 30pt 420pt 0}, clip, width=0.79\textwidth]{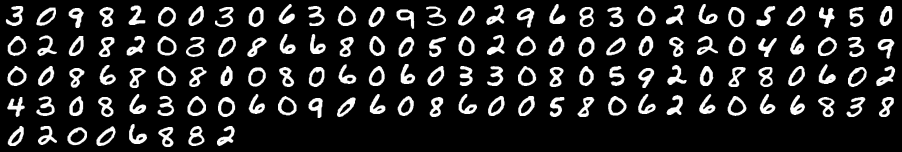}%
    \end{minipage}
  \end{subfigure}

  \begin{subfigure}[c]{\textwidth}
    \begin{minipage}[c]{\gencolPelength \textwidth}
      \centering
        0.02
    \end{minipage}%
    \begin{minipage}[c]{\gencolimglength \textwidth}
      \centering
      \includegraphics[trim={0 240pt 210pt 0}, clip, width=\genfacelength \textwidth]{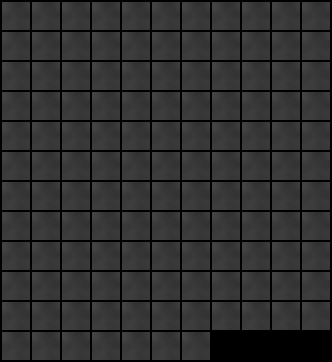}%
      \hfill
      \includegraphics[trim={0 30pt 420pt 0}, clip, width=0.79\textwidth]{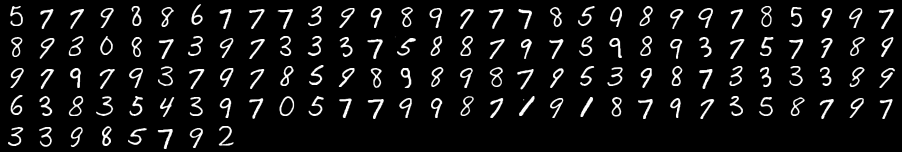}%
    \end{minipage}
  \end{subfigure}

  \begin{subfigure}[c]{\textwidth}
    \begin{minipage}[c]{\gencolPelength \textwidth}
      \centering
        0.04
    \end{minipage}%
    \begin{minipage}[c]{\gencolimglength \textwidth}
      \centering
      \includegraphics[trim={0 240pt 210pt 0}, clip, width=\genfacelength \textwidth]{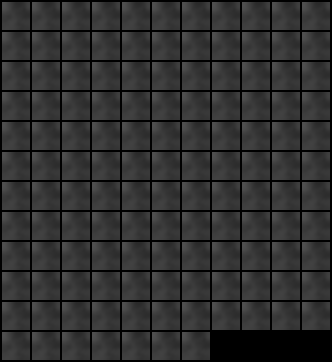}%
      \hfill
      \includegraphics[trim={0 30pt 420pt 0}, clip, width=0.79\textwidth]{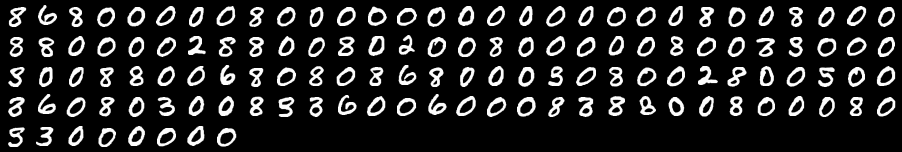}%
    \end{minipage}
  \end{subfigure}

  \begin{subfigure}[c]{\textwidth}
    \begin{minipage}[c]{\gencolPelength \textwidth}
      \centering
        0.06
    \end{minipage}%
    \begin{minipage}[c]{\gencolimglength \textwidth}
      \centering
      \includegraphics[trim={0 240pt 210pt 0}, clip, width=\genfacelength \textwidth]{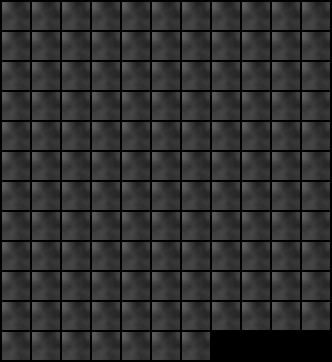}%
      \hfill
      \includegraphics[trim={0 30pt 420pt 0}, clip, width=0.79\textwidth]{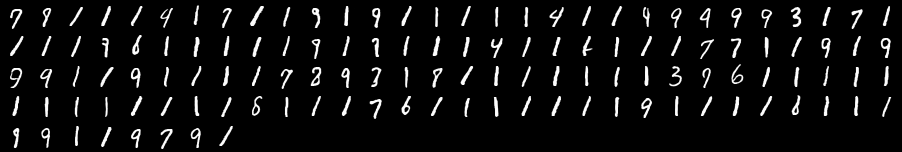}%
    \end{minipage}
  \end{subfigure}
  
  \begin{subfigure}[c]{\textwidth}
    \begin{minipage}[c]{\gencolPelength \textwidth}
      \centering
        0.08
    \end{minipage}%
    \begin{minipage}[c]{\gencolimglength \textwidth}
      \centering
      \includegraphics[trim={0 240pt 210pt 0}, clip, width=\genfacelength \textwidth]{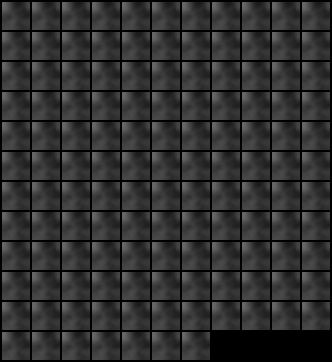}%
      \hfill
      \includegraphics[trim={0 30pt 420pt 0}, clip, width=0.79\textwidth]{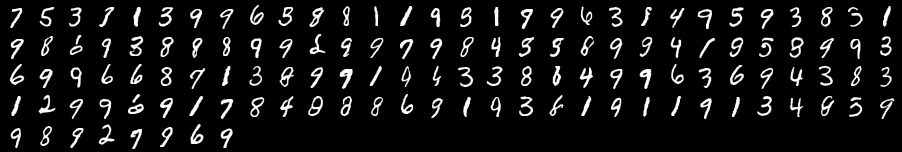}%
    \end{minipage}
  \end{subfigure}

  \begin{subfigure}[c]{\textwidth}
    \begin{minipage}[c]{\gencolPelength \textwidth}
      \centering
        0.10
    \end{minipage}%
    \begin{minipage}[c]{\gencolimglength \textwidth}
      \centering
      \includegraphics[trim={0 240pt 210pt 0}, clip, width=\genfacelength \textwidth]{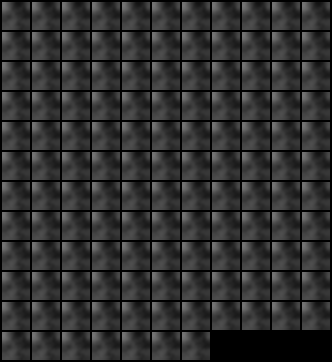}%
      \hfill
      \includegraphics[trim={0 30pt 420pt 0}, clip, width=0.79\textwidth]{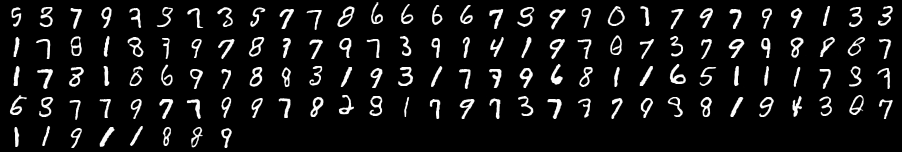}%
    \end{minipage}
  \end{subfigure}
  \caption{Additional samples with corresponding initial images from MNIST dataset, comparing different Peclet numbers. We can observe the impact of advection in forward process final step, that is the initial state for sampling.%
  }
  \label{fig:initial_state_sampling_mnist_sigma20}
\end{figure*}


\clearpage

\renewcommand{\genfacelength}{0.13} %
\renewcommand{\gencolPelength}{0.09} %
\renewcommand{\gencolimglength}{0.9} %

\begin{figure*}[h]
  \centering
  \small
  \begin{subfigure}[c]{\textwidth}
    \begin{minipage}[c]{0.085 \textwidth}%
    \centering
    \hspace{0pt} \textbf{Pe}
    \end{minipage}%
    \begin{minipage}[c]{\gencolimglength \textwidth}%
        \begin{minipage}[c]{0.19 \textwidth}%
        \centering
        \hspace{0pt} \textbf{Prior}
        \end{minipage}%
        \begin{minipage}[c]{0.75 \textwidth}%
        \centering 
        \hspace{0pt} \textbf{Generated samples}
        \end{minipage}%
    \end{minipage}%
  \end{subfigure}%
  \vspace{4pt}
  
  \begin{subfigure}[c]{\textwidth}
    \begin{minipage}[c]{\gencolPelength \textwidth}
    \centering
    0.00
    \end{minipage}%
    \begin{minipage}[c]{\gencolimglength \textwidth}
    \centering
    \includegraphics[width=\genfacelength \textwidth]{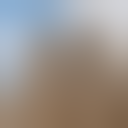}
    \hspace{2pt}
    \includegraphics[width=\genfacelength \textwidth]{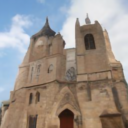}%
    \hspace{0.01pt}
    \includegraphics[width=\genfacelength \textwidth]{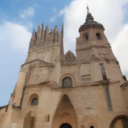}%
    \hspace{0.01pt}
    \includegraphics[width=\genfacelength \textwidth]{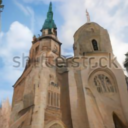}%
    \hspace{0.01pt}
    \includegraphics[width=\genfacelength \textwidth]{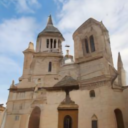}%
    \hspace{0.01pt}
    \includegraphics[width=\genfacelength \textwidth]{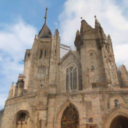}%
    \hspace{0.01pt}
    \includegraphics[width=\genfacelength \textwidth]{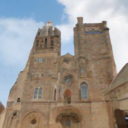}%
    \\[0.01pt]%
    \includegraphics[width=\genfacelength \textwidth]{figures/results/LSUN/grid/init_Pe_0e0.png}
    \hspace{2pt}
    \includegraphics[width=\genfacelength \textwidth]{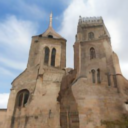}%
    \hspace{0.01pt}
    \includegraphics[width=\genfacelength \textwidth]{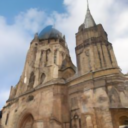}%
    \hspace{0.01pt}
    \includegraphics[width=\genfacelength \textwidth]{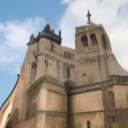}%
    \hspace{0.01pt}
    \includegraphics[width=\genfacelength \textwidth]{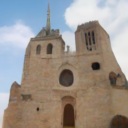}%
    \hspace{0.01pt}
    \includegraphics[width=\genfacelength \textwidth]{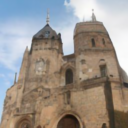}%
    \hspace{0.01pt}
    \includegraphics[width=\genfacelength \textwidth]{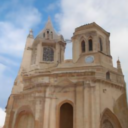}%
    \end{minipage}%
  \end{subfigure}%
  \vspace{4pt}
  
    \begin{subfigure}[c]{\textwidth}
    \begin{minipage}[c]{\gencolPelength \textwidth}
    \centering
    0.02
    \end{minipage}%
    \begin{minipage}[c]{\gencolimglength \textwidth}
    \centering
    \includegraphics[width=\genfacelength \textwidth]{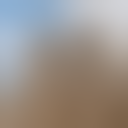}
    \hspace{2pt}
    \includegraphics[width=\genfacelength \textwidth]{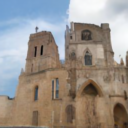}%
    \hspace{0.01pt}
    \includegraphics[width=\genfacelength \textwidth]{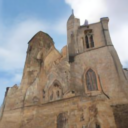}%
    \hspace{0.01pt}
    \includegraphics[width=\genfacelength \textwidth]{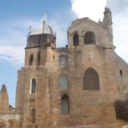}%
    \hspace{0.01pt}
    \includegraphics[width=\genfacelength \textwidth]{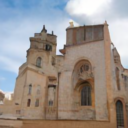}%
    \hspace{0.01pt}
    \includegraphics[width=\genfacelength \textwidth]{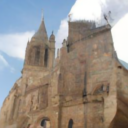}%
    \hspace{0.01pt}
    \includegraphics[width=\genfacelength \textwidth]{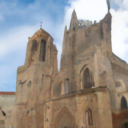}%
    \\[0.01pt]%
    \includegraphics[width=\genfacelength \textwidth]{figures/results/LSUN/grid/init_Pe_2e-2.png}
    \hspace{2pt}
    \includegraphics[width=\genfacelength \textwidth]{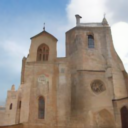}%
    \hspace{0.01pt}
    \includegraphics[width=\genfacelength \textwidth]{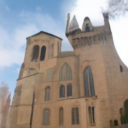}%
    \hspace{0.01pt}
    \includegraphics[width=\genfacelength \textwidth]{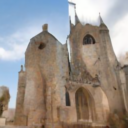}%
    \hspace{0.01pt}
    \includegraphics[width=\genfacelength \textwidth]{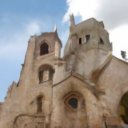}%
    \hspace{0.01pt}
    \includegraphics[width=\genfacelength \textwidth]{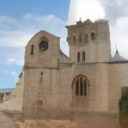}%
    \hspace{0.01pt}
    \includegraphics[width=\genfacelength \textwidth]{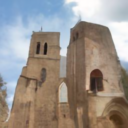}%
    \end{minipage}%
  \end{subfigure}%
  \vspace{4pt}
  
    \begin{subfigure}[c]{\textwidth}
    \begin{minipage}[c]{\gencolPelength \textwidth}
    \centering
    0.04
    \end{minipage}%
    \begin{minipage}[c]{\gencolimglength \textwidth}
    \centering
    \includegraphics[width=\genfacelength \textwidth]{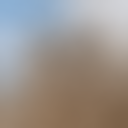}
    \hspace{2pt}
    \includegraphics[width=\genfacelength \textwidth]{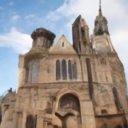}%
    \hspace{0.01pt}
    \includegraphics[width=\genfacelength \textwidth]{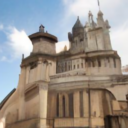}%
    \hspace{0.01pt}
    \includegraphics[width=\genfacelength \textwidth]{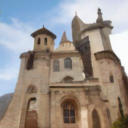}%
    \hspace{0.01pt}
    \includegraphics[width=\genfacelength \textwidth]{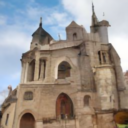}%
    \hspace{0.01pt}
    \includegraphics[width=\genfacelength \textwidth]{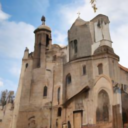}%
    \hspace{0.01pt}
    \includegraphics[width=\genfacelength \textwidth]{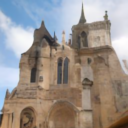}%
    \\[0.01pt]%
    \includegraphics[width=\genfacelength \textwidth]{figures/results/LSUN/grid/init_Pe_4e-2.png}
    \hspace{2pt}
    \includegraphics[width=\genfacelength \textwidth]{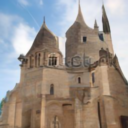}%
    \hspace{0.01pt}
    \includegraphics[width=\genfacelength \textwidth]{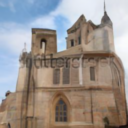}%
    \hspace{0.01pt}
    \includegraphics[width=\genfacelength \textwidth]{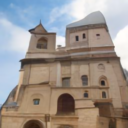}%
    \hspace{0.01pt}
    \includegraphics[width=\genfacelength \textwidth]{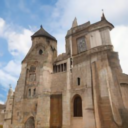}%
    \hspace{0.01pt}
    \includegraphics[width=\genfacelength \textwidth]{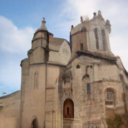}%
    \hspace{0.01pt}
    \includegraphics[width=\genfacelength \textwidth]{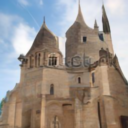}%
    \end{minipage}%
  \end{subfigure}%
  \vspace{4pt}
  
    \begin{subfigure}[c]{\textwidth}
    \begin{minipage}[c]{\gencolPelength \textwidth}
    \centering
    0.06
    \end{minipage}%
    \begin{minipage}[c]{\gencolimglength \textwidth}
    \centering
    \includegraphics[width=\genfacelength \textwidth]{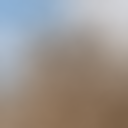}
    \hspace{2pt}
    \includegraphics[width=\genfacelength \textwidth]{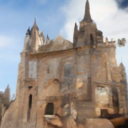}%
    \hspace{0.01pt}
    \includegraphics[width=\genfacelength \textwidth]{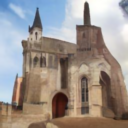}%
    \hspace{0.01pt}
    \includegraphics[width=\genfacelength \textwidth]{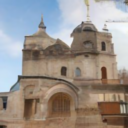}%
    \hspace{0.01pt}
    \includegraphics[width=\genfacelength \textwidth]{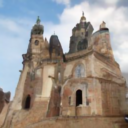}%
    \hspace{0.01pt}
    \includegraphics[width=\genfacelength \textwidth]{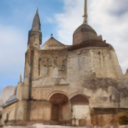}%
    \hspace{0.01pt}
    \includegraphics[width=\genfacelength \textwidth]{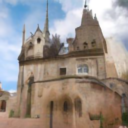}%
    \\[0.01pt]%
    \includegraphics[width=\genfacelength \textwidth]{figures/results/LSUN/grid/init_Pe_6e-2.png}
    \hspace{2pt}
    \includegraphics[width=\genfacelength \textwidth]{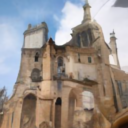}%
    \hspace{0.01pt}
    \includegraphics[width=\genfacelength \textwidth]{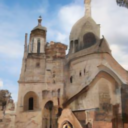}%
    \hspace{0.01pt}
    \includegraphics[width=\genfacelength \textwidth]{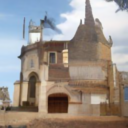}%
    \hspace{0.01pt}
    \includegraphics[width=\genfacelength \textwidth]{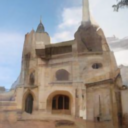}%
    \hspace{0.01pt}
    \includegraphics[width=\genfacelength \textwidth]{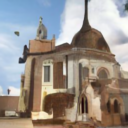}%
    \hspace{0.01pt}
    \includegraphics[width=\genfacelength \textwidth]{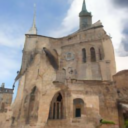}%
    \end{minipage}%
  \end{subfigure}%
    \vspace{4pt}
    
    \begin{subfigure}[c]{\textwidth}
    \begin{minipage}[c]{\gencolPelength \textwidth}
    \centering
    0.08
    \end{minipage}%
    \begin{minipage}[c]{\gencolimglength \textwidth}
    \centering
    \includegraphics[width=\genfacelength \textwidth]{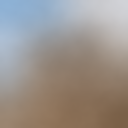}
    \hspace{2pt}
    \includegraphics[width=\genfacelength \textwidth]{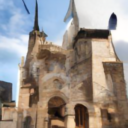}%
    \hspace{0.01pt}
    \includegraphics[width=\genfacelength \textwidth]{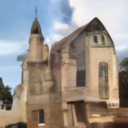}%
    \hspace{0.01pt}
    \includegraphics[width=\genfacelength \textwidth]{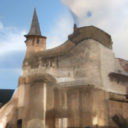}%
    \hspace{0.01pt}
    \includegraphics[width=\genfacelength \textwidth]{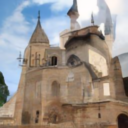}%
    \hspace{0.01pt}
    \includegraphics[width=\genfacelength \textwidth]{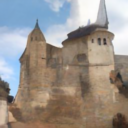}%
    \hspace{0.01pt}
    \includegraphics[width=\genfacelength \textwidth]{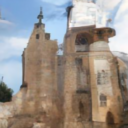}%
    \\[0.01pt]%
    \includegraphics[width=\genfacelength \textwidth]{figures/results/LSUN/grid/init_Pe_8e-2.png}
    \hspace{2pt}
    \includegraphics[width=\genfacelength \textwidth]{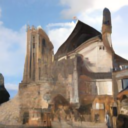}%
    \hspace{0.01pt}
    \includegraphics[width=\genfacelength \textwidth]{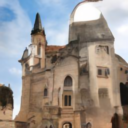}%
    \hspace{0.01pt}
    \includegraphics[width=\genfacelength \textwidth]{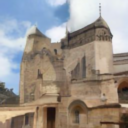}%
    \hspace{0.01pt}
    \includegraphics[width=\genfacelength \textwidth]{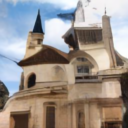}%
    \hspace{0.01pt}
    \includegraphics[width=\genfacelength \textwidth]{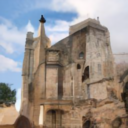}%
    \hspace{0.01pt}
    \includegraphics[width=\genfacelength \textwidth]{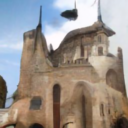}%
    \end{minipage}%
  \end{subfigure}%

  \caption{Results for $\sigma=20$, on the LSUN Church dataset, showing inverse processes with varying Pe numbers. The image prior is consistent across rows for visual comparison, preserving the color palette.}
  \label{fig:initial_state_sampling_lsun_sigma20}
\end{figure*}

\setlength{\cellspacetoplimit}{0.5\tabcolsep}
\setlength{\cellspacebottomlimit}{\cellspacetoplimit}

\clearpage

\section{Pseudocode}
\label{app:pseudocode}
We insert a pseudocode in Python to illustrate how the Spectral Turbulence Generator and Lattice Boltzmann Method (LBM) work internally.

\begin{lstlisting}
import torch as t

def tanh_limiter(x, min_val, max_val, sharpness=1.0):
    mid_val = (max_val + min_val) / 2
    range_val = (max_val - min_val) / 2
    return mid_val + range_val * t.tanh(sharpness * (x - mid_val) / range_val)

def limit_velocity_field(u, v, min_val, max_val):
    velocity_magnitude = t.sqrt(u**2 + v**2)
    limited_magnitude = tanh_limiter(velocity_magnitude, min_val, max_val)
    
    small_factor = 1E-9
    direction_factor = t.where(velocity_magnitude < small_factor, small_factor, limited_magnitude / velocity_magnitude)
    
    # Adjust u and v components to match the new limited magnitude while preserving direction
    upscale = 1. 
    direction_factor *=upscale
    u_limited = u * direction_factor
    v_limited = v * direction_factor

    return u_limited, v_limited
    
class SpectralTurbulenceGenerator(t.nn.Module):
    def __init__(self, std_dev_schedule):
        grid_size = (128, 128)
        self.std_dev = std_dev_schedule #turbulence intensity scaling factor 
        
        energy_slope = -2.0
        self.energy_spectrum = lambda k: t.where(t.isinf(k ** (energy_slope)), 0, k ** (energy_slope))
        self.frequency_range = {'k_min': 2.0 * t.pi / min(grid_size), 'k_max': 2.0 * t.pi / (min(self.domain_size) / 1024)}

        # Fourier transform wave numbers
        kx = (t.fft.fftfreq(grid_size[0], d=1/grid_size[0]) * 2 * t.pi).to('gpu')
        ky = (t.fft.fftfreq(grid_size[1], d=1/grid_size[1]) * 2 * t.pi).to('gpu')
        KX, KY = t.meshgrid(kx, ky)
        self.K = t.sqrt(KX**2 + KY**2).to('gpu')

        # Initialize the phases once and use them in each run
        self.phase_u = (t.rand(grid_size) * 2 * t.pi).to('gpu')
        self.phase_v = (t.rand(grid_size) * 2 * t.pi).to('gpu')
        
        self.amplitude = (t.where(self.K != 0, (self.energy_spectrum(self.K)), 0)).to('gpu')
        self.amplitude = (t.where((self.K >= self.frequency_range['k_min']) & (self.K <= self.frequency_range['k_max']), self.amplitude, 0.0)).to("gpu")

        dt_turb = 1E-4
        self.omega = dt_turb*self.K
        
    def generate_turbulence(self, time: int):
        u_hat = self.amplitude * t.exp(1j * (self.phase_u + self.omega * time))
        v_hat = self.amplitude * t.exp(1j * (self.phase_v + self.omega * time))
        u = t.real(t.fft.ifft2(u_hat))
        v = t.real(t.fft.ifft2(v_hat))
       
        if self.std_dev[time]< 1E-14:
            u,v = 0*self.K, 0*self.K #avoid division by 0 in t.std(u)
        else:
            u *= self.std_dev[time] / t.std(u)
            v *= self.std_dev[time] / t.std(v)
            
        u, v = limit_velocity_field(u, v, min_val=-1E-3, max_val=1E-3)

        return u.float(), v.float() 

\end{lstlisting}

\begin{lstlisting}
import taichi as ti
import taichi.math as tm

# Fluid solver based on lattice boltzmann method using taichi language
# Inspired by: https://github.com/hietwll/LBM_Taichi

@ti.data_oriented
class LBM_ADE_Solver():
    def __init__(self, config, turbulenceGenerator):
        self.nx, self.ny = config.domain_size  
        self.turbulenceGenerator = turbulenceGenerator

        self.cs2 = ti.field(ti.f32)(1./3.)
        self.omega_kin = ti.field(ti.f32, shape=self.max_iter[None])
        self.omega_kin.from_numpy(1.0 / (3.0* config.kin_visc + 0.5))
            
        self.rho = ti.field(float, shape=(self.nx, self.ny))
        self.vel = ti.Vector.field(2, float, shape=(self.nx, self.ny))

        self.f = ti.Vector.field(9, float, shape=(self.nx, self.ny))
        self.f_new = ti.Vector.field(9, float, shape=(self.nx, self.ny))
    
        self.Force = ti.Vector.field(2, float, shape=(self.nx, self.ny))
        self.w = ti.types.vector(9, float)(4, 1, 1, 1, 1, 1 / 4, 1 / 4, 1 / 4, 1 / 4) / 9.0
        self.e = ti.types.matrix(9, 2, int)(
            [0, 0],  [1, 0], [0, 1], [-1, 0],[0, -1],[1, 1], [-1, 1], [-1, -1], [1, -1])
        
    def init(self, np_image): 
        self.rho.from_numpy(np_image)
        self.vel.fill(0)
  
    def solve(self, iterations):
        for iteration in range(iterations):                
            self.stream()
            self.update_macro_var()
            self.collide_srt()      
            self.vel = self.turbulenceGenerator.generate_turbulence(iteration)
            self.apply_bounceback_boundary_condition()

    @ti.kernel
    def stream(self):
        for i, j in ti.ndrange(self.nx, self.ny):
            for k in ti.static(range(9)):
                ip = i - self.e[k, 0]
                jp = j - self.e[k, 1]
                self.f[i, j][k] = self.f_new[ip, jp][k] 
                
    @ti.kernel
    def update_macro_var(self): 
        for i, j in ti.ndrange((1, self.nx-1), (1,self.ny-1)):
            self.rho[i, j] = 0
            for k in ti.static(range(9)):
                self.rho[i, j] += self.f[i, j][k] 

    @ti.kernel
    def collide_srt(self):
        omega_kin = self.omega_kin[self.iterations_counter[None]]
        for i, j in ti.ndrange((1, self.nx - 1), (1, self.ny - 1)):
            for k in ti.static(range(9)):
                feq = self.f_eq(i, j)
                self.f_new[i, j][k] = (1. - omega_kin) * self.f[i, j][k] + feq[k] * omega_kin
        
    @ti.func 
    def f_eq(self, i, j):
        eu = self.e @ self.vel[i, j]
        uv = tm.dot(self.vel[i, j], self.vel[i, j])
        return self.w * self.rho[i, j] * (1 + 3 * eu + 4.5 * eu * eu - 1.5 * uv)

    @ti.func
    def apply_bounceback_core(self, i: int, j: int):
        tmp = ti.f32(0.0)          
        for k in ti.static([1,2,5,6]):
            tmp = self.f[i, j][k]
            self.f[i, j][k] = self.f[i, j][k+2]
            self.f[i, j][k+2] = tmp    

        for k in ti.static(range(9)):
             self.f_new[i, j][k] =  self.f[i, j][k]

    @ti.kernel
    def apply_bounceback_boundary_condition(self):
        for i in range(0, self.nx):
            self.apply_bounceback_core(i, 0) 
            self.apply_bounceback_core(i, self.ny-1) 
                 
        for j in range(1, self.ny-1):
            self.apply_bounceback_core(0, j) 
            self.apply_bounceback_core(self.nx-1, j)         
\end{lstlisting}

\end{document}